\documentclass[a4paper,11pt]{article}
 \pdfoutput=1
 \usepackage{lscape}
 \usepackage[all]{xy} 
 \usepackage{longtable} 
 \usepackage{jheppub} 
 \usepackage{graphicx}
 \usepackage{amssymb}
 \usepackage{amsmath}
\usepackage{mathtools}
\usepackage{listings}
\usepackage{dcolumn}
\usepackage{bm}
\usepackage{color}
\usepackage{multirow}
\usepackage{upquote} 
 
\newcommand{\met}{{/\!\!\! E_T}} 
 
\allowdisplaybreaks

\newcommand{\lsim}{{\;\raise0.3ex\hbox{$<$\kern-0.75em\raise-1.1ex\hbox{$\sim$}}\;}}
\newcommand{\gsim}{{\;\raise0.3ex\hbox{$>$\kern-0.75em\raise-1.1ex\hbox{$\sim$}}\;}}
\newcommand{\beq}{\begin{equation}}
\newcommand{\eeq}{\end{equation}}
\newcommand{\bea}{\begin{eqnarray}}
\newcommand{\eea}{\end{eqnarray}}
\mathchardef\minus="002D

\title{\boldmath  Identifying Phase Space Boundaries with Voronoi Tessellations}
\author[a]{Dipsikha Debnath,}
\author[b]{James~S.~Gainer,}
\author[c]{Can Kilic,}
\author[a,d]{Doojin Kim{${}^1$}\note{Corresponding author: {\tt
      immworry@gmail.com}},}
\author[a]{Konstantin~T.~Matchev,}
\author[c]{and Yuan-Pao Yang}

\affiliation[a]{Physics Department, University of Florida, Gainesville, FL
  32611, U.S.A.}
\affiliation[b]{Department of Physics and Astronomy, University of Hawaii,
  Honolulu, HI 96822, U.S.A.}
\affiliation[c]{Theory Group, Department of Physics and Texas Cosmology Center,
  The University of Texas at Austin, Austin, TX 78712, U.S.A.} 
\affiliation[d]{CERN, Theory Division, CH-1222 Geneva 23, Switzerland}

\abstract{
  Determining the masses of new physics particles appearing in decay chains is an
  important and longstanding problem in high energy phenomenology.  Recently
  it has been shown that these mass measurements can be improved by utilizing
  the boundary of the allowed region in the fully differentiable phase space in its full dimensionality.  
  Here we show that the practical challenge of identifying this boundary can be
  solved using techniques based on the geometric properties of the cells
  resulting from Voronoi tessellations of the relevant data. The robust detection of such 
  phase space boundaries in the data could also be used to corroborate a new physics
  discovery based on a cut-and-count analysis.
}

\date{June 8, 2016}
\preprint{
\begin{flushleft} 
UTTG-04-16 \\
UH511-1256-2015
\end{flushleft} 
}

\begin{document} 
\maketitle
\flushbottom

\section{Introduction}
\label{sec:introduction}

Voronoi tessellations~\cite{Voronoi} are useful in a wide variety of fields, from
biology~\cite{biology} to astronomy~\cite{Ramella:2001yt,Cappellari:2003ir} 
to condensed matter physics~\cite{bowick}. In high energy physics, 
they have been used rather sporadically, e.g., as an optional approach to 
QCD jet-finding and area determination in {\sc FastJet}~\cite{Cacciari:2011ma} 
and in the model-independent definition of search regions in {\sc SLEUTH}
~\cite{Abbott:2000fb, Abbott:2001ke, Aaltonen:2007dg, Aaltonen:2008vt}.
In Ref.~\cite{Debnath:2015wra}, some of us pointed out that Voronoi methods can 
be applied directly to the analysis of data from high energy physics 
experiments, e.g., when trying to detect the presence of a new physics 
signal in the data or to perform parameter measurements.

In most Voronoi-based approaches, the goal is to use Voronoi tessellations 
to identify ``neighbors'' of data points. The tessellation then automatically 
provides a number of cell-based attributes for each data point.
Ref.~\cite{Debnath:2015wra} argued that using the geometric properties of 
Voronoi cells, and, in particular, functions of the geometric properties of 
Voronoi cells and their neighbors, gives valuable additional information 
and can allow for relatively model-independent searches for targeted 
``features'' in the data. As briefly discussed in
Ref.~\cite{Debnath:2015wra}, a particularly useful application is the study of
kinematic edges when investigating cascade decays in new physics models such
as supersymmetry (SUSY)~\cite{Martin:1997ns}.  

To understand the importance of edge-finding in multidimensional spaces
for SUSY mass measurement, we first note that 
many extensions of the standard model (SM) are characterized by
a $\mathbb{Z}_2$ symmetry under which new physics particles (NPPs) are 
charged but the SM particles are uncharged.
Such a symmetry ensures that the lightest NPP will be
stable and hence may constitute the dark matter.
With the assumption of such a symmetry, a typical collider event involving
NPPs proceeds as follows:
\begin{enumerate}
\item NPPs are pair produced.
\item Each NPP goes through a series of (generally two and three-body) decays
called a ``decay chain''.
In each decay, an NPP decays to another, lighter, NPP,
and one or more SM particles.
The NPPs generally have a small intrinsic width compared with their mass.
Hence it is generally a good approximation to view the decay chain
as consisting of a series of on-shell decays of NPPs.
\item Eventually the lightest particle charged under the $\mathbb{Z}_2$ is
  reached.  It is stable, and, if a dark matter candidate, uncharged and
  uncolored.  Hence it will escape the detector without being detected.
\end{enumerate}
Popular new physics models within this paradigm include SUSY,
where the $\mathbb{Z}_2$ symmetry is called ``$R$-parity'';
Universal Extra Dimensions (UED)~\cite{Appelquist:2000nn}, in which the
$\mathbb{Z}_2$ symmetry is called ``KK-parity''; and 
Little Higgs models, in which the $\mathbb{Z}_2$ symmetry is called 
``$T$-parity''~\cite{Cheng:2003ju}.

As the lightest NPP escapes detection, we are not able to determine directly 
the masses of the {\em initial} new physics particles produced in
the collision, nor the masses of any {\em intermediate} particles in the decay, 
as we would if we were studying a resonance decaying to visible particles.  
However, we \textit{can} determine the masses of the NPPs by studying the 
distributions of (functions of) the momenta of observed 
particles~\cite{Barr:2010zj}.

Much effort has gone into determining the best way to actually perform
this mass measurement. The simplest methods involve finding an edge or an 
endpoint in the one-dimensional distribution of the invariant mass of two 
(or more) reconstructed objects\footnote{From a theorist's point of view, 
those represent Standard Model (SM) particles visible in the detector.} 
~\cite{Hinchliffe:1996iu, Bachacou:1999zb, Allanach:2000kt,
Lester:2001zx, Gjelsten:2004ki, Gjelsten:2005aw, Lester:2005je, Miller:2005zp, 
Lester:2006yw, Ross:2007rm,Barr:2008ba, Barr:2008hv,Cho:2012er}. 
If one is able to measure enough of these kinematic endpoints, it is then 
possible to solve for the unknown masses, possibly up to discrete ambiguities
~\cite{Gjelsten:2005sv,Burns:2009zi,Matchev:2009iw}. 
This approach naturally evolved into the so-called ``polynomial method''
~\cite{Hinchliffe:1998ys,Nojiri:2003tu,Kawagoe:2004rz,Cheng:2007xv, 
Nojiri:2008ir, Cheng:2008mg, Cheng:2009fw, Webber:2009vm, Autermann:2009js, 
Kang:2009sk, Nojiri:2010dk, Kang:2010hb, Hubisz:2010ur, Cheng:2010yy, 
Gripaios:2011jm},
where one attempts to solve explicitly for the momenta of the invisible 
particles in a given event, possibly using additional information from prior 
measurements of kinematic invariant mass endpoints.
Since at hadron colliders the longitudinal momenta of the initial state partons 
are unknown, much effort went into the development of suitable 
``transverse variables''
\cite{Lester:1999tx, Barr:2003rg, Baumgart:2006pa,
Lester:2007fq, Tovey:2008ui, Serna:2008zk, Nojiri:2008vq, Cho:2008tj, 
Cheng:2008hk, Burns:2008va, Choi:2009hn, Matchev:2009ad,Polesello:2009rn, 
Konar:2009wn, Cho:2009ve, Nojiri:2010mk},
which are Lorentz-invariant under longitudinal boosts\footnote{For an 
alternative approach, see Refs.~\cite{Agashe:2012bn, Agashe:2013eba,Agashe:2015wwa,Agashe:2015ike}.}.
In principle, the optimal approach to mass measurement is provided
by the so-called ``Matrix Element Method'' (MEM),
\cite{Kondo:1988yd,Dalitz:1991wa, Abazov:2004cs, Alwall:2009sv, 
Artoisenet:2010cn, Alwall:2010cq, Fiedler:2010sg, Gainer:2013iya}.
However, its use is often computationally prohibitive, especially
when dealing with complicated final states and/or large reducible
backgrounds.
Many of the approaches described above have been extended in various ways, 
e.g., the $M_{T2}$ kink method allows the measurement of the mass of the lightest NPP 
~\cite{Cho:2007qv, Gripaios:2007is, Barr:2007hy, Cho:2007dh, Nojiri:2008hy, 
Barr:2009jv,Matchev:2009fh,Konar:2009qr},
and useful $3+1$-dimensional analogues of the ``transverse" invariant mass 
variables have been suggested~\cite{Konar:2008ei, Konar:2010ma, 
Barr:2011xt, Robens:2011zm, Mahbubani:2012kx, Bai:2013ema, Cho:2014naa, 
Cho:2014yma, Cho:2015laa, Konar:2015hea}.

\begin{figure}[t]
\begin{center}
\includegraphics[width=0.5\textwidth]{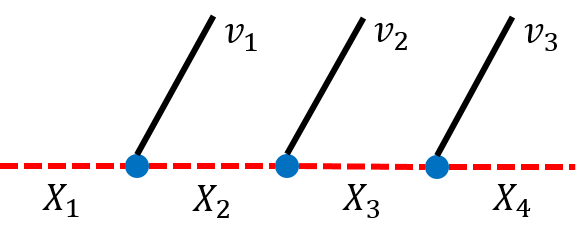} 
\end{center}
\caption{Decay process of a heavy resonance $X_1$ into three visible 
particles, $v_1$, $v_2$ and $v_3$, along with an invisible particle, $X_4$, via two on-shell 
intermediate states, $X_2$ and $X_3$. The NPPs, $X_i$, are denoted by red 
dashed lines while visible SM particles are denoted by black solid lines.}
\label{fig:topology}
\end{figure}

The approach to mass measurement taken here seeks to improve on those 
described in the existing literature in the following ways:
\begin{enumerate}
\item Instead of finding edges or endpoints in the one-dimensional distribution 
of a single variable, we will attempt to determine the boundary of the signal 
region in a higher-dimensional phase space.
This improves on one-dimensional methods, in increasing greatly
the amount of information that can be extracted from the data \cite{Agrawal:2013uka}.\footnote{At the same time, 
this approach avoids the drawbacks of 
some of the more model-dependent, process-specific, and computationally intensive methods, 
of which the MEM is perhaps the paradigmatic example.} 
To be specific, we shall consider the classic SUSY decay chain of 
three successive two-body decays as shown in Fig.~\ref{fig:topology}.
From the measured four-momenta, $p_i~ (i=1,2,3)$, of the three visible particles, 
$v_i$, in the decay, one can form three two-body invariant mass combinations, 
$m_{ij}=\sqrt{(p_i+p_j)^2}$, for $i<j$. Signal events will then 
populate the interior of a compact region, ${\cal V}_3$, 
in the three-dimensional phase space, $(m_{12}, m_{23}, m_{13})$, 
of invariant masses~\cite{Costanzo:2009mq,Kim:2015bnd}. 
The size and the shape of the {\em two-dimensional} surface boundary
of ${\cal V}_3$, which we term ${\cal S}_2$, contains the complete 
information about the underlying mass spectrum, $m_{X_i} (i=1,2,3,4),$ 
of the NPPs in Fig.~\ref{fig:topology}.
Therefore, we shall focus on methods for detecting ${\cal S}_2$ 
directly.\footnote{One can also project the allowed 
three-dimensional phase space ${\cal V}_3$ 
onto the subspace of two variables, say $(m_{12},m_{13})$, and obtain 
a corresponding {\em two-dimensional} allowed phase space ${\cal V}_2$ 
whose {\em one-dimensional} boundary ${\cal S}_1$ can be similarly used 
for mass measurements and disambiguation~\cite{Costanzo:2009mq, 
Burns:2009zi, Matchev:2009iw, Matchev:2009ad}.
With regards to utilizing the full amount of information contained 
in the data, this approach stands midway between the traditional method 
of using one-dimensional distributions of single variables and 
the three-dimensional approach considered here.}
One can imagine doing this in two ways:
\begin{itemize}
\item By defining a kinematic variable which takes the same (constant) value 
(e.g., zero) everywhere along the phase space boundary.\footnote{Compare 
to the ``singularity coordinate" defined in Ref.~\cite{Kim:2009si}.}
This approach will be discussed below in section~\ref{sec:phase-space}, 
where we review the relevant variable, $\Delta_4$, introduced in Ref.~\cite{
Agrawal:2013uka}. 
\item By analyzing the measured density of events in phase space and 
locating the boundary, ${\cal S}_2$, directly using techniques inspired 
by spatial analyses performed in other fields of science.
This will be the main subject of this paper, and will be discussed in 
sections~\ref{sec:3dtoy}, \ref{sec:enhanced3dtoy}, 
\ref{sec:sphere}, and \ref{sec:phase-space-with-voronoi}.
\end{itemize}
\item We build on the idea of Ref.~\cite{Debnath:2015wra} that Voronoi 
tessellations provide a powerful and model-independent tool for identifying 
edges (for a brief introduction to Voronoi tessellations, see section~\ref{sec:2d} below). 
While the analysis of Ref.~\cite{Debnath:2015wra} 
was limited to data in two dimensions, here we extend the procedure 
to the three-dimensional case and try to delineate the region, ${\cal V}_3$, 
in the phase space of the three variables, $(m_{12}, m_{23}, m_{13})$.
Before tackling a SUSY physics example in 
section~\ref{sec:phase-space-with-voronoi}, we consider several analogous 
toy examples of increasing complexity in sections~\ref{sec:voronoi} and
\ref{sec:phase-space}.  This helps develop
the reader's intuition and motivates some of our analysis choices.
Following Ref.~\cite{Debnath:2015wra}, in order to select ``edge" cells 
in the Voronoi tessellation of the data, we
consider the relative standard deviation\footnote{In statistics, 
the RSD is also known as the coefficient of variation (CV).} 
(RSD), $\bar\sigma_i$, of the volumes of neighboring cells, which is defined 
as follows.  Let $N_i$ be the set of neighbors of the $i$-th Voronoi cell, 
with volumes, $\{v_j\}$, for $j\in N_i$.  The RSD, $\bar\sigma_i$, is now defined by
\begin{equation}
\label{defvar}
\bar{\sigma}_i \equiv 
\frac{1}{\langle {v}(N_i)\rangle}\, \sqrt{\sum_{j\in N_i} \frac{\left(v_j-\langle {v}(N_i)\rangle \right)^2}{|N_i|-1}},
\end{equation}
where we have normalized by the average volume of the set of neighbors, $N_i$, of the $i$-th cell
\begin{equation}
\langle {v}(N_i)\rangle \equiv \frac{1}{|N_i|}\sum_{j\in N_i} v_j.
\label{ave-vol-neighbors}
\end{equation}

The variable defined in eq.~(\ref{defvar}) is a straightforward
extension to three dimensions of the ``scaled standard deviation'' of neighbor
areas found to be helpful in Ref.~\cite{Debnath:2015wra}.
\item We make crucial use of the recent observation of 
Ref.~\cite{Agrawal:2013uka} that for sufficiently many-body final states there 
is an enhancement
(in fact a singularity) in the phase space density 
near the boundary, ${\cal S}_2$, of the allowed phase space, ${\cal V}_3$.
Due to the enhancement in the density of signal events near the boundary
of phase space, we can alternatively target the boundary points of ${\cal V}_3$
as being points in a densely populated region. This motivates
us to consider, in addition to $\bar\sigma_i$, a second variable related to volume.
We choose
\begin{equation}
\label{eq:sv-scaled-def}
    \bar{v}_i = \frac{v_i} {\langle {v}(N_i)\rangle},
\end{equation}
where again we normalize by the average volume (\ref{ave-vol-neighbors}) 
of the set of neighbors, $N_i$.
\end{enumerate}

In what follows, we will therefore focus on the two Voronoi-based dimensionless
variables (\ref{defvar}) and (\ref{eq:sv-scaled-def}).
The former is motivated by the discontinuity in the density of events 
at the boundary~\cite{Debnath:2015wra},
while the latter is motivated by the enhancement in the density of signal 
events at the boundary~\cite{Agrawal:2013uka}.
We will find that the judicious combination of these two variables yields 
a significant increase in sensitivity as compared with either variable in 
isolation.  As a result we find that we are able to identify the 
boundary, ${\cal S}_2$, of the allowed signal phase space, ${\cal V}_3$,
with a high degree of accuracy, even when the ratio of signal to 
background events, $S/B$, is relatively small.

To support these conclusions, as well as to explain in detail to the reader
the methods employed, we proceed as follows.  In section~\ref{sec:voronoi} we
will provide a brief, but sufficient, review of Voronoi tessellation methods
and the use of the geometric properties of Voronoi cells for identifying features
in high energy physics data.  Section~\ref{sec:4body} will review the
consideration of events in multi-dimensional phase space, and, especially the
observation that, for sufficiently many dimensions, the differential phase space 
volume is highly peaked near the boundary.  
Then in sections~\ref{sec:enhanced3dtoy} and \ref{sec:sphere}
we study the efficacy of Voronoi methods for finding a densely populated 
{\em spherical} boundary in a generalized ``phase space'',
while section~\ref{sec:phase-space-with-voronoi} will examine the application 
of these methods to an actual benchmark point.  
We present our conclusions in section~\ref{sec:conclusions}.
Throughout our studies, we will use ROC curves to quantify the sensitivity of the
variables we define; we briefly review and discuss this approach in
Appendix~\ref{sec:ROC-and-ROLL}.  

\section{Voronoi Methods for Finding Boundaries}
\label{sec:voronoi}

Voronoi tessellation~\cite{voronoi} refers to the procedure,
previously proposed by Dirichlet~\cite{dirichlet} and hinted at
by Descartes~\cite{descartes}, through which an $n$-volume
containing a set of $N_d$ data points, $\{d_i\}$, is divided into $N_d$
non-overlapping regions, $\{R_i\}$, such that $d_i \in R_i, \forall i$.  The boundaries
of $R_i$ are chosen such that, for every point in some region $R_j$,
the corresponding data point, $d_j$, is the nearest data point.

For applications in high energy physics, we consider the data points to be events 
in a suitably chosen phase space\footnote{Here we are following a slightly 
confusing (but standard) usage and using the term ``phase space'' to refer to the space of 
$n$-tuple values of $n$ observables used to categorize events.  This set of
observables may not be sufficient to totally specify the kinematics of the
event.  In section~\ref{sec:phase-space}, ``phase space'' will have a more
precise meaning.}. It is important to make a judicious choice of 
phase space --- on the one hand, it should be of low enough dimensionality 
to keep the problem tractable in practice, yet the dimensional reduction should not result 
in the loss of any useful information, e.g., the washing out of interesting 
features in the 
underlying phase space distributions. Consider, as an example, the decay chain
of Fig.~\ref{fig:topology}. In general, the inclusive production of the $X_1$ resonance 
will be described by a 9-dimensional phase space, consisting, e.g., of the nine
momentum components of the visible particles, $v_i$. However, three of those 
degrees of freedom correspond to uninteresting Lorentz boosts of $X_1$,
another three degrees of freedom are angular variables which are sensitive 
to spins but not the $X_i$ mass spectrum, leaving only the three  
degrees of freedom relevant to a mass measurement. As already mentioned in the 
introduction, we can take these three degrees of freedom to be the invariant mass quantities
$(m_{12}, m_{23}, m_{13})$. We shall present the results from our analysis of 
this physics example in section~\ref{sec:phase-space-with-voronoi},
but we first begin with a few toy studies.

\subsection{Voronoi tessellations in two dimensions}
\label{sec:2d}

In order to make contact with Ref.~\cite{Debnath:2015wra} and to introduce 
our notation, 
we begin by studying several simplified scenarios in two dimensions. 
In the next section,~(\ref{sec:3dtoy}), we will generalize the approaches
taken and the lessons learned here to the case of three dimensions.

\subsubsection{A linear boundary in two dimensions}
\label{sec:2dlinear}

\begin{figure}[t]
\centering
\includegraphics[height=7cm]{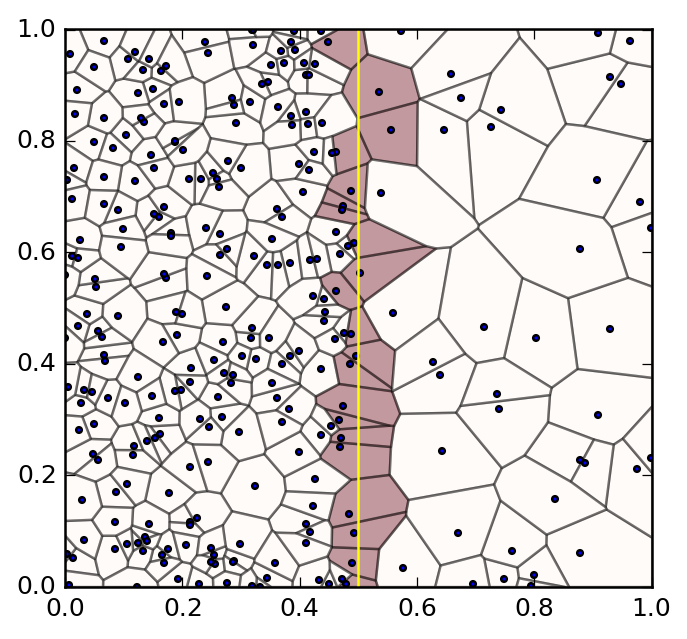}
\caption{\label{fig:2DVoronoi} 
$N_{events}=280$ events distributed according to (\ref{fstep}) with $\rho=6$ 
and the respective Voronoi tessellation.
The shaded cells are those crossed by the boundary (vertical yellow line)
and are defined to be the ``edge cells''. 
}
\end{figure}

In Fig.~\ref{fig:2DVoronoi}, we consider the unit square in the first quadrant 
($x\ge 0, y\ge 0$) and simulate $N_{events}=280$ events (data points) 
according to the probability distribution
\beq
f(x,y) = 
\frac{2}{1+\rho}
\left[ \rho H(0.5-x) + H(x-0.5) \right] ,
\label{fstep}
\eeq
where $H(x)$ is the Heaviside step function and $\rho$ is a constant density 
ratio, taken in Fig.~\ref{fig:2DVoronoi} to be $\rho=6$. The meaning of 
the distribution (\ref{fstep}) is very simple: the unit square is divided 
into two equal halves by the vertical boundary at $x=0.5$
(the yellow line in Fig.~\ref{fig:2DVoronoi}). Within each half, the density 
is constant (on average), but the left region is denser by a factor of $\rho$. 
This setup produces an edge at $x=0.5$, where the density changes by a factor 
of $\rho$. Our goal will be to detect this edge by tagging the 
Voronoi cells that are crossed by the boundary line --- such cells from now 
on will be referred to as ``edge cells'' and in Fig.~\ref{fig:2DVoronoi} 
they are shaded in brown.  The remaining Voronoi cells away from the edge 
will be referred to as ``bulk" cells, and in Fig.~\ref{fig:2DVoronoi} 
they are left white.

\begin{figure}[t]
\centering
\includegraphics[height=5.6cm]{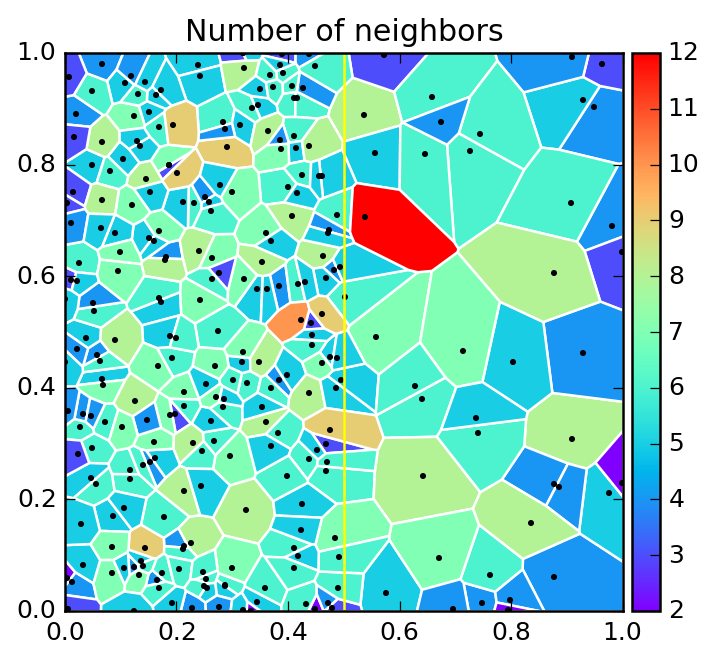}
\includegraphics[height=5.6cm]{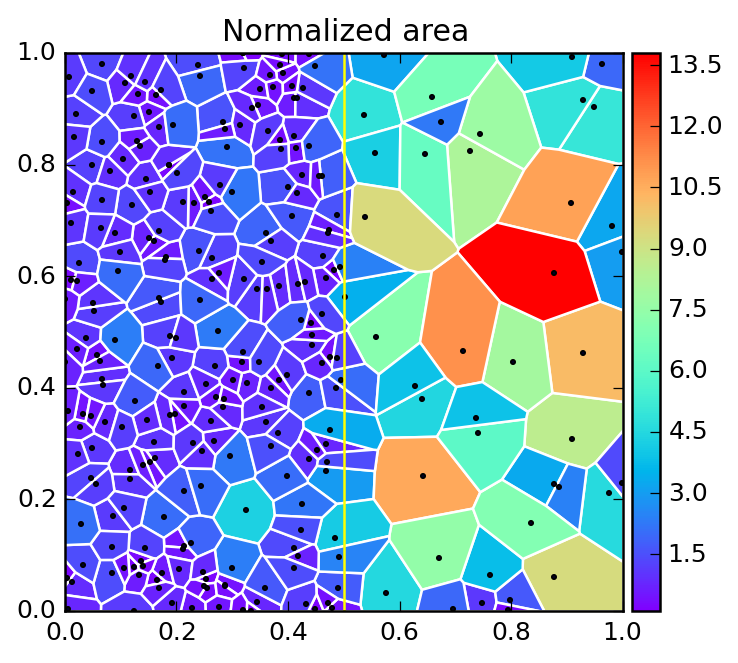}  \\
\includegraphics[height=5.6cm]{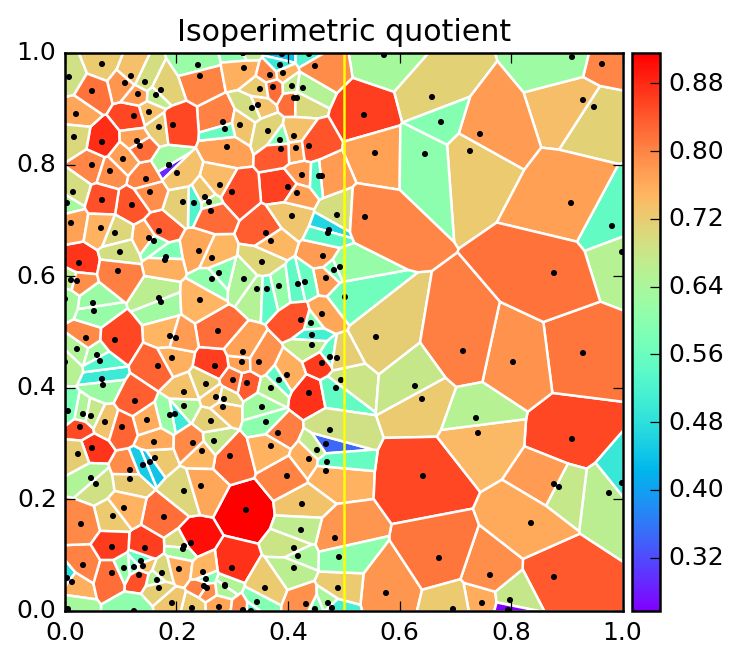}
\includegraphics[height=5.6cm]{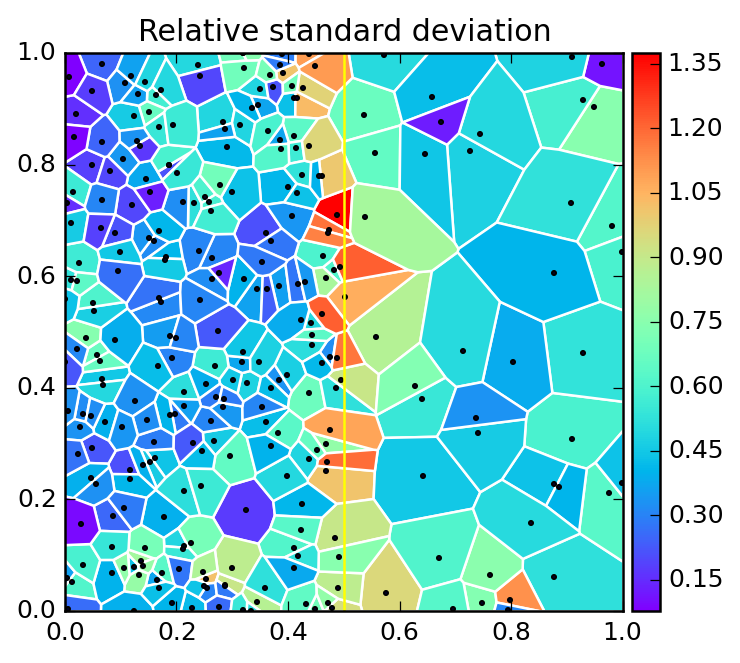}
\caption{\label{fig:plots2D} 
The Voronoi tessellation shown in Fig.~\ref{fig:2DVoronoi}, with cells color-coded according to
the number of neighboring polygons (upper left), normalized area (\ref{area_norm}) (upper right),
isoperimetric quotient (\ref{isoqdef}) (lower left), and RSD (\ref{defvar_area}) (lower right).
}
\end{figure}

\begin{figure}[t]
\centering
\includegraphics[width=5.6cm]{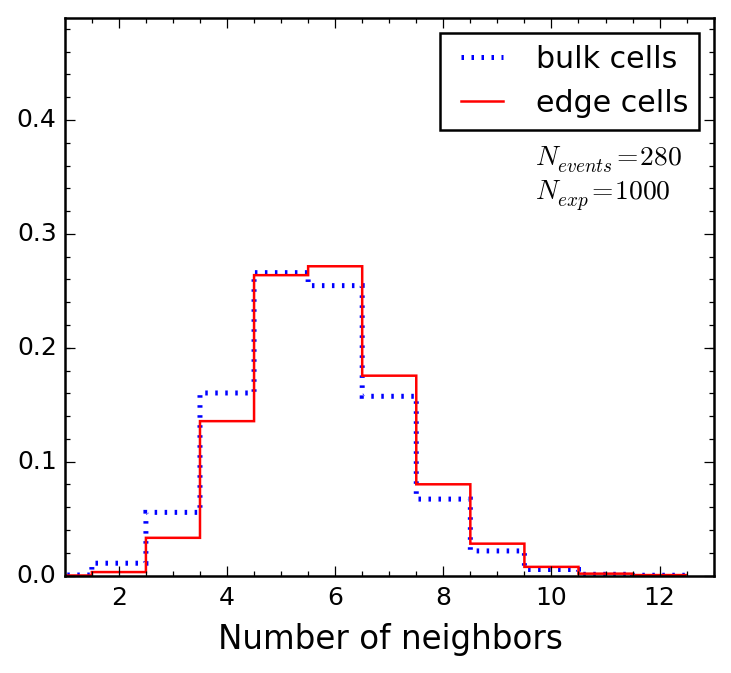} 
\includegraphics[width=5.65cm]{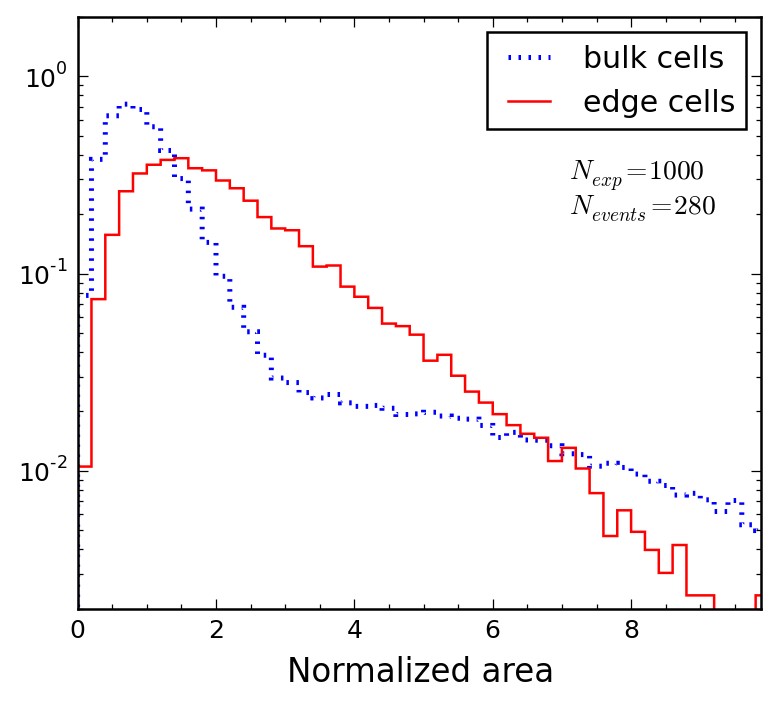} \\
\includegraphics[width=5.6cm]{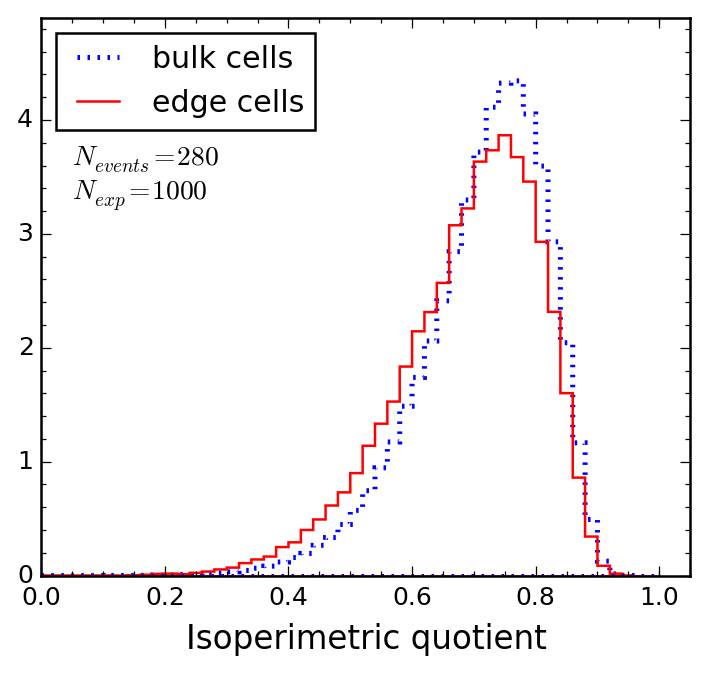} 
\includegraphics[width=5.8cm]{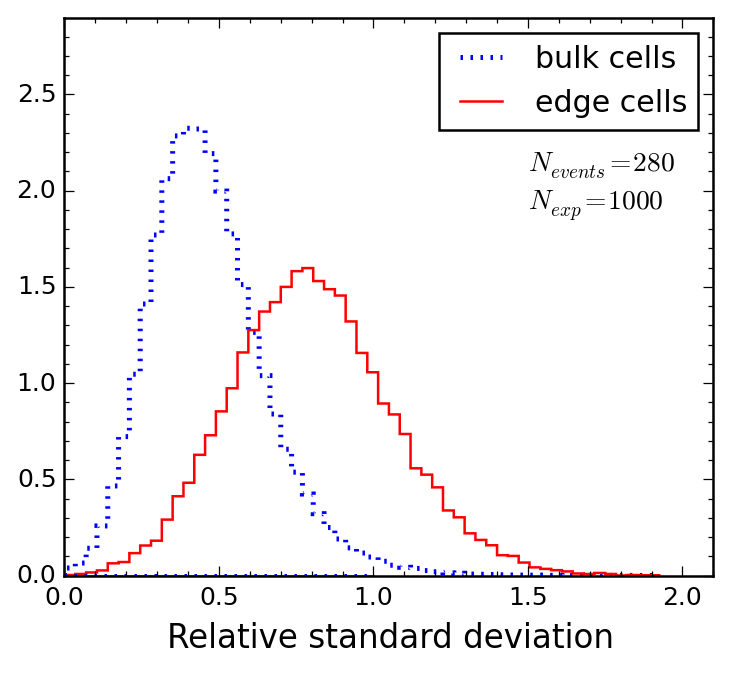}
\caption{\label{fig:plots1D} 
Unit-normalized distributions of the four Voronoi cell properties depicted in Fig.~\ref{fig:plots2D}.
Blue dotted (red solid) histograms refer to bulk (edge) cells. In order to increase the statistics,
we show results from $N_{exp}=1000$ pseudo-experiments with $N_{events}=280$ each.
}
\end{figure}

The basic idea put forth in Ref.~\cite{Debnath:2015wra} was to study the 
resulting Voronoi tessellation and identify edge cells from their 
geometric properties (as well as from the geometric properties of 
neighboring cells within the immediate vicinity). The Voronoi cells 
in two dimensions are planar polygons, for which one can investigate 
the usual geometric properties like number of sides, area, perimeter, etc. 
Fig.~\ref{fig:plots2D} shows four possibilities, where the Voronoi cells 
are color-coded according to the value of the corresponding geometric 
quantity. Then, in Fig.~\ref{fig:plots1D}, we plot the probability 
distributions for these geometric quantities separately for the edge 
cells (red solid lines) and the bulk cells (blue dotted lines).
As can be seen in Fig.~\ref{fig:2DVoronoi}, the edge cells, by 
construction, represent a very small fraction of the 
total number of Voronoi cells in the tessellation.  Thus, in order to 
increase the statistical precision of the distributions in 
Fig.~\ref{fig:plots1D}, we generated $N_{exp}=1000$ pseudo-experiments 
analogous to the one shown in Fig.~\ref{fig:2DVoronoi}.

In the upper left panels of Figs.~\ref{fig:plots2D} and \ref{fig:plots1D}, 
we study the number of elements, 
$|N_i|$, in the set of neighbors, $N_i$.  This is equivalent to the number of 
sides of the $i$-th Voronoi polygon.
This variable has been studied in the literature \cite{MilesMaillardet1982}: 
for example, it is known that the Voronoi polygons 
most commonly have 5 or 6 sides, which is confirmed in Figs.~\ref{fig:plots2D} 
and \ref{fig:plots1D}.
There is also a long tail of polygons with many sides, which is conjectured to 
behave asymptotically as
$|N_i|^{-2|N_i|}$ \cite{Drouffe:1983wr}. Indeed, Fig.~\ref{fig:2DVoronoi} 
contains an example of a polygon with as many as 12 sides!
The upper left panel of Fig.~\ref{fig:plots1D} demonstrates that, as expected, 
the $|N_i|$ distributions for bulk and edge cells are rather similar, 
and are thus not suitable for tagging edge cells \cite{Debnath:2015wra}.

The upper right panels of Figs.~\ref{fig:plots2D} and \ref{fig:plots1D} 
illustrate a different geometric quantity 
related to the areas, $a_i~(i=1,2,\ldots,N_{events})$, of the Voronoi cells.
The areas of the Voronoi polygons are meaningful because they provide an 
estimate of the value of the underlying distribution $f(x,y)$ (\ref{fstep}) 
at the corresponding data point $(x_i,y_i)$:
\beq
f(x_i,y_i) \approx \frac{1}{N_{events}\times a_i},
\eeq
so that $f(x,y)$ is still unit-normalized:
\beq
\sum_{i=1}^{N_{events}} f(x_i,y_i) \times a_i = \frac{1}{N_{events}} 
\sum_{i=1}^{N_{events}} \frac{a_i}{a_i} = 1.
\eeq
In Figs.~\ref{fig:plots2D} and \ref{fig:plots1D}
we choose to normalize the cell areas not locally as in (\ref{eq:sv-scaled-def}), 
but by the {\em expected} 
average area in the dense region.
Thus, a typical bulk cell to the left of the vertical boundary has 
normalized area of approximately 1, while a typical bulk cell to the right of the 
boundary has a normalized area
of approximately $\rho=6$. Note that while we fix the total number of events, 
$N_{events}$, the fraction which ends up
on one side of the boundary varies --- for example, in the single 
pseudo-experiment depicted in 
Figs.~\ref{fig:2DVoronoi} and \ref{fig:plots2D}, there happen to be 243 
events on the left side 
and 37 events on the right side (to be compared with the expectation of 
$\rho N_{events}/(\rho+1)=240$ 
events on the left side and $N_{events}/(\rho+1)=40$ events on the right side). 
If the total area is $A=\sum_i a_i$,
the expected average size of a bulk cell in the dense region on the left is 
given by $A(\rho+1)/(2\rho N_{events})$,
hence in the upper right panels of Figs.~\ref{fig:plots2D} and 
\ref{fig:plots1D} we plot the cell areas, $a_i$, normalized as
\beq
\bar{a}_i = \frac{2\rho N_{events}}{\rho+1}\, \frac{a_i}{A}.
\label{area_norm}
\eeq
The distribution of the Voronoi cell areas when the points have been 
randomly selected (in a ``Poisson process'') is not known analytically, 
and is typically approximated with a three-parameter generalized gamma 
function, where the parameters are fitted to results from Monte Carlo 
simulations~\cite{HindeMiles1980}.
For our purposes, we are not interested in the form of the actual distributions 
but in the question of
whether the distributions for bulk and edge cells show any appreciable 
differences.  As seen in the upper right panel of Fig.~\ref{fig:plots1D}, 
the area distribution for bulk cells is 
nearly bimodal; with the normalization (\ref{area_norm}) two peaks are 
expected near $\bar{a}_i\sim 1$ and $\bar{a}_i\sim\rho=6$\footnote{
As there are $\rho$ times fewer cells in the low density region, the
corresponding peak is $\rho$ times smaller and hence appears to be more of a shoulder
than a true peak in the upper right panel of Fig.~\ref{fig:plots1D}.  The 
same considerations hold for the analogous plot in Fig.~\ref{fig:plots3D_1d}.}. 
The area distribution for edge cells, on the other hand,
is unimodal, peaking relatively close to $\bar{a}_i\sim 1$. After all, 
we expect a larger fraction, namely $\rho/(\rho+1)$
of the edge cells, to have their centers on the ``dense" side of the boundary 
and only $1/(\rho+1)$
of the edge cells to have their centers on the ``sparse" side of the boundary. 
A close inspection of
Fig.~\ref{fig:2DVoronoi} confirms this expectation: out of the 21 edge cells, 
18 (3) have 
their centers to the left (to the right) of the vertical boundary, 
which is consistent with our expectations for $\rho = 6$.
In conclusion, it is clear that in this case, the Voronoi cell area by itself 
is not a very good candidate for 
an edge-tagging variable \cite{Debnath:2015wra}. We expect that in the more
general situation, where the densities on each side 
of the boundary are not uniform, this variable will be even more unsuitable.

Having investigated a variable describing the size of the Voronoi polygon, 
we now examine a variable 
characterizing the {\em shape} of the polygon, e.g., the isoperimetric quotient
\beq
q_i \equiv \frac{4\pi a_i}{p_i^2},
\label{isoqdef}
\eeq
where $a_i$ is the area and $p_i$ is the perimeter of the $i$-th Voronoi polygon.
The variable (\ref{isoqdef}) is a measure of ``roundedness" --- it is equal to 
zero for infinitely thin (pencil-like) polygons, 
and is equal to 1 in the limit of a perfectly symmetric polygon with infinitely 
many sides (i.e., a circle).
The corresponding results for the isoperimetric quotient are shown in the lower 
left panels of Figs.~\ref{fig:plots2D} and \ref{fig:plots1D}. We observe that 
the edge cells tend to be slightly more ``squashed",
but the difference is very minor and not useful in practice.

In a similar vein, one could continue to study other geometric characteristics 
of a single Voronoi cell, e.g., perimeter, average side length, 
etc.~\cite{Debnath:2015hva}, but, similarly, this is unlikely to lead to any 
success in identifying edge cells. The reason is that we are trying to detect 
a discontinuity and therefore we need to study the {\em relative} 
properties of cells on both sides of the boundary. 
One possibility is to compute a derivative quantity, e.g., the gradient 
at each cell location \cite{Debnath:2015wra}.
Another option is to compare the spread in the areas of the neighboring 
cells, e.g., by computing the
RSD, $\bar{\sigma}_i$, of the areas of the cells in $N_i$ 
(the set of neighbors of the $i$-th Voronoi polygon) 
in analogy to (\ref{defvar}) \cite{Debnath:2015wra}:
\begin{equation}
\label{defvar_area}
\bar{\sigma}_i \equiv 
\frac{1}{\langle {a}(N_i)\rangle}\, \sqrt{\sum_{j\in N_i} 
\frac{\left(a_j-\langle {a}(N_i)\rangle \right)^2}{|N_i|-1}},
\end{equation}
where the normalization now is done by the average {\em area} of the neighbors
\begin{equation}
\langle {a}(N_i)\rangle \equiv \frac{1}{|N_i|}\sum_{j\in N_i} a_j.
\label{ave-area-neighbors}
\end{equation}
The idea is very simple --- the neighbors of edge cells are 
typically quite diverse --- 
some happen to be on the dense side and are therefore relatively small, 
while others are on the sparse side and are relatively large. As a result, 
the RSD of neighbor areas
for edge cells is expected to be enhanced. On the other hand, for bulk cells, 
all neighbors are roughly similar, 
and the RSD of their areas should be small. These expectations are 
confirmed in the
lower right panels of Figs.~\ref{fig:plots2D} and \ref{fig:plots1D}.
In the temperature plot of Fig.~\ref{fig:plots2D}, the edge cells are clearly 
distinguished by the different color,
and the $\bar{\sigma}_i$ distributions for bulk and edge cells in 
Fig.~\ref{fig:plots1D} are visibly displaced from each other.
We see that, in agreement with the conclusions from 
Ref.~\cite{Debnath:2015wra},
the RSD, $\bar{\sigma}_i$, is a promising variable for edge detection\footnote{
Ref.~\cite{Debnath:2015wra}
also considered a few other variables related to derivatives --- 
the magnitude of the gradient at each data point,
the correlation between the directions of the gradients computed at two 
neighboring cells, the scalar 
product of the gradients at two neighboring cells, etc. The conclusion, 
drawn on the basis of ROC curves (see below 
appendix~\ref{sec:ROC-and-ROLL}) was that the RSD variable, $\bar{\sigma}_i$, 
was optimal among all those choices.}.

\subsubsection{A circular boundary in two dimensions}

\begin{figure}[t]
\centering
\includegraphics[height=5.5cm]{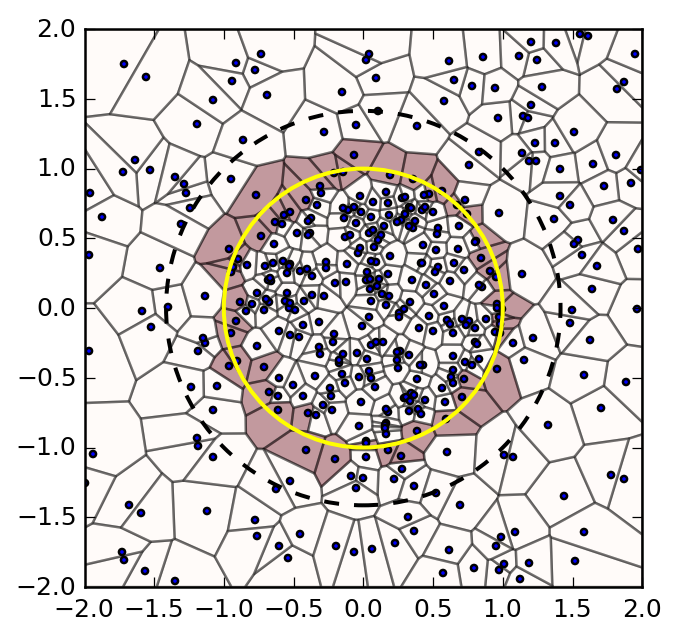} 
\includegraphics[height=5.5cm]{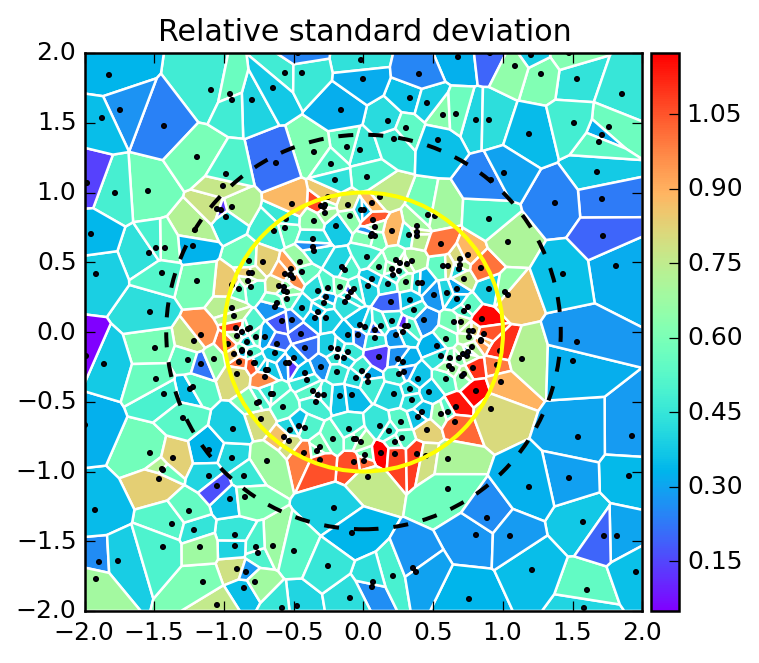} 
\caption{\label{fig:circleplots2D} 
The analogue of Fig.~\ref{fig:2DVoronoi} (left panel) and 
the analogue of the lower right panel in Fig.~\ref{fig:plots2D} (right panel),
for the radially symmetric distribution (\ref{fstepr}).
In order to keep the statistics the same as in Figs.~\ref{fig:2DVoronoi} 
and \ref{fig:plots2D} we place $N_{events}=280$ events inside the dashed 
circle with radius $r=\sqrt{2}$.}
\end{figure}

Before concluding our discussion in two dimensions, we perform one more toy 
exercise.  In the example of the previous section~\ref{sec:2dlinear}, 
the boundary was a straight line; 
in a more realistic situation we will encounter a boundary which is an 
arbitrary curved line. In anticipation of the physics example discussed
in section~\ref{sec:phase-space-with-voronoi}, we now consider a 
two-dimensional example with a curved boundary in the shape of a circle. 
Instead of the rectangular pattern given by (\ref{fstep}), we
consider the radially symmetric distribution
\beq
f(\vec{r}) \sim 
\rho H(1-r) + H(r-1) H(\sqrt{2}-r),
\label{fstepr}
\eeq
where $\vec{r}=(x,y)$ is the position vector in 2 dimensions and 
$r\equiv |\vec{r}|$ is its magnitude.  As in (\ref{fstep}), 
the distribution (\ref{fstepr}) describes two regions, 
the inner region is a unit circle, while the outer region 
is a hollow disk extending up to $r=\sqrt{2}$ (the circular dashed line in Fig.~\ref{fig:circleplots2D}).
The regions are separated by a circular boundary at $r=1$, marked with 
the solid yellow curve in Fig.~\ref{fig:circleplots2D}.
Similarly to the example from section \ref{sec:2dlinear},
the two regions have equal areas, each region has a constant density,
and one region is $\rho$ times denser than the other, see Fig.~\ref{fig:circleplots2D}.
Just as in section~\ref{sec:2dlinear}, we choose $\rho=6$ and generate 
$N_{events}=280$ events according to (\ref{fstepr}); 
they are distributed so that the ratio of the
bulk events on the two sides of the boundary is equal to $\rho$.  
Thus, out of the $N_{events}=280$ events inside the dashed circle with $r=\sqrt{2}$, 
on average we will have $\rho N_{events}/(\rho+1)=240$ events in 
the dense interior region (the unit circle) and $N_{events}/(\rho+1)=40$ 
events within the sparse exterior hollow disk.\footnote{Since our plots are rectangular, 
in Fig.~\ref{fig:circleplots2D} we have extended the exterior region beyond $r=\sqrt{2}$,
populating the additional real estate with the same density as the hollow disk. 
This was done to avoid spurious, but visually distracting empty areas on the plots. }

In the left panel of Fig.~\ref{fig:circleplots2D}, the brown-shaded 
polygons are by definition the edge cells (those crossed by the yellow 
circular boundary). The right panel in Fig.~\ref{fig:circleplots2D} 
demonstrates that, once again, the edge cells can be effectively 
selected by the RSD, $\bar{\sigma}_i$, of the areas of the neighboring cells.

\subsection{Voronoi tessellations in three dimensions}
\label{sec:3dtoy}

Since the relevant physics example we treat in 
section~\ref{sec:phase-space-with-voronoi} is in a three-dimensional 
phase space, $(m_{12}, m_{23}, m_{13})$, we shall now generalize our
previous discussion to  three dimensions. For this purpose, we consider 
the three-dimensional analogue of 
(\ref{fstepr}):
\beq
f(\vec{R}) \sim 
\rho H(1-R) + H(R-1) H(\sqrt[3]{2}-R) ,
\label{fstepr3d}
\eeq
where now $\vec{R}=(x,y,z)$ is the position vector in {\em three} 
dimensions and $R\equiv |\vec{R}|$.  The distribution, (\ref{fstepr3d}), 
again describes two regions of constant density, except now
the dense region is a three-dimensional spherical core of radius 1. 
Again, we choose $\rho=6$ and generate $N_{events}=4200$
events according to (\ref{fstepr3d}). The events populate
a ball of radius $R=\sqrt[3]{2}$ centered at the origin, $(x,y,z)=(0,0,0)$.
On average, we expect to have $\rho N_{events}/(\rho+1)=3600$ events in 
the core and $N_{events}/(\rho+1)=600$ events in the outer hollow 
spherical shell ($1\le R\le \sqrt[3]{2}$).\footnote{Once again, to avoid
misleading voids on the plots, we extend the exterior region beyond 
$R=\sqrt[3]{2}$, populating this outer region with 
points having the same density as the hollow sphere.}

\begin{figure}[t]
\centering
\includegraphics[height=5.6cm]{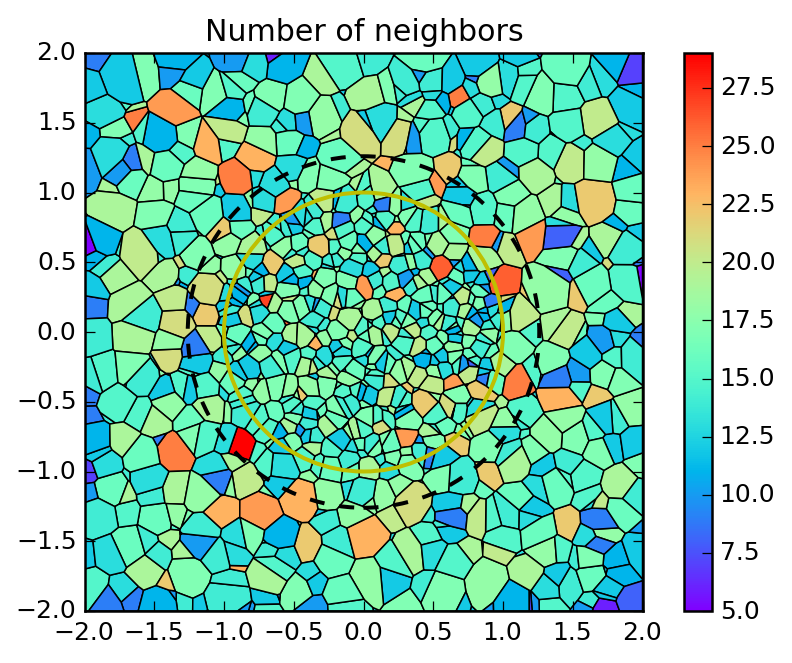}
\includegraphics[height=5.6cm]{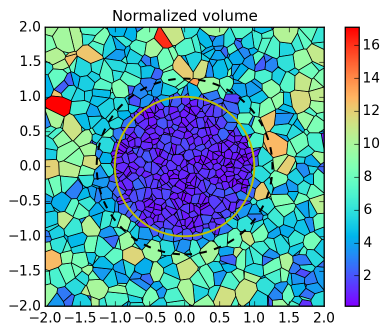} \\ 
\includegraphics[height=5.6cm]{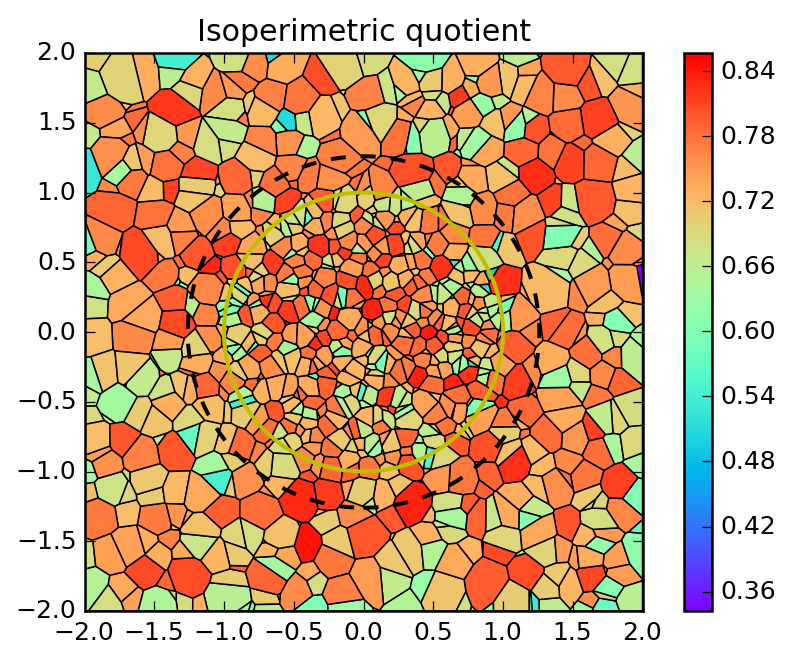}
\includegraphics[height=5.6cm]{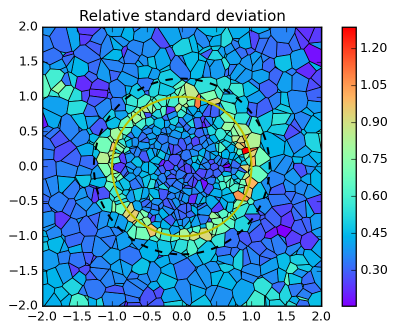}
\caption{\label{fig:plots3D_voronoi} 
Two-dimensional slices at $z=0$ through phase space for the 
three-dimensional toy example studied in section~\ref{sec:3dtoy}.
We distribute $N_{events}=4200$ points according to the three-dimensional 
probability distribution (\ref{fstepr3d}) 
within a sphere of radius $\sqrt[3]{2}$ centered at the origin 
$(x,y,z)=(0,0,0)$. The Voronoi tessellation is done {\em before}
taking the two-dimensional slice, i.e., the cell boundaries seen 
on these four plots are obtained by intersecting
the three-dimensional Voronoi cell boundaries with the plane at $z=0$. 
The yellow circle marks the boundary of the dense core. 
The resulting cells in the two-dimensional slice are color coded by 
a certain attribute of the corresponding
three-dimensional Voronoi cell: number of neighbors (upper left); 
normalized volume (upper right); isoperimetric ratio 
(\ref{isoq3def}) (lower left), and RSD of 
the neighboring volumes (\ref{defvar}) (lower right).
}
\end{figure}

\begin{figure}[t]
\centering
\includegraphics[width=5.8cm]{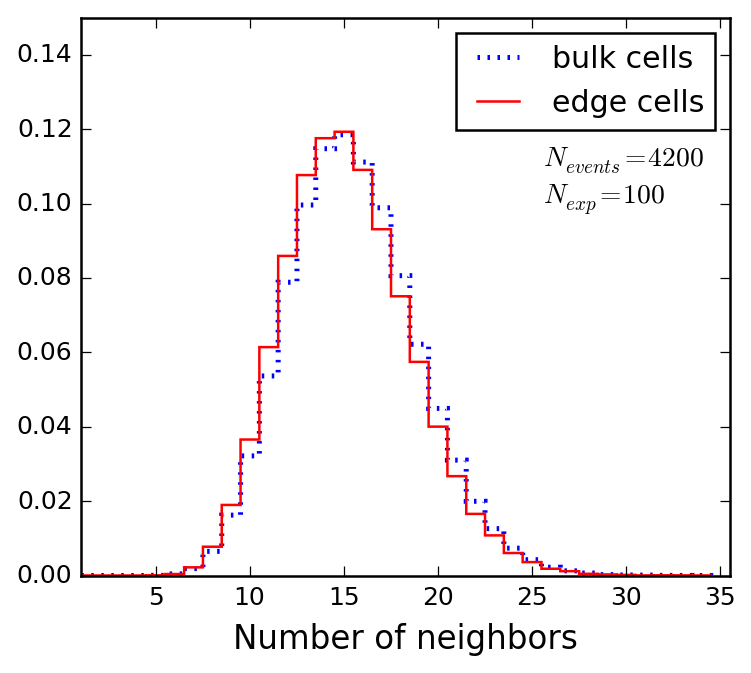} 
\includegraphics[width=5.8cm]{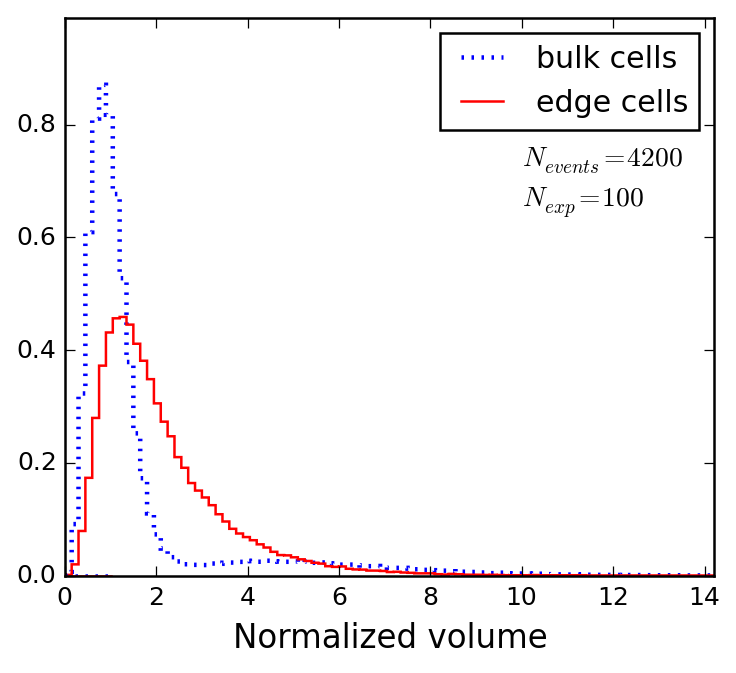} \\ 
\includegraphics[width=5.6cm]{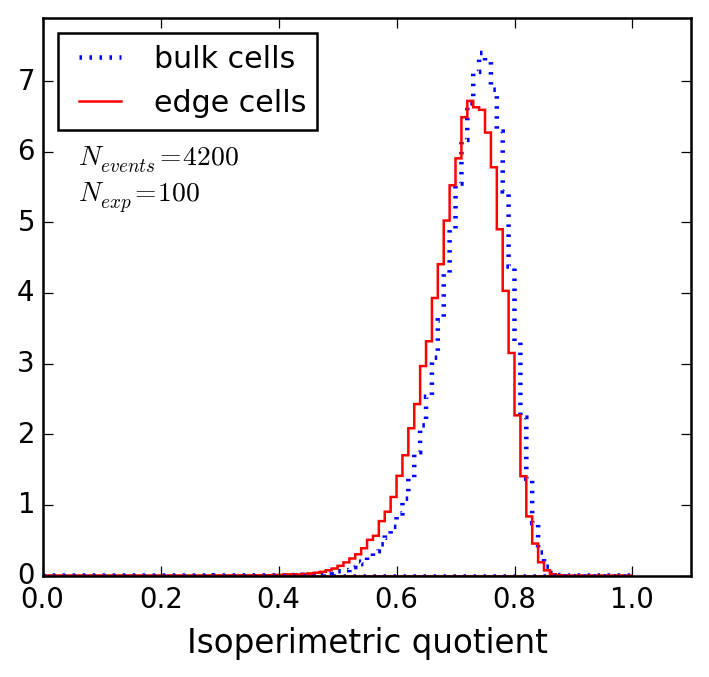} 
\includegraphics[width=5.8cm]{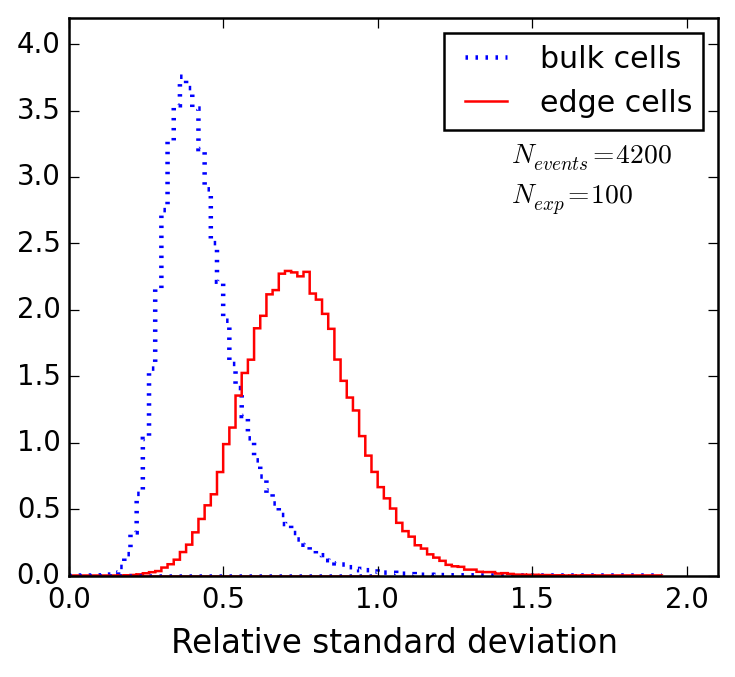}
\caption{\label{fig:plots3D_1d} 
The same as Fig.~\ref{fig:plots1D}, but for the three-dimensional toy example 
depicted in Fig.~\ref{fig:plots3D_voronoi}.
Edge cells are defined to be those Voronoi cells which are crossed by the 
boundary of the unit sphere ($r=1$).
}
\end{figure}

Figs.~\ref{fig:plots3D_voronoi} and \ref{fig:plots3D_1d}  
illustrate this three-dimensional simplified scenario 
in analogy to Figs.~\ref{fig:plots2D} and \ref{fig:plots1D}. 
Since the Voronoi cells in three dimensions are polyhedra, it is difficult 
to visualize them on a planar plot.  Thus, 
Fig.~\ref{fig:plots3D_voronoi} shows only a slice at a fixed $z=0$,
i.e., an equatorial plane view. The cell boundaries seen in the figure are 
the intersections of the equatorial plane with the walls of the Voronoi 
polyhedra. The interiors of those cells are color-coded according 
to the value of the geometric property (number of faces, volume, etc.) 
of the corresponding {\em three-dimensional} polyhedron\footnote{
We caution the reader that the color bars in Fig.~\ref{fig:plots3D_voronoi}
refer to the three-dimensional Voronoi polyhedra and not the polygons seen 
in the plots --- the latter result from the intersection with the 
equatorial plane, and, in general, have different properties.}. 
For example, the upper left panel in Fig.~\ref{fig:plots3D_voronoi} 
shows that the Voronoi polyhedra typically have a relatively large number 
of faces (or equivalently, neighbors); the corresponding distribution 
for bulk cells, shown in the upper left panel in Fig.~\ref{fig:plots3D_1d},
is known to peak at 15 \cite{Lorz1991}. We also observe that the edge cells 
are very similar in that regard, i.e., there is no appreciable difference 
in the number of neighbors as we move across the boundary.

In the upper right panels of Figs.~\ref{fig:plots3D_voronoi} and 
\ref{fig:plots3D_1d} we show the corresponding
result for the normalized volumes, $\bar{v}_i$, of the Voronoi polyhedra, 
where, in analogy to (\ref{area_norm}), we scale each volume, $v_i$, by the 
expected average volume in the dense core, 
$\frac{4}{3}\pi (\rho+1)/(\rho N_{events})$.
As expected, the distribution for bulk cells is bimodal, while edge cells 
behave somewhat similarly to the interior bulk cells (as we already saw 
in the two-dimensional example of section~\ref{sec:2dlinear}). 

In the lower left panels of Figs.~\ref{fig:plots3D_voronoi} and 
\ref{fig:plots3D_1d} 
we plot the analogous ``isoperimetric quotient" for the three-dimensional case, 
\beq
Q_i \equiv \frac{6\sqrt{\pi}\, v_i}{s_i^{3/2}},
\label{isoq3def}
\eeq
where $v_i$ ($s_i$) is the volume (surface area) of the Voronoi polyhedron
and the normalization is chosen so that $Q_i=1$ for a perfect sphere.
Figs.~\ref{fig:plots3D_voronoi} and \ref{fig:plots3D_1d} show that the shapes 
of the Voronoi polyhedra, as measured by (\ref{isoq3def}), are very similar 
in the two bulk regions and not much different near the boundary either.

This leaves us with the RSD, $\bar\sigma_i$, of the volumes for the set of 
neighbors, $N_i$.  This quantity was already defined in (\ref{defvar}) 
and our results are shown in the lower right panels of 
Figs.~\ref{fig:plots3D_voronoi} and \ref{fig:plots3D_1d}. We see that 
$\bar{\sigma}_i$ can efficiently identify edge cells; the circular 
boundary is clearly seen in the lower right plot of 
Fig.~\ref{fig:plots3D_voronoi}. The $\bar{\sigma}_i$ distributions 
for bulk and edge cells are quite distinct, as shown in 
Fig.~\ref{fig:plots3D_1d}. Thus we verify that $\bar{\sigma}_i$ 
remains a promising variable for edge detection beyond the two-dimensional 
examples studied in Ref.~\cite{Debnath:2015wra}.

\section{Phase Space Considerations}
\label{sec:phase-space}

While two and three-body phase space is discussed at length in most 
standard lectures and textbooks on quantum field theory, a Lorentz-invariant 
formulation of the general case with an arbitrary number of final state 
particles is often omitted.  This is in part because processes with more than 
three final state particles can, in almost all circumstances, be analyzed as a 
sequence of on-shell production and decay stages and in part because the 
level of formalism required to describe the general case is significantly 
more involved.  Nevertheless, as was shown in Ref.~\cite{Agrawal:2013uka}, 
even when a cascade decay proceeds through multiple on-shell stages, 
treating the phase space in its fully-differential form captures important 
correlations that cannot be inferred from more traditional one-dimensional 
observables such as kinematic edges and endpoints. 
In this context, we briefly review the geometry of four-body phase 
space,\footnote{The description of more than four-body phase space can be 
given in a very similar fashion although there are additional subtleties 
due to non-linear constraints. 
We refer the interested reader to Ref.~\cite{ByersYang}.}
concentrating on the equation describing the boundary of the kinematically 
available region and on the differential volume element.
In section~\ref{sec:enhanced3dtoy},
we shall apply kinematic features obtained from the phase space considerations 
in section~\ref{sec:4body} to our uniform sphere example from 
section~\ref{sec:3dtoy} and use the resulting toy example to study our 
Voronoi methods for three-dimensional data.

\subsection{Review of the four-body phase space of a cascade decay}
\label{sec:4body}

Let us consider the process where a heavy resonance, $X$, decays 
into four particles.  We first focus on presenting the form of 
four-body phase space in full generality and will further specialize 
to the case where the decay proceeds via three-step two-body cascade 
decays as in Fig.~\ref{fig:topology}.  Following the argument in 
Ref.~\cite{ByersYang}, we begin by introducing a $4\times 4$ matrix 
defined as
\bea
{\mathcal Z}=\left\{ z_{ij}\right\}\quad {\rm with}\quad z_{ij}=
p_{i}\cdot p_{j}\, ,
\eea
where the $\{p_i\}$ denote four momenta of final state particles, including the NPP $X_4$.\footnote{Note 
that in this section, Latin indices refer to the final state particles and not data points (Voronoi cells).}
We then define the characteristic polynomial of $\mathcal{Z}$ as
\bea
\det \left[\lambda I_{4\times4}-\mathcal{Z}\right]\equiv 
\lambda^4-\left(\sum_{i=1}^{4} \Delta_{i}\lambda^{4-i}\right)=0,
\eea
where $\lambda$ represents the relevant eigenvalues and the $\Delta_i$ 
identify the coefficients of the above polynomial. 
Specifically, one finds that $\Delta_{1}={\rm Tr}[{\mathcal Z}]=
\sum_{i=1}^{4}m_{i}^{2}$ and $\Delta_{4}=-\det\left[{\mathcal Z}\right]$.
It turns out that the kinematically allowed region is given by 
$\Delta_{1,2,3,4}>0$~\cite{ByersYang}; the boundary of this region is 
formed by
\bea
\Delta_4 =0,\qquad \Delta_{1,2,3}>0\,. \label{eq:boundary}
\eea

What makes four-body (and higher) phase space particularly interesting 
is the form of the volume element. In terms of 
$m_{ij}^{2}=\left(p_{i}+p_{j}\right)^2=2z_{ij}+m_{i}^{2}+m_{j}^{2}$, 
the four-body phase space, $\Pi_4$, can be written as 
\begin{equation} 
d\Pi_{4}=\left(\prod_{i<j}dm_{ij}^{2}\right)\frac{8}{(4\pi)^{10}M_{X}^{2}
\Delta_{4}^{1/2}}\ \delta\left(\sum_{i<j}m_{ij}^{2}-\left(M_{X}^{2} + 2 
\sum_{i=1}^{4} m_{i}^{2}\right)\right),
\end{equation}
where the normalization has been chosen to reproduce the PDG 
convention~\cite{Agashe:2014kda} for the well-known expression with 
non-Lorentz invariant quantities
\begin{equation}
d\Pi_{4}=\delta\left(p_{X}-\sum_{i=1}^{4}p_{i}\right)\prod_{i=1}^{4}
\frac{d^{3}p_{i}}{(2\pi)^{3}2E_{i}}.
\end{equation}
We remark that $d\Pi_4$ is inversely proportional to $\Delta_4$, 
and, given the fact that $\Delta_4=0$ defines the kinematic boundary, 
as in~\eqref{eq:boundary}, the phase space has a singular structure 
near $\Delta_4=0$.  While being an integrable singularity, this implies 
that events are more likely to be populated close to the boundary rather 
than far away from it.  This observation is ideal for mass measurements 
which ultimately rely on the determination of this phase space boundary. 

Given the generic formalism for the phase space with four  particles 
in the final state, we now specialize to the case where the decay proceeds 
through the three consecutive two-body decays shown in Fig.~\ref{fig:topology}. 
The $X_{i}$'s are NPPs represented by red dashed lines, 
while the $v_{i}$'s are SM particles represented by black solid lines. 
For simplicity, we assume that all SM particles are mass{\it less} unless 
specified otherwise.  $X_{1,2,3}$ are assumed to be narrow resonances, 
while $X_{4}$ is collider-stable and invisible.

We point out that the presence of the intermediate particles does 
{\it not} affect the enhancement near the boundary of phase space 
discussed above.  Within the narrow width approximation, each internal 
propagator squared can be replaced by a delta function, whose argument 
is linear in the $z_{ij}$ or, equivalently, in the $m_{ij}^{2}$ variables. 
Therefore, integrating over those delta functions does {\it not} 
introduce any non-trivial Jacobian factors which would ruin the enhancement. 

To quantify the enhancement near the boundary for this event topology, 
we derive the analytic form of the $\Delta_4$ probability distribution 
and show that it is completely independent of $m_{X_i}$ for the massless 
limit, i.e., $m_{v_i}=0$.  We start by writing $\Delta_{4}$ in terms of 
the experimental observables $m_{v_{i}v_{j}}^{2}$ which are denoted by 
$m_{12}^{2}$, $m_{13}^{2}$, and $m_{23}^{2}$.
These dimensionful variables can be traded for dimensionless, 
unit-normalized variables $\xi_{ij}$ as
 \beq
 m_{ij}^{2}\equiv\xi_{ij}\,m_{ij,{\rm max}}^{2},
\label{defxi}
 \eeq 
where $0\le\xi_{ij}\le 1$ and the maximal values, 
$m_{ij,{\rm max}}^{2}$, are given by the well-known kinematic 
endpoint formulae (see, e.g., \cite{Allanach:2000kt}):
\bea
m_{12,{\rm max}}^{2}&=&\frac{(m_{X_{1}}^{2}-m_{X_{2}}^{2})(m_{X_{2}}^{2}-
m_{X_{3}}^{2})}{m_{X_{2}}^{2}}\,,   \label{m12max} \\
m_{13,{\rm max}}^{2}&=&\frac{(m_{X_{1}}^{2}-m_{X_{2}}^{2})(m_{X_{3}}^{2}-
m_{X_{4}}^{2})}{m_{X_{3}}^{2}}\,,   \label{m13max}\\
m_{23,{\rm max}}^{2}&=&\frac{(m_{X_{2}}^{2}-m_{X_{3}}^{2})(m_{X_{3}}^{2}-
m_{X_{4}}^{2})}{m_{X_{3}}^{2}}\,. \label{m23max}
\eea
We also trade the dimension-8 quantity $\Delta_4$ for a dimensionless 
and unit-normalized quantity $q$ defined as 
\beq
\Delta_{4}\equiv q\,\Delta_{4,{\rm max}}, \quad 0\le q\le 1.
\label{defq}
\eeq 
Here the maximum value of $\Delta_4$ is given by
\begin{equation}
\Delta_{4,{\rm max}}=\left(\frac{(m_{X_{1}}^{2}-m_{X_{2}}^{2})(m_{X_{2}}^{2}-
m_{X_{3}}^{2})(m_{X_{3}}^{2}-m_{X_{4}}^{2})}{8m_{X_{2}}m_{X_{3}}}\right)^{2}.
\end{equation}
As shown in Ref.~\cite{Agrawal:2013uka}, for any given set of masses, 
$\left\{m_{X_{i}}\right\}$, in this topology, the probability of obtaining 
any given event near the point, $\left\{m_{ij}^{2}\right\}$, is expressed as
\begin{equation}
dP=\frac{1}{4\pi m_{X_{1}}^{2}}\left(1-\frac{m_{X_{2}}^{2}}
{m_{X_{1}}^{2}}\right)^{-1} \left(1-\frac{m_{X_{3}}^{2}}
{m_{X_{2}}^{2}}\right)^{-1} \left(1-\frac{m_{X_{4}}^{2}}
{m_{X_{3}}^{2}}\right)^{-1} \frac{H\left(\Delta_{4} \right)}
{\sqrt{\Delta_{4}}} dm_{12}^{2}\,dm_{13}^{2}\,dm_{23}^{2}, \label{eq:samosa}
\end{equation}
or equivalently, in terms of the dimensionless quantities, $\xi_{ij}$ 
and $q$, defined in (\ref{defxi}) and (\ref{defq}), this can be rewritten as
\begin{equation}
dP=\frac{2}{\pi}\frac{m_{X_{3}}}{m_{X_{2}}} \frac{H\left(q \right)}
{\sqrt{q}} d\xi_{12}\,d\xi_{13}\,d\xi_{23}.
\label{eq:delta4PDF}
\end{equation}
Here $H(x)$ is the usual Heaviside step function.

\begin{figure}[t]
\centering
\includegraphics[width=6.5cm]{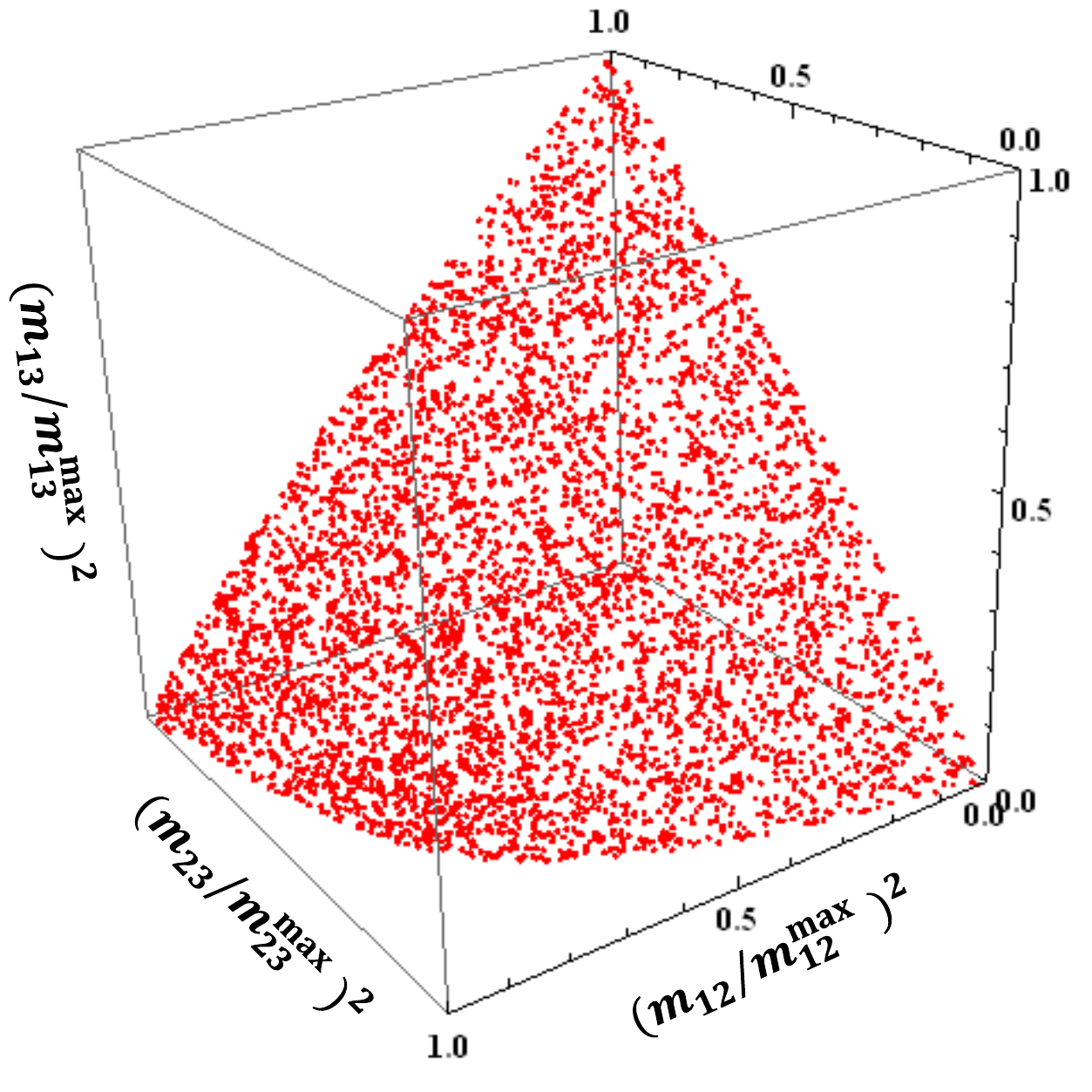} \hspace{0.25cm}
\includegraphics[width=6.5cm]{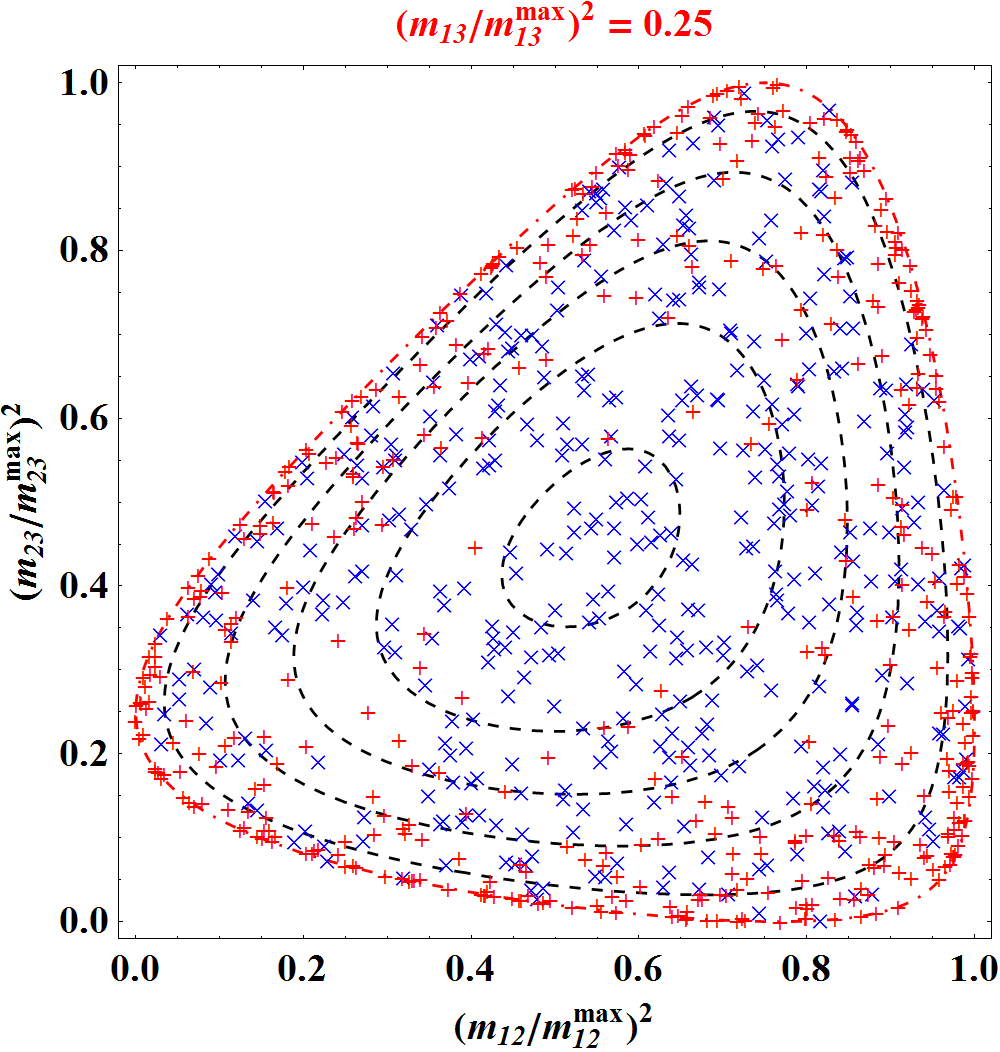}\\ \vspace{0.25cm}
\includegraphics[width=6.5cm]{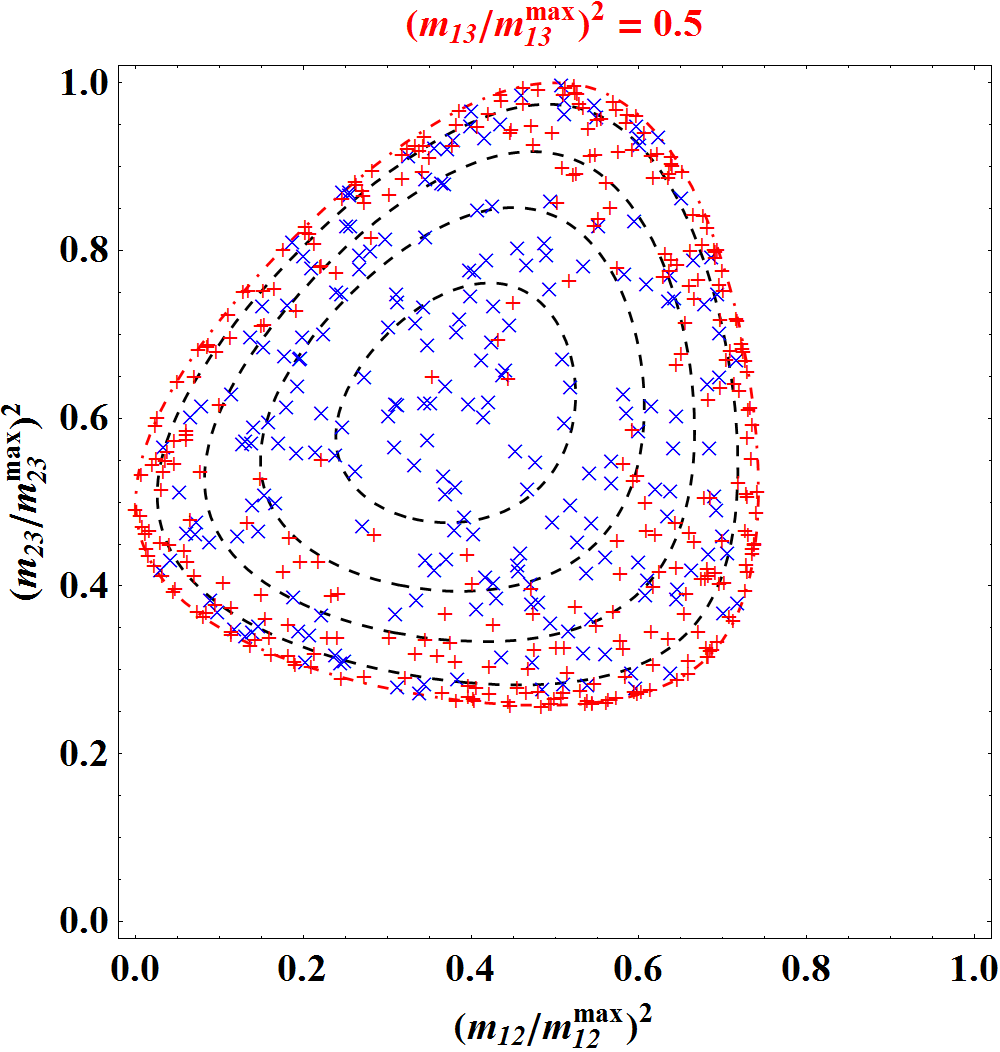} \hspace{0.25cm}
\includegraphics[width=6.5cm]{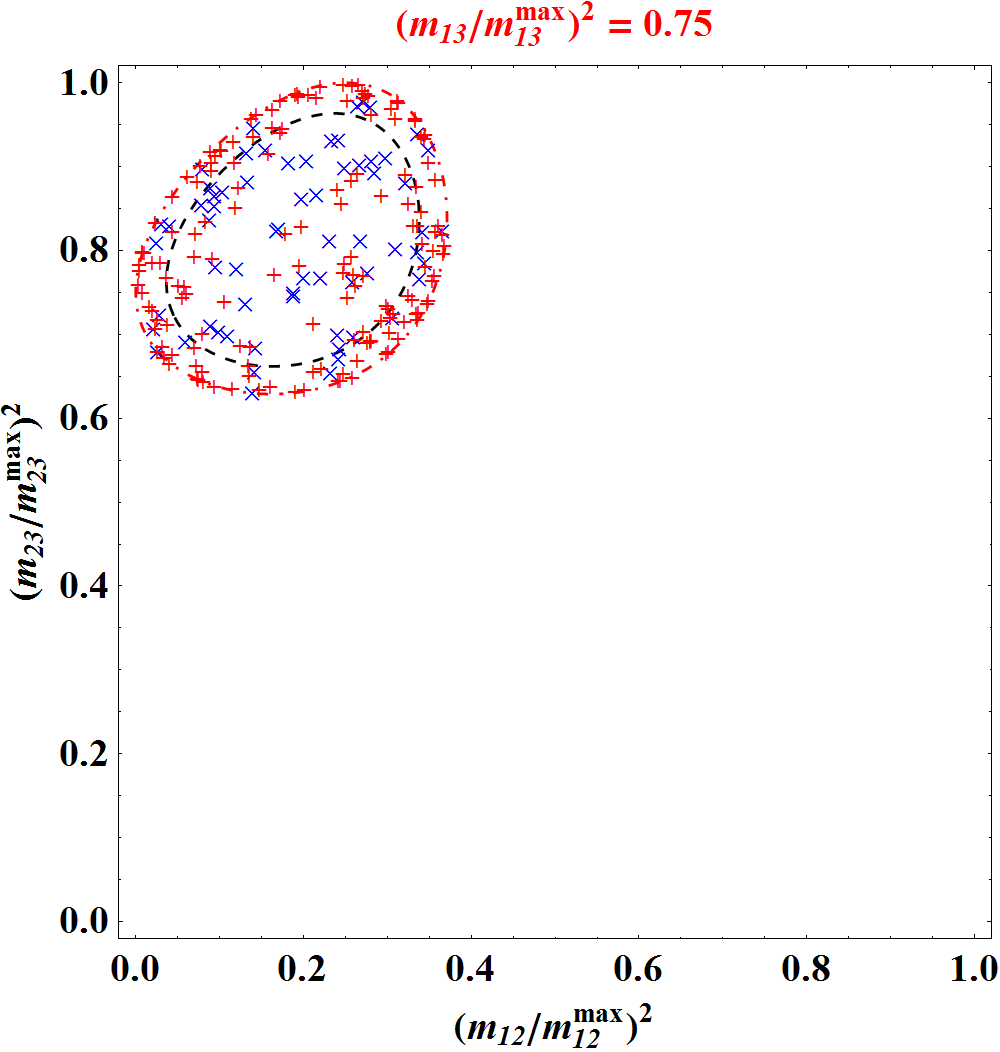}
\caption{\label{fig:samosa_scatter} The phase space structure implied by 
eq.~\eqref{eq:samosa}. The data points are generated with the event topology in Fig.~\ref{fig:topology},
using a constant matrix element. 
The mass spectrum is $(m_{X_1},m_{X_2},m_{X_3},m_{X_4}) = (500,350,200,100)$ GeV.
A three-dimensional scatter plot (upper left) and 
three phase space slices at fixed $\xi_{13}$: $\xi_{13}=0.25$ (upper right),
$\xi_{13}=0.5$ (lower left), and $\xi_{13}=0.75$ (lower right).
The red dot-dashed (outermost) curve is the contour for $\Delta_4=0$, 
while the black dashed curves correspond to $\Delta_4$ contours for 
10\%, 30\%, 50\%, 70\%, and 90\% of $\Delta_{4,\max}$. 
The data points which would have emerged via the flat component in eq.~\eqref{eq:piecePDF} 
are represented by blue ``$\times$'' symbols, whereas the data points from the remaining enhanced
component $\sim \frac{1}{\sqrt{q}}-1$ are represented by red ``$+$'' symbols. }
\end{figure}

Obviously, the expression in eq.~\eqref{eq:delta4PDF} diverges 
for $q\rightarrow0$, as expected from the general discussion earlier, and 
has a non-zero finite value at $q_{\max}=1$.  In order to visualize the 
enhancement near $q\sim 0$, it is useful to partition the probability 
density in~\eqref{eq:delta4PDF} into two components: a flat piece, 
proportional to $1$, and an enhanced piece, containing the $q^{-1/2}$ 
singularity:
\bea
\frac{dP}{dV_{\xi}}\sim \frac{1}{\sqrt{q_{\max}}}  + 
\left(\frac{1}{\sqrt{q}}-\frac{1}{\sqrt{q_{\max}}} \right) = 
1+ \left(\frac{1}{\sqrt{q}}-1 \right) , 
\label{eq:piecePDF}
\eea
where $dV_{\xi}\equiv d\xi_{12}\,d\xi_{23}\,d\xi_{13}$ is a shorthand 
notation.  If events were uniformly distributed over the entire phase space 
in $\xi_{ij}$, their probability density would simply be proportional to 
the first (constant) term in eq.~\eqref{eq:piecePDF}. Hence, all non-trivial 
effects in the phase space density distribution are due to the second term 
(inside the parentheses) in~\eqref{eq:piecePDF}.

Fig.~\ref{fig:samosa_scatter} helps us develop some useful intuition about 
the probability distribution (\ref{eq:piecePDF}).
The upper left panel shows a scatter plot of physical events in the 
dimensionless $\xi_{ij}$-space,
generated according to (\ref{eq:piecePDF}). We used a mass spectrum of 
$(m_{X_1},m_{X_2},m_{X_3},m_{X_4}) = (500,350,200,100)$ GeV. 
The events populate a compact region whose shape has been likened to that of a 
``samosa'' \cite{Kim:2015bnd}.
Since it is difficult to visualize the enhancement near the phase space 
boundary in this three-dimensional view, in the next three panels of 
Fig.~\ref{fig:samosa_scatter} we take a few slices at fixed $\xi_{13}$:  
$\xi_{13}=0.25$ (upper right), $\xi_{13}=0.5$ (lower left), and 
$\xi_{13}=0.75$ (lower right).
For each slice at a fixed $\xi_{13}$, we show all data points whose 
$m_{13}^2$ values fall within $0.5$ GeV$^2$ of the nominal value for that 
slice, i.e., within $\xi_{13} m_{13,max}^2 \pm 0.5$ GeV$^2$.
Then we project those points onto the plane of $\xi_{12}$ vs. $\xi_{23}$ 
and divide them into two (color-coded) groups.
The data points which would have emerged from the flat piece 
in~\eqref{eq:piecePDF} are denoted with blue ``$\times$'' symbols, 
whereas the points arising from the enhanced piece in~\eqref{eq:piecePDF} 
are identified by red ``$+$'' symbols. In addition, we also show several 
theoretical contours of constant $\Delta_4$ values, starting with the 
outermost red dot-dashed curve at $\Delta_4=0$ (i.e., $q=0$)
representing the phase space boundary. The internal, black dashed 
curves mark the contours for $\Delta_4=0.1\, \Delta_{4,\max}$, $\Delta_4=0.3\, 
\Delta_{4,\max}$, $\Delta_4=0.5\, \Delta_{4,\max}$,
$\Delta_4=0.7\, \Delta_{4,\max}$, and $\Delta_4=0.9\, \Delta_{4,\max}$, 
respectively.  Note that some of these contours are absent from the 
bottom panels because the relevant hyper-surfaces, 
corresponding to large $\Delta_4$ values do not intersect those slices.  

Comparing the densities of red and blue data points, we get an idea about 
the effect of the enhancement in the vicinity of the phase space boundary. 
The blue points are more or less uniformly distributed,
which is by design. In contrast, the distribution of red points is highly 
irregular, and their density peaks at the phase space boundary. 
For a more quantitative understanding, we derive the 
analytic expression for the probability density function in $q$ and 
obtain~\cite{del4short} 
\begin{equation}
\frac{dP}{dq}=\frac{\arcsin(\sqrt{1-q})}{2\sqrt{q}}. 
\label{eq:dPdq}
\end{equation}
As previously advertised, this probability density function is completely 
independent of all $\left\{m_{X_{i}}\right\}$ 
and is enhanced near $q\approx 0$.  In other words, the fraction of events 
that lie in a fixed $q$-interval is {\it universal}, 
and it is enhanced near the boundary of the phase space region. 
For example, roughly $5\%$ of events have $q\le 10^{-3}$, i.e., 
less than 0.1\% of $\Delta_{4,\max}$. 

\begin{figure}[t]
\centering
\includegraphics[scale=0.7]{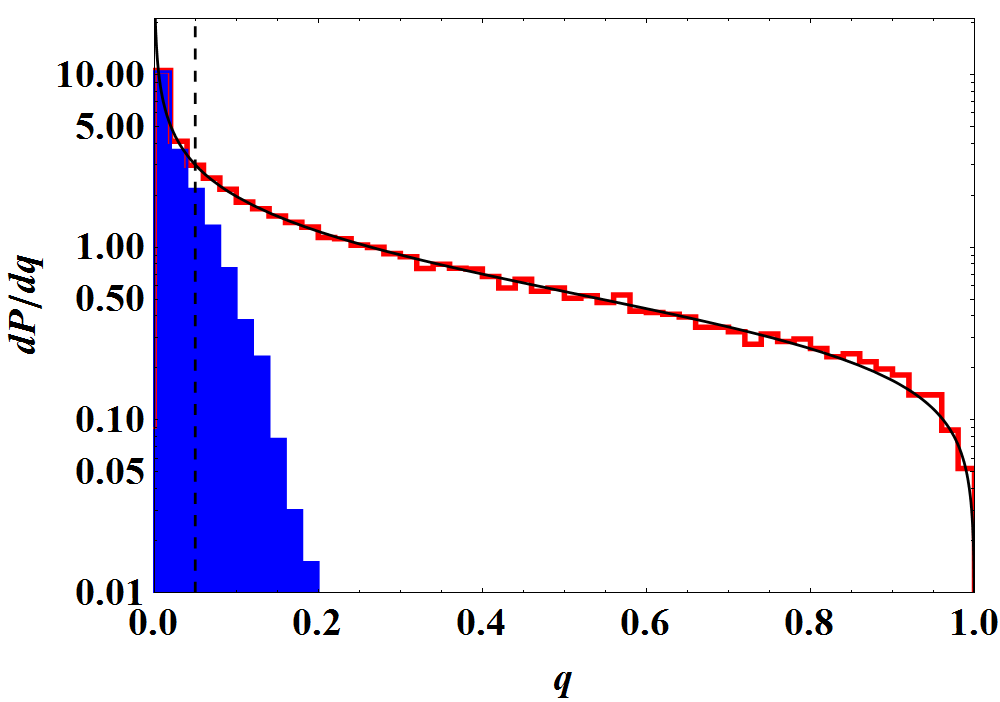}
\caption{\label{fig:del4_dist} Probability density distribution in the $q$ 
variable (\ref{defq}) 
using the same event sample as in Fig.~\ref{fig:samosa_scatter}.
The blue shaded histogram is contributed by the boundary data points which 
are tagged by the Voronoi tessellation.  The black solid curve is the theory 
prediction for $dP/dq$ in eq.~\eqref{eq:dPdq}. 
}
\end{figure}

In Fig.~\ref{fig:del4_dist}, we plot the distribution of the $q$ variable from (\ref{defq}),
taking 20,000 events out of the same event 
sample as the one used for Fig.~\ref{fig:samosa_scatter}. 
If we define any phase space point whose $q$ value is less than 5\% of $q_{\max}$ 
as a boundary point, we find that $\sim 33\%$ of the events are then categorized 
as boundary points.  The red histogram represents the $q$ distribution with 
respect to the full data set; the black dashed, vertical line denotes 
the location of $0.05\,q_{\max}$. 
The black solid curve shows the theoretical prediction from eq.~\eqref{eq:dPdq}. 
One can easily see that the $q$ distribution (red histogram) is fully 
consistent with the theory expectation. Indeed, the value of $q$ 
(or equivalently, $\Delta_4$) is not an experimental observable, since 
it requires a model assumption (the input of a mass spectrum for $X_i$). 
What is needed then is a practical way of tagging the boundary data points 
with such low values of $q$ by some other means; 
we employ Voronoi tessellations as an available tool.
We Voronoi tessellate our phase space using the full data set. 
If a given Voronoi cell has vertices on both sides of the
``samosa'' surface defined by $q=0$ (or equivalently, $\Delta_4=0$), 
then the associated data point is tagged as a boundary point (recall the 
definition of ``edge" cells in Fig.~\ref{fig:2DVoronoi}). 
The contribution from the boundary points extracted with the above algorithm is 
shown by the blue shaded histogram in Fig.~\ref{fig:del4_dist}, which we find 
represents $\sim 38\%$ of the events in the sample.
As Fig.~\ref{fig:del4_dist} demonstrates, the set of boundary cells which can be tagged
by placing a cut on $q$ is essentially the same as the set of boundary cells 
identified with the Voronoi tessellation. In what follows, therefore, instead of 
using the variable $q$, which is experimentally inaccessible, we shall focus on the
Voronoi cells belonging to the blue-shaded histogram in Fig.~\ref{fig:del4_dist}
and try to develop a tagging method based on their geometric properties, since they {\em are}
experimentally observable.

\subsection{Density-enhanced sphere boundaries}
\label{sec:enhanced3dtoy}

\begin{figure}[t]
\centering
\includegraphics[height=5.5cm]{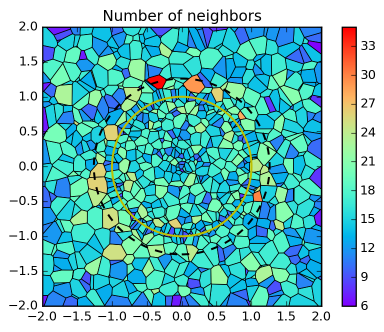} 
\includegraphics[height=5.5cm]{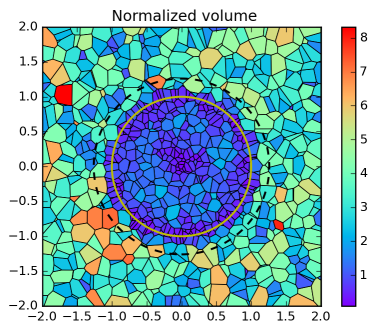} \\ 
\includegraphics[height=5.5cm]{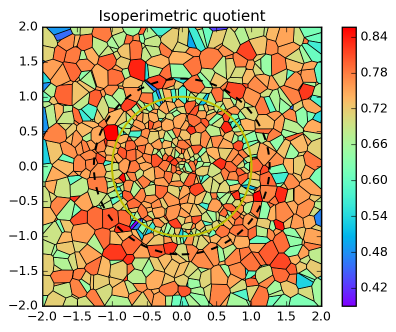} 
\includegraphics[height=5.5cm]{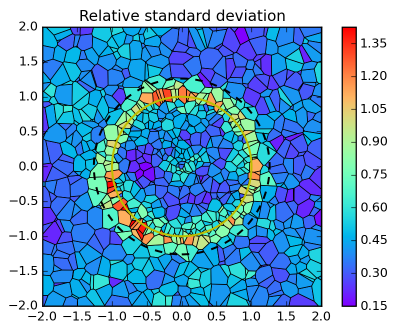}
\caption{\label{fig:sphereenhanced} 
The same as Fig.~\ref{fig:plots3D_voronoi}, but for a toy example in which 
the dense core has a non-uniform distribution given by (\ref{fenhr3d}).
}
\end{figure}

\begin{figure}[t]
\centering
\includegraphics[height=5.5cm]{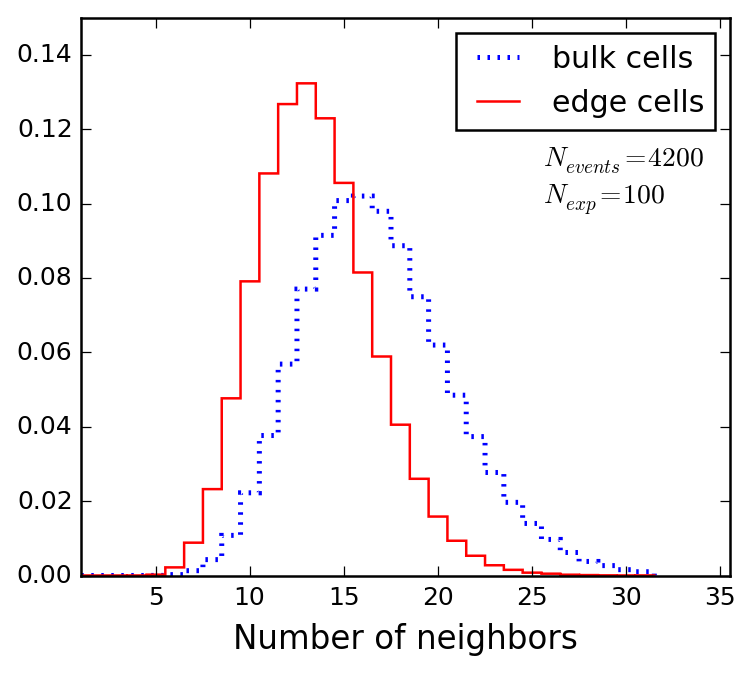} 
\includegraphics[height=5.5cm]{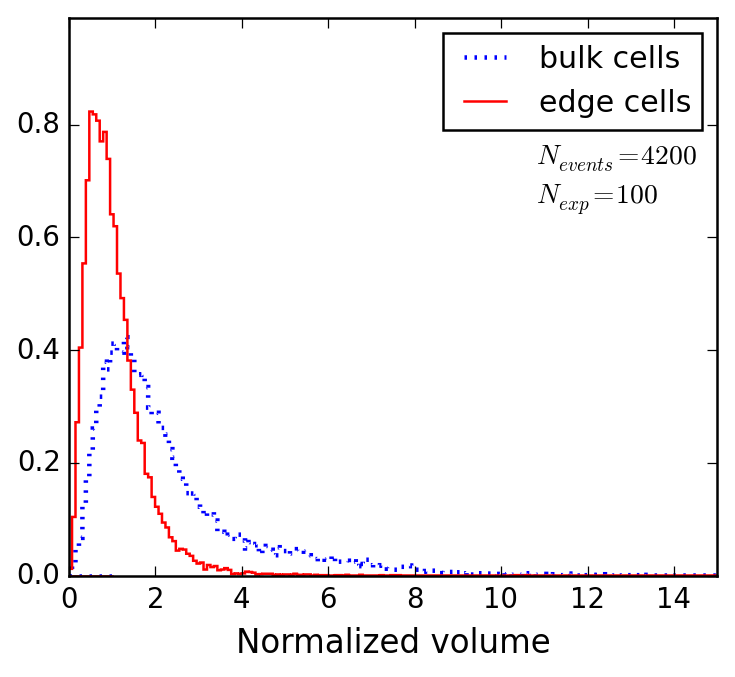} \\ 
\includegraphics[height=5.5cm]{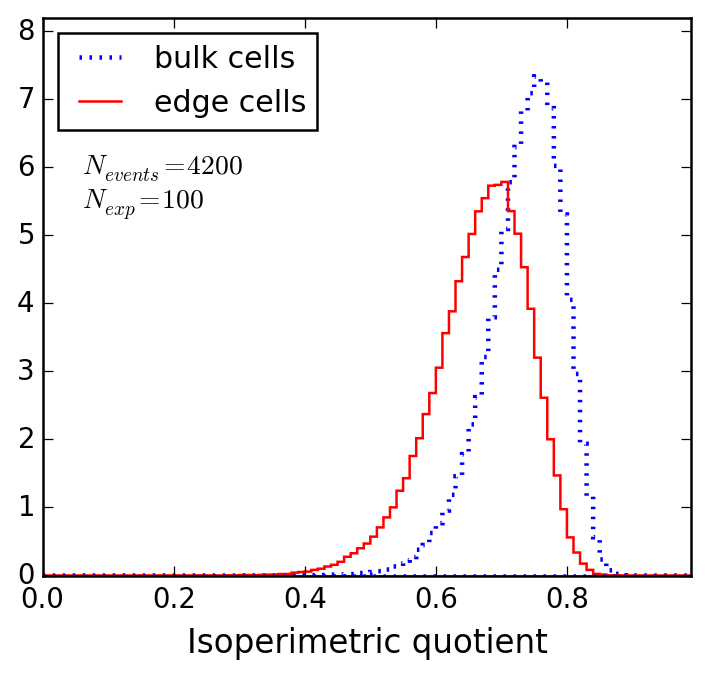} 
\includegraphics[height=5.5cm]{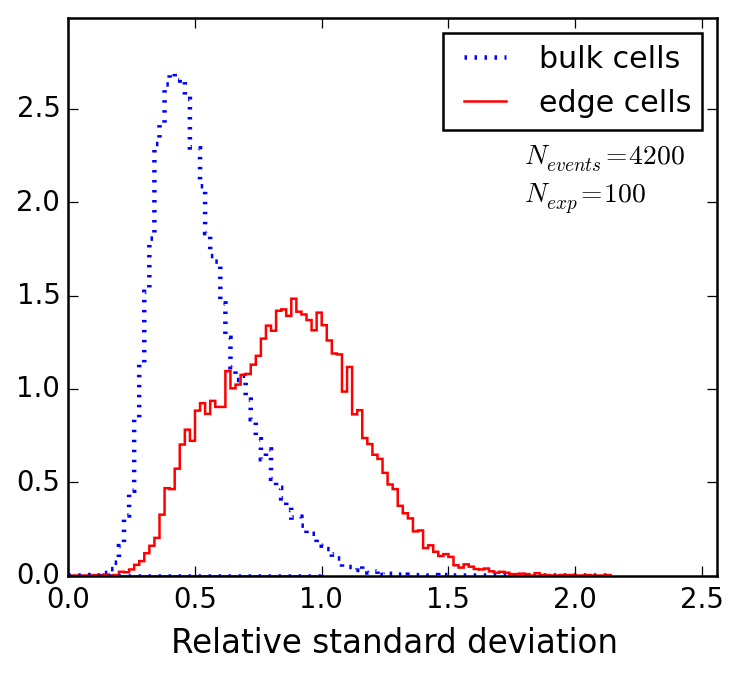}
\caption{\label{fig:sphereenhanced1d} 
The same as Fig.~\ref{fig:plots3D_1d}, but for the example shown in 
Fig.~\ref{fig:sphereenhanced}.
}
\end{figure}

Inspired by the behavior of the phase space density near the boundary, 
we deform the density of data points from the sphere example considered previously in  
section~\ref{sec:3dtoy}.  Performing Voronoi tessellations and studying the
properties of the resulting Voronoi cells, we can develop our insight on 
what is expected from physical examples.  Note that $\Delta_4$ vanishes on 
the phase space boundary and takes its maximum 
somewhere in the bulk.  In other words, the $\Delta_4$ value increases as the 
distance between the data point of interest and the boundary surface 
increases (see also contours in Fig.~\ref{fig:samosa_scatter}).
Although specifying the value of distance does {\it not} determine the
 $\Delta_4$  value, 
it turns out that there exists a positive correlation between the 
two quantities~\cite{del4short}.
To proceed, we make the simplifying \textit{Ansatz} that the distribution of 
the data points inside a unit sphere depends only on the radius, $R$, with an 
enhancement at $R=1$.  Motivated by the form of the probability density 
in~\eqref{eq:delta4PDF}, we introduce the following volume density function 
for the data points inside the unit sphere
\bea
\frac{dP}{dV}\sim\frac{1}{\sqrt{1-R}}\,.
\eea
Now in analogy to (\ref{fstepr3d}), we consider the three-dimensional distribution
\beq
f(\vec{R}) \sim 
\frac{\rho}{\sqrt{1-R}} H(1-R) + H(R-1)H(\sqrt[3]{2}-R).
\label{fenhr3d}
\eeq
Following the example from section~\ref{sec:3dtoy}, we again take the density ratio $\rho=6$
and generate $N_{events}=4200$ events according to (\ref{fenhr3d}). 
Our results are shown in Figs.~\ref{fig:sphereenhanced} and \ref{fig:sphereenhanced1d},
which are the analogues of Figs.~\ref{fig:plots3D_voronoi} and \ref{fig:plots3D_1d}, respectively.

We see that, in principle, all four variables plotted in Fig.~\ref{fig:sphereenhanced}
show some potential for discriminating edge cells. For example, a careful 
inspection of the lower left panel of Fig.~\ref{fig:sphereenhanced} reveals that 
the edge cells appear somewhat elongated, which results in a lower 
isoperimetric quotient, as confirmed by the lower left panel in Fig.~\ref{fig:sphereenhanced1d}.
On the other hand, due to the density enhancement near the boundary, we would also expect 
the edge cells to have smaller normalized volumes. This expectation is also confirmed --- 
in the upper right panels of Figs.~\ref{fig:sphereenhanced} and \ref{fig:sphereenhanced1d}.
Finally, the lower right panels of Figs.~\ref{fig:sphereenhanced} and \ref{fig:sphereenhanced1d}
again demonstrate that the RSD of the neighboring areas is a good discriminator, 
in agreement with our observations from the earlier toy examples.
In order to compare the performance of the four variables investigated in
Figs.~\ref{fig:sphereenhanced} and \ref{fig:sphereenhanced1d}, we 
use the concept of a ROC curve, which is reviewed in Appendix~\ref{sec:ROC-and-ROLL}.

\subsection{Finding density-enhanced sphere boundaries with Voronoi tessellations}
\label{sec:sphere}

\begin{figure}[t]
\centering
\includegraphics[height=5.5cm]{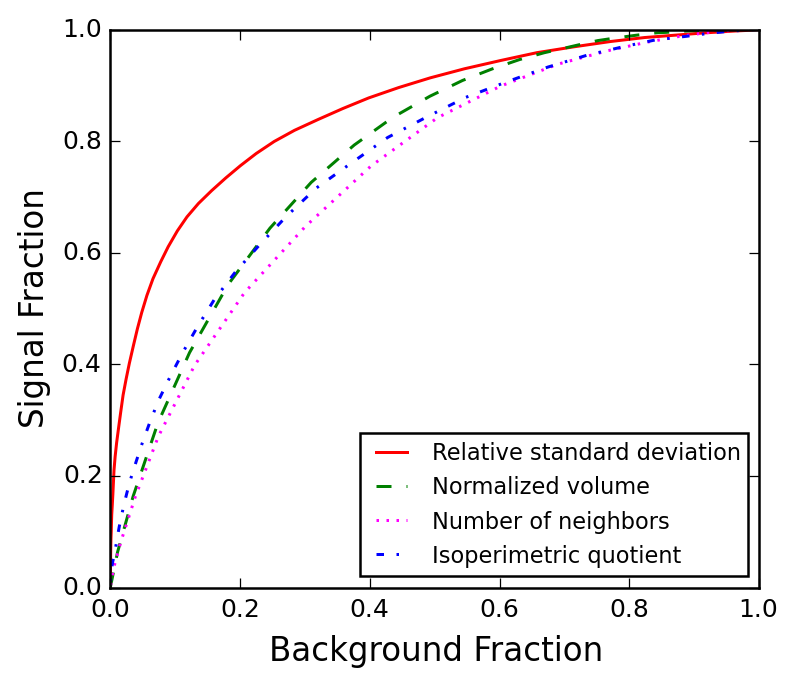} 
\includegraphics[height=5.5cm]{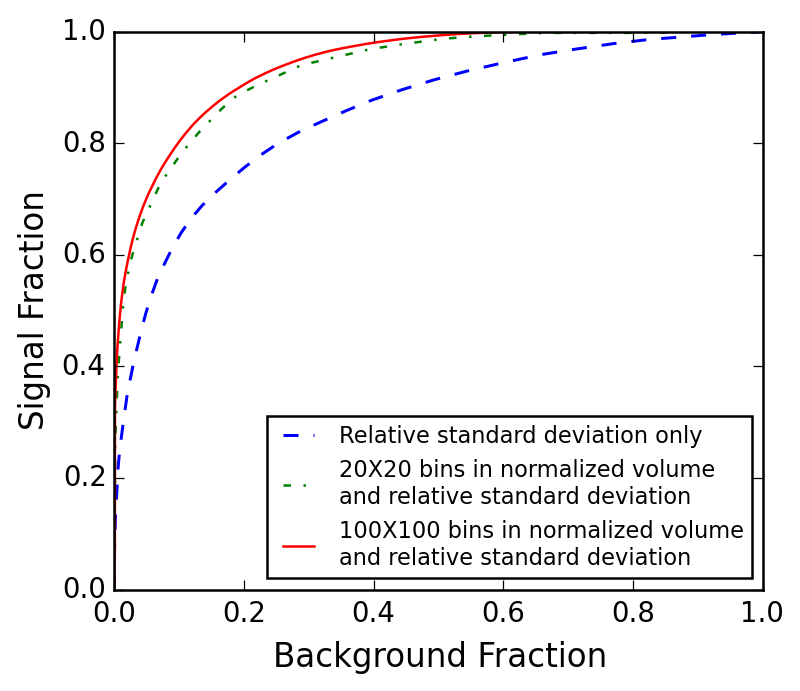} 
\caption{\label{fig:ROC_curves} ROC curves for the toy example 
depicted in Fig.~\ref{fig:sphereenhanced}.
Left: The four different ROC curves resulting from each of the four 
variables shown in Fig.~\ref{fig:sphereenhanced1d}.
Right: Improved ROC curves with optimal two-dimensional cuts in the 
$(\bar v, \bar\sigma)$ plane as illustrated in Fig.~\ref{fig:ROC_bin}: 
with $20\times 20$ binning (green dot-dashed) or $100\times 100$ binning 
(dotted blue).  The blue dashed line is the ROC curve based on the 
$\bar\sigma$ variable alone and is identical to the solid red line in the left panel. }
\end{figure}

\begin{figure}[t]
\centering
\includegraphics[height=5.5cm]{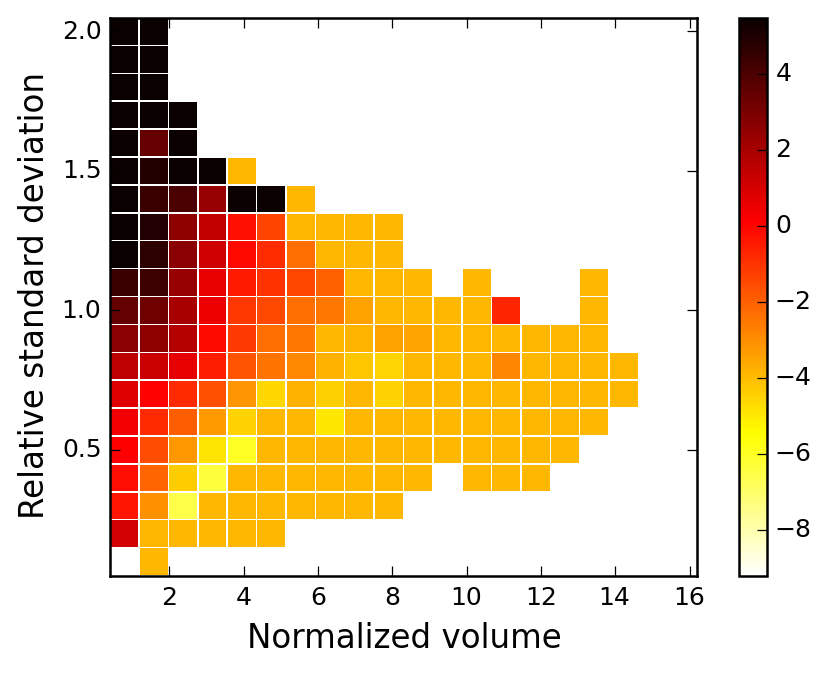} 
\includegraphics[height=5.5cm]{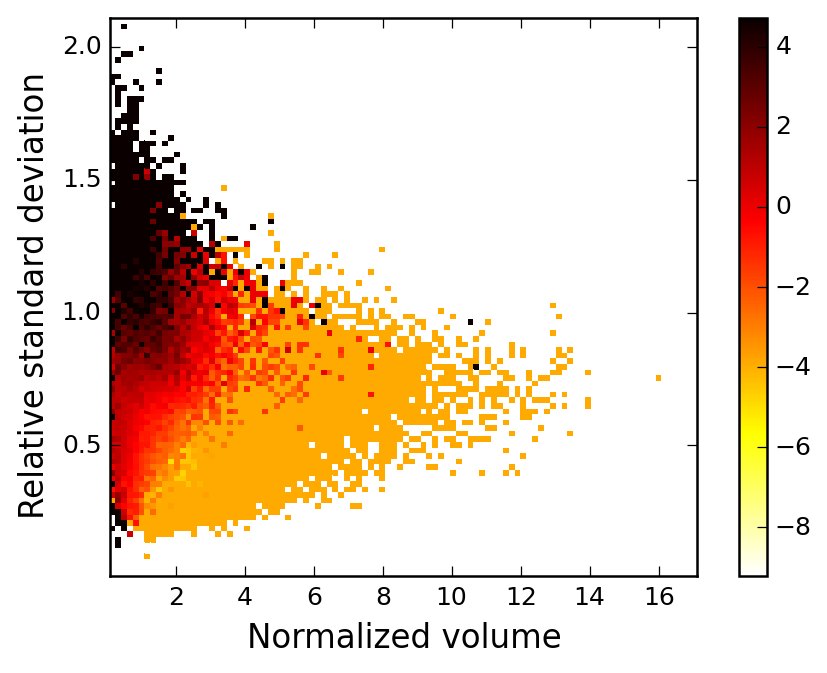}
\caption{\label{fig:ROC_bin} 
Two-dimensional histograms of the expected signal to background 
ratio in the $(\bar v, \bar\sigma)$ plane:
for $20\times 20$ bin (left) and $100\times 100$ bins (right). 
The ROC curves in the right panel of Fig.~\ref{fig:ROC_curves}
were built by successively cutting away the bin with the lowest 
signal-to-background ratio among all remaining bins.
(Alternatively, one could start from zero and successively keep 
adding the bin with the highest signal-to-background ratio 
among all remaining bins.)}
\end{figure}

We now analyze the example of a sphere with an enhanced density 
near the boundary considered in section~\ref{sec:enhanced3dtoy}, in terms of ROC curves.
In the left panel of Fig.~\ref{fig:ROC_curves}, we show the ROC curve for each of the four variables depicted
in Figs.~\ref{fig:sphereenhanced} and \ref{fig:sphereenhanced1d}:
number of neighbors (magenta dotted line), normalized volume (green dashed line),
isoperimetric quotient (blue dot-dashed line) and RSD of neighbor areas (red solid line).
We observe that the RSD outperforms the other three variables, 
in agreement with the conclusions from \cite{Debnath:2015wra} for the two-dimensional case.
However, the other three variables also have a certain degree of discriminating power, as seen in Fig.~\ref{fig:sphereenhanced1d}.
The natural question then is how much additional sensitivity can be gained by considering not just one,
but {\em two} variables simultaneously. We studied the correlations between 
the RSD, $\bar\sigma$, and each of the other three variables, and generally find that they are {\em not} perfectly correlated.
(This makes sense intuitively because the RSD is computed from the
neighbor set, $N_i$, while the other three variables are properties of the individual cell.)
We concluded that, among the three options, 
the normalized volume, $\bar v$, is the most promising, since it appears least correlated with $\bar\sigma$.
Therefore, we expect that the sensitivity will improve once we incorporate the normalized volume, $\bar v$,
in the analysis. This expectation is confirmed in the right panel of Fig.~\ref{fig:ROC_curves},
where we show ``improved" ROC curves based on binning in both $\bar v$ and $\bar \sigma$. 
The procedure, illustrated in Fig.~\ref{fig:ROC_bin}, is as follows.
We consider the $(\bar v, \bar\sigma)$ plane divided into $20\times 20$ bins (left panel of Fig.~\ref{fig:ROC_bin})
or $100\times 100$ bins (right panel of Fig.~\ref{fig:ROC_bin}). 
We expect the signal, i.e., the boundary Voronoi cells, to populate the bins with small volume and relatively large RSD,
while the background, i.e., the bulk cells, are distributed more uniformly throughout the $(\bar v, \bar\sigma)$ plane.
In order to build the optimal ROC curve, we need to determine the signal to background ratio, $S/B$, in each bin,
and design the cuts so that we remove successively the bins with the smallest $S/B$. 
The bins in Fig.~\ref{fig:ROC_bin} are color-coded according to the corresponding 
value of $\log(S/B)$\footnote{``log'' refers to the natural logarithm 
throughout this work.}.
Given the finite statistics, there are bins which have no events (neither signal nor background); they are left uncolored.
For definiteness, the bins which have some signal events, but no background events, are assigned the same value as the 
maximal $\log(S/B)$ value among the bins containing both signal and background events.
Similarly, the bins which had some background events, but no signal events, were assigned the same value as the 
minimal $\log(S/B)$ value among the bins containing both signal and background events.

Fig.~\ref{fig:ROC_bin} shows that, as expected, the bins with the largest $S/B$ (colored in black)
are located in the upper left corner of the plot, corresponding to small $\bar v$ and large $\bar\sigma$.
The spread in the cluster of black-colored bins is indicative of the gain in sensitivity due to simultaneous
consideration of the two variables, $\bar v$ and $\bar\sigma$. According to the right panel
of Fig.~\ref{fig:ROC_curves}, the bulk of the gain is already obtained with a $20\times 20$ grid; 
increasing the number of bins 25 times to a $100\times 100$ grid does not lead to substantial 
improvement. Therefore, in practice, one might want to consider grids of even smaller size,
especially since the true ranking of the bins in terms of $S/B$ depends on the parameter values, e.g., the 
density enhancement on the boundary and the value of $\rho$, which are not known {\em a priori}.
This is why when we consider the physics example in the next section, we shall utilize a smaller 
grid of $15\times 15$ bins in the $(\bar v, \bar\sigma)$ plane (see Fig.~\ref{fig:susy_rank}).

\section{Finding Phase Space Boundaries with Voronoi Tessellations}
\label{sec:phase-space-with-voronoi}

We now use Voronoi tessellations to find the phase space boundary 
for SUSY events at the $14$ TeV LHC. We consider the $2+2+2$ topology from Fig.~\ref{fig:topology},
where, as usual, a (left-handed) squark $X_1=\tilde q$ undergoes a cascade decay through a heavy neutralino,
$X_2=\tilde \chi^0_2$; a slepton, $X_3=\tilde \ell$; and a light neutralino, $X_4=\tilde\chi^0_1$. 
As in Ref.~\cite{Agrawal:2013uka}, we consider the production of a squark in association with 
a neutralino LSP ($\tilde\chi^0_1$). Events were generated with {\sc MadGraph5} \cite{Alwall:2011uj}.  
The mass spectrum that we used was $m_{\tilde q}=350$ GeV, $m_{\tilde\chi^0_2}=300$ GeV, 
$m_{\tilde \ell}=250$ GeV, and $m_{\tilde\chi^0_1}=200$ GeV.\footnote{Despite the relatively low
squark mass, this study point does not seem to be ruled out by the current LHC data. 
Since $\tilde\chi^0_1$ is mostly Bino, the cross-section for squark-neutralino associated production 
is suppressed by the left-handed squark hypercharge, and is only $\sim 5\ {\rm fb}$ at $8$ TeV.
Furthermore, there is no dedicated search for such asymmetric event topologies (squark-LSP production).
If we nevertheless test against a standard SUSY search, e.g.~a signature with a
same-flavour opposite-sign lepton pair, jets and large missing transverse momentum \cite{Aad:2015wqa},
we find a rather low selection efficiency ($\lsim 1\%$), due to the softness of the visible decay products and the 
tendency of the two final state neutralinos to be back to back, thus reducing the amount 
of missing transverse energy.}
The particles visible in the detector are:
a quark jet $v_1=j$, a ``near" lepton $v_2=\ell_n$,
and a ``far" lepton $v_3=\ell_f$. The relevant phase space is then
$(m_{12}, m_{23}, m_{13})\equiv (m_{j\ell_n}, m_{\ell\ell}, m_{j\ell_f})$.
For SUSY signal events, each of these three variables exhibits an upper 
kinematic endpoint. The three endpoint values are given by eqs.~(\ref{m12max}-\ref{m23max}):
\bea
m^2_{j\ell_n,max} &=&  9931\ {\rm GeV}^2, \\ [2mm]
m^2_{\ell\ell,max} &=& 9900\ {\rm GeV}^2,  \\ [2mm]
m^2_{j\ell_f,max} &=& 11700\ {\rm GeV}^2.
\eea
From the three-dimensional point of view, the signal events 
populate the interior of a compact region in the $(m_{j\ell_n}, m_{\ell\ell}, m_{j\ell_f})$
space, whose boundary is given by the constraint \cite{Costanzo:2009mq,Kim:2015bnd}
\beq
\hat{m}^2_{j\ell_f} = 
\left[ \sqrt{ \hat{m}^2_{\ell\ell} \left(1- \hat{m}^2_{j\ell_n}\right) }
\pm  \frac{m_{\tilde \ell}}{m_{\tilde\chi^0_2}} 
\sqrt{ \hat{m}^2_{j\ell_n} \left(1-\hat{m}^2_{\ell\ell}\right)} \right]^2 ,
\label{eq:samosa2}                               
\eeq
which, for convenience, is written in terms of unit-normalized variables (see also (\ref{defxi}))
\bea
\hat{m}_{j\ell_n} &=&  \frac{m_{j\ell_n}}{ m_{j\ell_n,max}}, \\ [2mm]  
\hat{m}_{\ell\ell} &=&  \frac{m_{\ell\ell}}{ m_{\ell\ell,max}}, \\ [2mm]  
\hat{m}_{j\ell_f} &=&  \frac{m_{j\ell_f}}{ m_{j\ell_f,max}}.  
\eea
Our main goal in this section will be to test the algorithms from the previous sections 
for tagging the Voronoi cells in the vicinity of the boundary surface (\ref{eq:samosa2}).
In addition to the signal events from squark-neutralino associated production with the squark decaying as in Fig.~\ref{fig:topology},
we shall also consider a representative number of background events. 
In order to make contact with the results from the previous sections, 
in section~\ref{sec:idealbknd} we first take the background events to be uniformly distributed in 
mass-squared phase-space, and we ignore the combinatorial background.
Then in section~\ref{sec:ttbarcomb} we study a more realistic case, where the SM background 
is comprised of dilepton $t\bar{t}$ events and we also account for the combinatorial problem with the two leptons.

\subsection{An example with uniform background and no combinatorics}
\label{sec:idealbknd}

As in the other two three-dimensional examples considered in sections~\ref{sec:3dtoy} and \ref{sec:sphere}, 
in this section we include ``SM physics background" events, 
which we take to be uniformly distributed everywhere throughout the mass-squared phase-space
$(m^2_{j\ell_n}, m^2_{\ell\ell}, m^2_{j\ell_f})$ 
and normalized so that the density contrast across the boundary
(\ref{eq:samosa2}) is equal to $\rho=4$.
Note that in this scenario the interior ``bulk" events and the 
``edge" cells on the surface boundary (\ref{eq:samosa2})
consist of both SUSY signal and SM background events.

\begin{figure}[t]
\centering
\includegraphics[width=4.9cm]{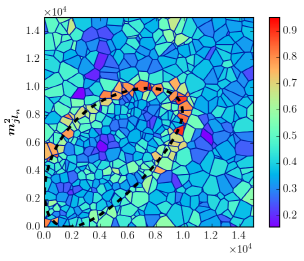} 
\includegraphics[width=4.7cm]{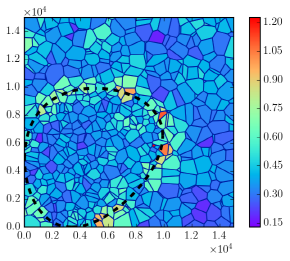} 
\includegraphics[width=4.6cm]{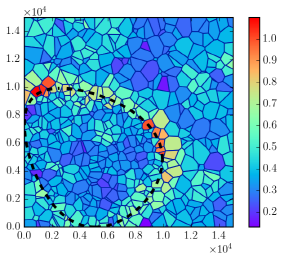} \\
\includegraphics[width=5.0cm]{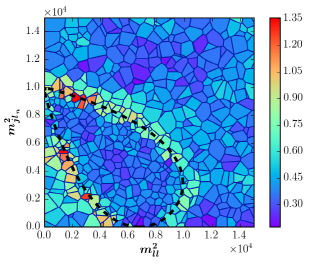} 
\includegraphics[width=4.7cm]{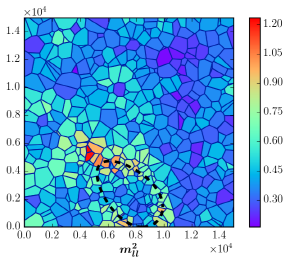} 
\includegraphics[width=4.7cm]{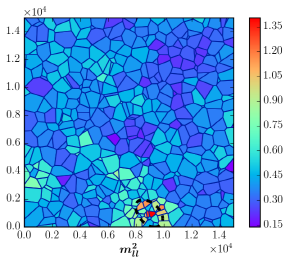} 
\caption{\label{fig:susy_sigma} Two-dimensional slices of the relevant 
three-dimensional phase space of the SUSY-like cascade decay in Fig.~\ref{fig:topology}.
Each slice is in the $(m_{\ell\ell}^2,m_{j\ell_n}^2)$
plane at a fixed value of 
$m_{j\ell_f}^2=2000$ ${\rm GeV}^2$ (upper left panel);
$m_{j\ell_f}^2=4000$ ${\rm GeV}^2$ (upper middle panel);
$m_{j\ell_f}^2=6000$ ${\rm GeV}^2$ (upper right panel);
$m_{j\ell_f}^2=8000$ ${\rm GeV}^2$ (lower left panel);
$m_{j\ell_f}^2=10000$ ${\rm GeV}^2$ (lower middle panel); and
$m_{j\ell_f}^2=11000$ ${\rm GeV}^2$ (lower right panel).
As in Figs.~\ref{fig:plots3D_voronoi} and \ref{fig:sphereenhanced}, 
the two-dimensional cells seen in the plots result from the intersection of 
the projective plane with the three-dimensional Voronoi cells, and are 
color-coded by the value of $\bar\sigma_i$ for the corresponding 
three-dimensional Voronoi cell.
}
\end{figure}

\begin{figure}[t]
\centering
\includegraphics[width=5.2cm]{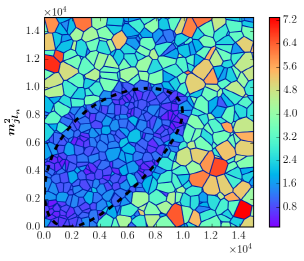} 
\includegraphics[width=4.7cm]{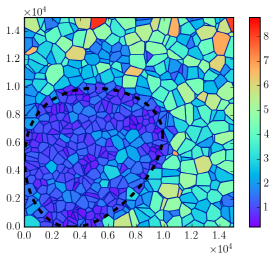} 
\includegraphics[width=4.8cm]{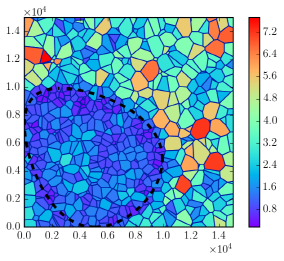} \\
\includegraphics[width=5.0cm]{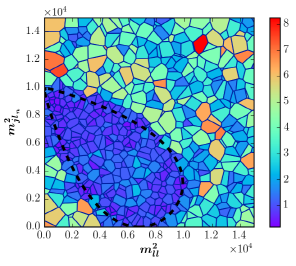} 
\includegraphics[width=4.7cm]{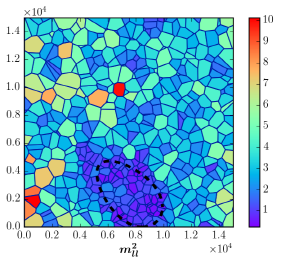} 
\includegraphics[width=4.8cm]{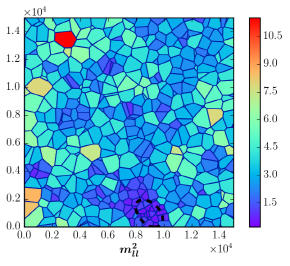} 
\caption{\label{fig:susy_volume} 
The same as Fig.~\ref{fig:susy_sigma}, but color-coding the cells according to the 
normalized volume, $\bar{v}_i$, defined in (\ref{eq:sv-scaled-def}).
}
\end{figure}

\begin{figure}[t]
\centering
\includegraphics[height=5.5cm]{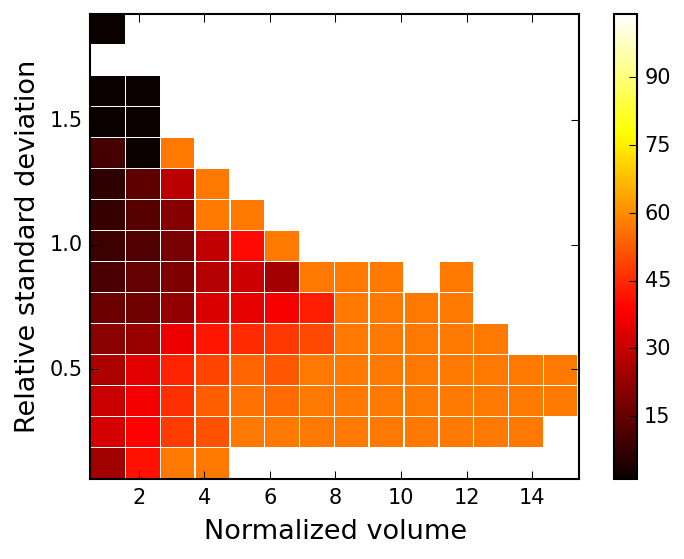} 
\includegraphics[height=5.5cm]{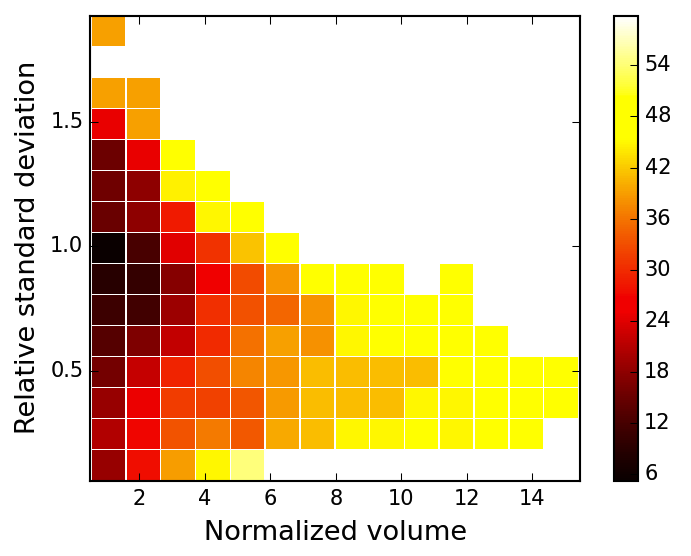} 
\caption{\label{fig:susy_rank} The integer ranking of the $15\times 15$ bins 
in the $(\bar v, \bar\sigma)$ plane according to
their signal-to-background ratio. Left: The ranking for the case of $\rho=4$. 
Right: the average ranking for the 
cases of $\rho=1.2$, $\rho=1.5$, $\rho=2.0$, $\rho=3.0$, and $\rho=4.0$. 
}
\end{figure}

\begin{figure}[t]
\centering
\includegraphics[height=5.5cm]{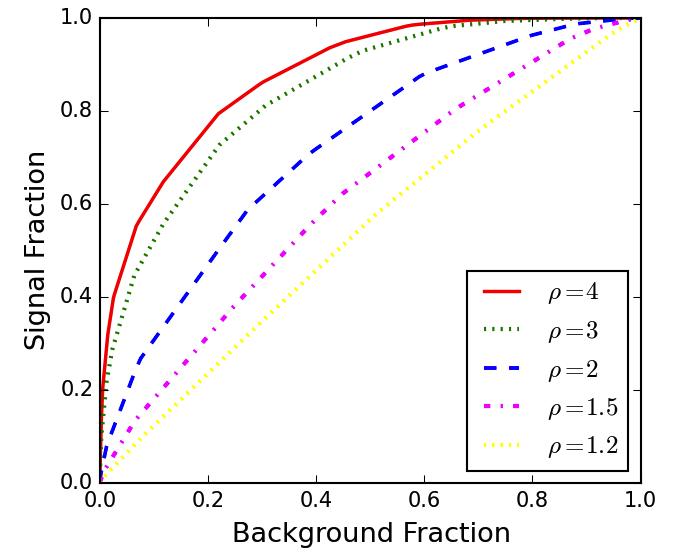} 
\includegraphics[height=5.5cm]{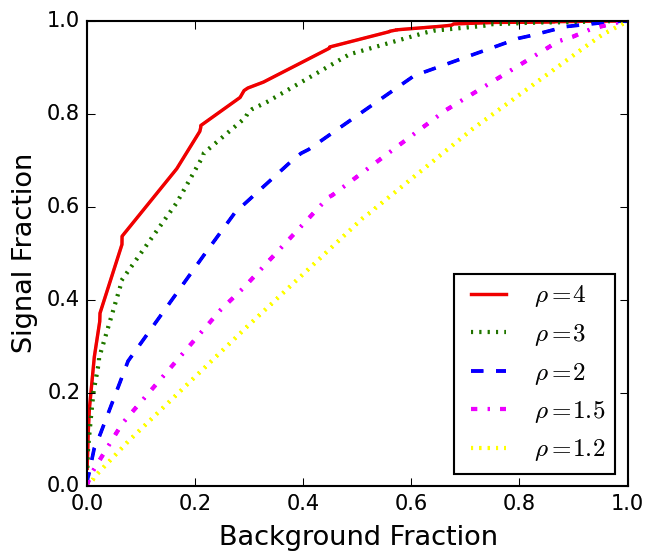} 
\caption{\label{fig:ROC_rank} The same as the ROC curves shown in the right 
panel of Fig.~\ref{fig:ROC_curves}, 
but for a $15\times 15$ grid. Left: the ranking of the bins in constructing 
the ROC curve was done with the 
correct value of $\rho$ (shown on each curve) used in generating the ``data". 
Right: the ranking of the bins
was always done according to the ``average" ranking shown in the right panel 
of Fig.~\ref{fig:susy_rank}.
}
\end{figure}

As before, we visualize the resulting Voronoi tessellation by presenting two-dimensional slices of the relevant 
three-dimensional phase space, in this case\footnote{It is known that working in terms of the {\em squared} masses 
is more convenient and intuitive \cite{Costanzo:2009mq,Kim:2015bnd}.} $(m^2_{j\ell_n}, m^2_{\ell\ell}, m^2_{j\ell_f})$. 
In Figs.~\ref{fig:susy_sigma} and \ref{fig:susy_volume} we show six slices
in the $(m_{\ell\ell}^2,m_{j\ell_n}^2)$ plane at a fixed value of $m^2_{j\ell_f}$ as follows:  
$m_{j\ell_f}^2=2000$ ${\rm GeV}^2$ (upper left panel);
$m_{j\ell_f}^2=4000$ ${\rm GeV}^2$ (upper middle panel);
$m_{j\ell_f}^2=6000$ ${\rm GeV}^2$ (upper right panel);
$m_{j\ell_f}^2=8000$ ${\rm GeV}^2$ (lower left panel);
$m_{j\ell_f}^2=10000$ ${\rm GeV}^2$ (lower middle panel);
$m_{j\ell_f}^2=11000$ ${\rm GeV}^2$ (lower right panel).
As in Figs.~\ref{fig:plots3D_voronoi} and \ref{fig:sphereenhanced}, 
the two-dimensional cells seen in the plots result from the intersection of 
the projective plane with the three-dimensional Voronoi cells and are 
color-coded by the value of the RSD, $\bar\sigma_i$, defined in (\ref{defvar}) 
(in Fig.~\ref{fig:susy_sigma}) 
or the normalized volume $\bar{v}_i$ defined in (\ref{eq:sv-scaled-def}) 
(in Fig.~\ref{fig:susy_volume}) of the corresponding three-dimensional Voronoi cell.
Just as in the case of the density-enhanced sphere considered in section~\ref{sec:sphere},
Figs.~\ref{fig:susy_sigma} and \ref{fig:susy_volume} suggest that the edge cells
near the phase space boundary (\ref{eq:samosa2})
are characterized both by a large value of $\bar\sigma_i$
and a small value of $\bar{v}_i$. Therefore, in designing a selection cut to pick up edge cells, 
it makes sense to consider both of these two variables at the same time.
This is illustrated in the left panel of Fig.~\ref{fig:susy_rank}, which is the analogue of 
Fig.~\ref{fig:ROC_bin} for this case.
We consider a moderately large $15\times 15$ grid in the
$(\bar v, \bar\sigma)$ plane and rank the resulting bins according to their
signal-to-background ratio\footnote{We remind the reader that for the purpose of
the boundary detection analysis performed here, ``signal" refers to edge cells, 
while ``background" refers to bulk cells. The interior bulk cells and the edge cells
arise from both SUSY signal and SM background events, 
while the exterior bulk cells are only due to SM background events.} as follows.
The bin with the highest $S/B$ is assigned rank $1$, while the bin with the lowest 
$S/B$ is assigned rank $15\times 15=225$. In case of a tie between several bins, 
each bin is assigned the same average rank. Finally, bins with no events at all are 
ranked at the very bottom.\footnote{This scheme is analogous to a college football 
ranking poll among 225 universities, some of which do not have a football program.} 
We observe that, similarly to Fig.~\ref{fig:ROC_bin}, the highest ranked bins in terms of
$S/B$ appear at large values of $\bar\sigma$ and small values of $\bar{v}$.
Using the obtained bin ranking, we can build the corresponding ROC curve, shown 
with the red solid line in the left panel of Fig.~\ref{fig:ROC_rank}, which in some sense 
is the ``ideal" ROC curve that could be achieved, if $\bar\sigma$ and $\bar{v}$
were the only discriminating variables under consideration.

One can now repeat the same procedure for different values of $\rho$.
We show four more examples in the left panel of Fig.~\ref{fig:ROC_rank},
with increasingly pessimistic values for the density contrast $\rho$:
$3$, $2$, $1.5$ and $1.2$. As expected, the ROC curves become 
progressively worse, as quantified in the figure. With regards to the bin ranking in each case, 
we notice the following trend --- the highest ranked bins remain at the lowest possible values 
of $\bar{v}$, but slide down the $\bar\sigma$ axis to slightly lower values of the RSD, 
near, or even below $\bar\sigma\sim 1$. This is easy to understand intuitively --- 
as $\rho$ is decreased, the number densities on both sides of the surface boundary 
become more similar, and there is less variation between the sizes of bulk cells on the inside 
and on the outside. The fact that the bin ranking derived from our Monte Carlo simulations 
depends on the value of $\rho$ poses an important conceptual problem with this procedure --- 
when the analysis is performed on real data, we will not know the actual value of $\rho$,
and, hence, we will not be certain which particular ordering of bins to use.
Nevertheless, the fact that the highest ranked bins are clustered, more or less, in the same location, 
suggests a possible resolution: we can simply average our results obtained for several different 
values of $\rho$ and use the resulting average rank for each bin, 
which is shown in the right panel of Fig.~\ref{fig:susy_rank}. The corresponding
ROC curves derived with the help of this ``average" bin ranking procedure are shown 
in the right panel of Fig.~\ref{fig:ROC_rank}. Comparing the two panels of Fig.~\ref{fig:ROC_rank},
we see that the ROC curves based on the average ranking are only slightly degraded compared 
to the ``ideal" case.

\begin{figure}[t]
\centering
\includegraphics[width=4.9cm]{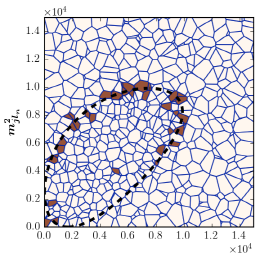} 
\includegraphics[width=4.5cm]{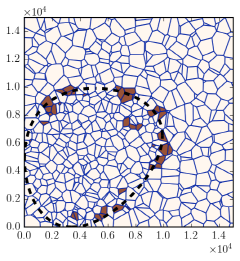} 
\includegraphics[width=4.5cm]{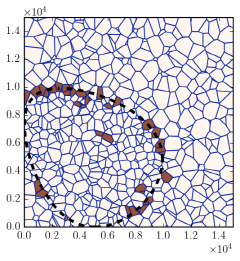} \\
\includegraphics[width=4.9cm]{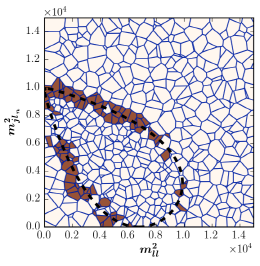} 
\includegraphics[width=4.5cm]{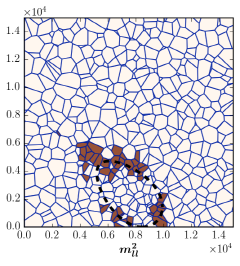} 
\includegraphics[width=4.5cm]{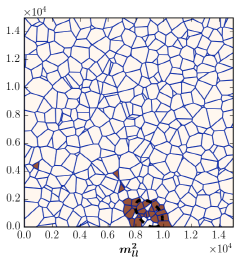} 
\caption{\label{fig:susy_cut} The Voronoi cells which pass the two-dimensional 
selection cut requiring the cell to belong to one of the top 5 bins in terms 
of signal-to-background ratio for the correct choice of $\rho=4$
(see the left panel in Fig.~\ref{fig:susy_rank}).
}
\end{figure}

We are now ready to start designing selection cuts for the edge cells.
One possibility is to select the cells which fall into the a predetermined number, 
$N_{top\ bins}$, of the highest ranked bins in the $(\bar v, \bar\sigma)$ plane. 
If we ``cheat", i.e., use the correct value of $\rho$ for the ranking 
(as in the left panel of  Fig.~\ref{fig:susy_rank}), we obtain the result shown in
Fig.~\ref{fig:susy_cut}, where we have chosen $N_{top\ bins}=5$. 
In general, the tagged cells are distributed throughout 
the volume of the three dimensional phase space, so for illustration purposes we again
use the same six two-dimensional slices as in Figs.~\ref{fig:susy_sigma} and \ref{fig:susy_volume}.
We observe that the procedure is pretty efficient in tagging edge cells, and occasionally 
we tag an isolated bulk cell. Of course, for such a low value of $N_{top\ bins}$,
not all edge cells will pass the cut, which will cause the boundary contours 
(marked with black dashed lines) to appear segmented and incomplete.
By increasing the value of $N_{top\ bins}$, we can obviously tag more edge cells and 
eventually ``close" those contours, but at the cost of 
more mistagged bulk cells.

\begin{figure}[t]
\centering
\includegraphics[width=4.9cm]{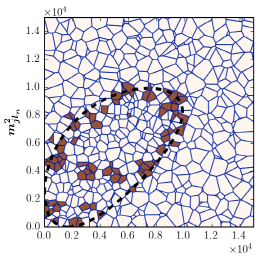} 
\includegraphics[width=4.5cm]{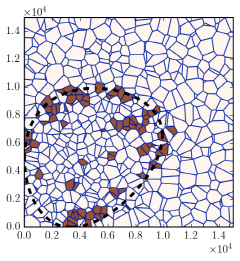} 
\includegraphics[width=4.5cm]{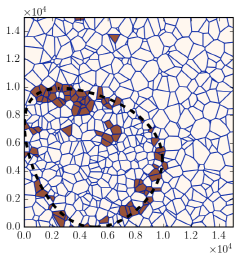} \\
\includegraphics[width=4.9cm]{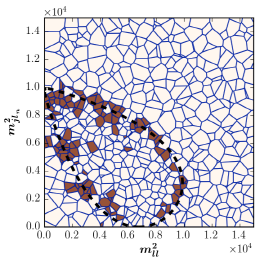} 
\includegraphics[width=4.5cm]{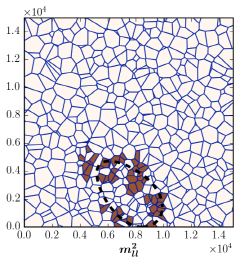} 
\includegraphics[width=4.5cm]{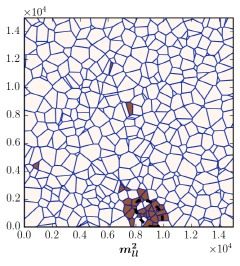} 
\caption{\label{fig:susy_cut_ave} The same as Fig.~\ref{fig:susy_cut}, but 
using the {\em average} 
ranking of the bins shown in the right panel of Fig.~\ref{fig:susy_rank}.
}
\end{figure}

Since the true value of $\rho$ will be unknown, the plots in Fig.~\ref{fig:susy_cut}
are for academic purposes only. The more realistic situation is depicted in the
analogous Fig.~\ref{fig:susy_cut_ave}, where we again choose $N_{top\ bins}=5$,
only this time we use the average bin ranking from the right panel of Fig.~\ref{fig:susy_rank}.
The result in Fig.~\ref{fig:susy_cut_ave} is slightly worse than Fig.~\ref{fig:susy_cut} ---
while we do find a higher rate for mistagging bulk cells (typically in the interior), the majority
of the tagged cells are very close to the surface boundary, suggesting that 
this is a promising technique for identifying edge cells.

\subsection{A realistic example with $t\bar{t}$ background and combinatorics}
\label{sec:ttbarcomb}

We now repeat the exercise from the previous section~\ref{sec:idealbknd} with two improvements.
First, we have to address the combinatorial problem of distinguishing the ``near" and ``far" lepton.
The standard approach in the literature is to trade the original variables $m_{j\ell_n}$ and $m_{j\ell_f}$ 
for the ordered pair \cite{Allanach:2000kt,Gjelsten:2004ki,Matchev:2009iw}
\begin{subequations}
\label{lowhigh}
\begin{align}
m_{j\ell}(high) &\equiv  \max\left\{ m_{j\ell_n}, m_{j\ell_f} \right\}   ,  \\
m_{j\ell}(low) &\equiv  \min\left\{ m_{j\ell_n}, m_{j\ell_f} \right\}  .
\end{align}
\end{subequations}

\begin{figure}[t]
\centering
\includegraphics[width=3.6cm]{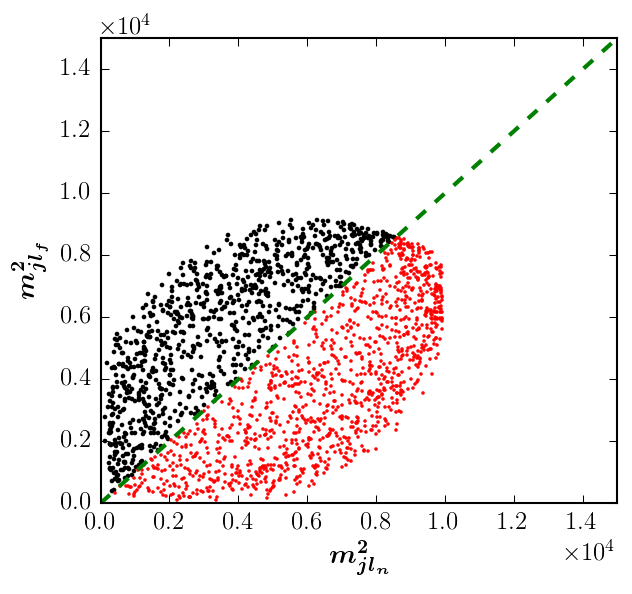} 
\includegraphics[width=3.6cm]{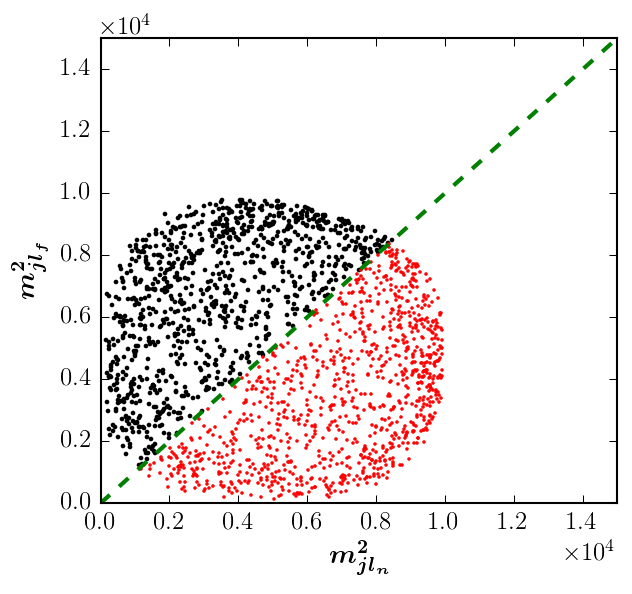} 
\includegraphics[width=3.6cm]{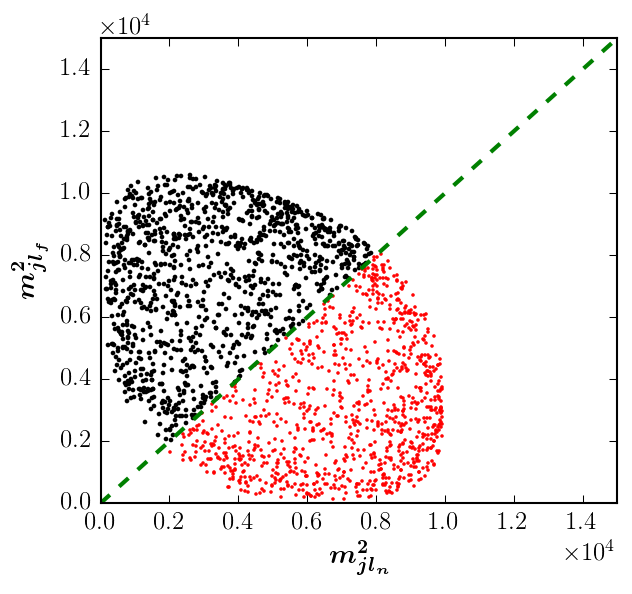} 
\includegraphics[width=3.6cm]{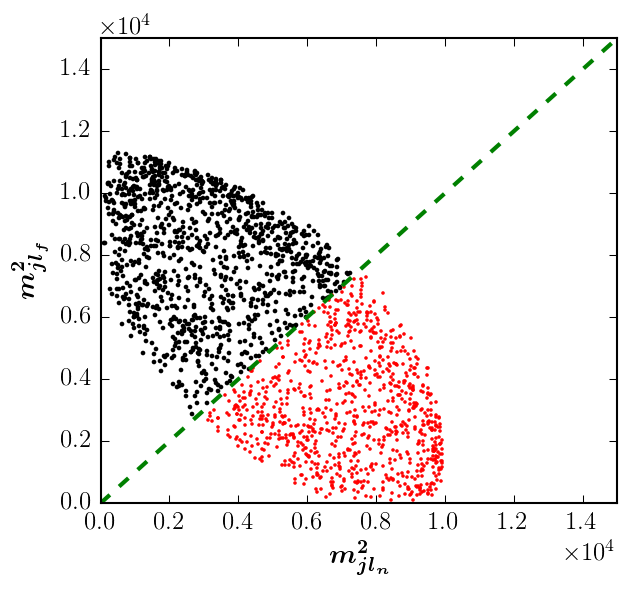} \\
\includegraphics[width=3.6cm]{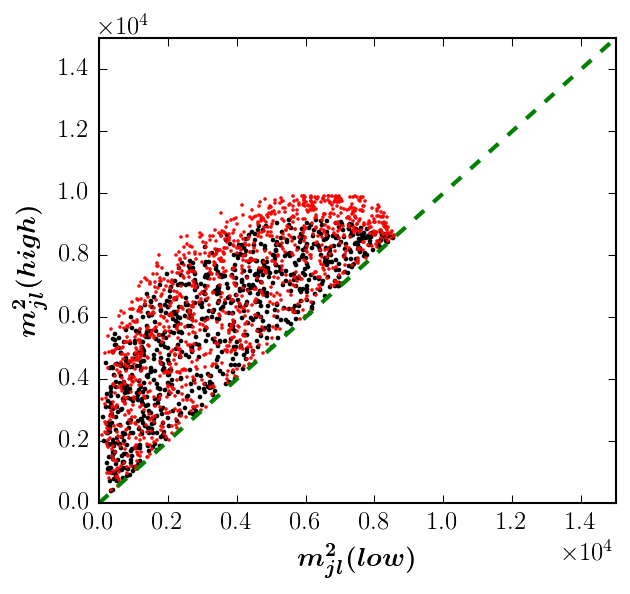} 
\includegraphics[width=3.6cm]{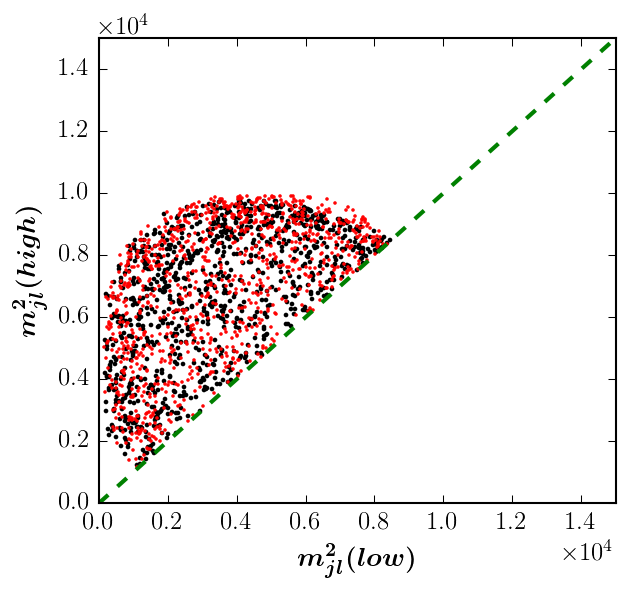} 
\includegraphics[width=3.6cm]{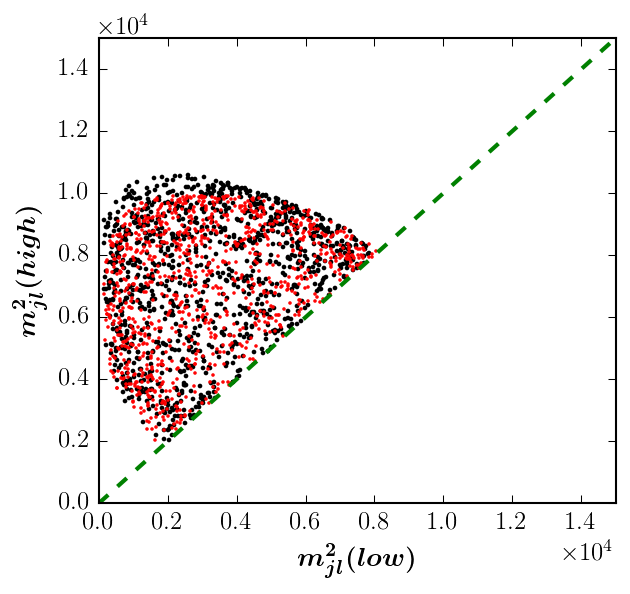} 
\includegraphics[width=3.6cm]{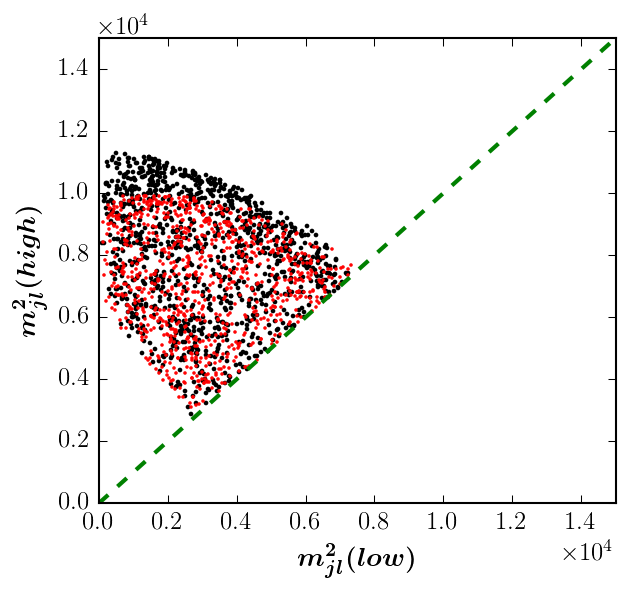}\\
\includegraphics[width=3.6cm]{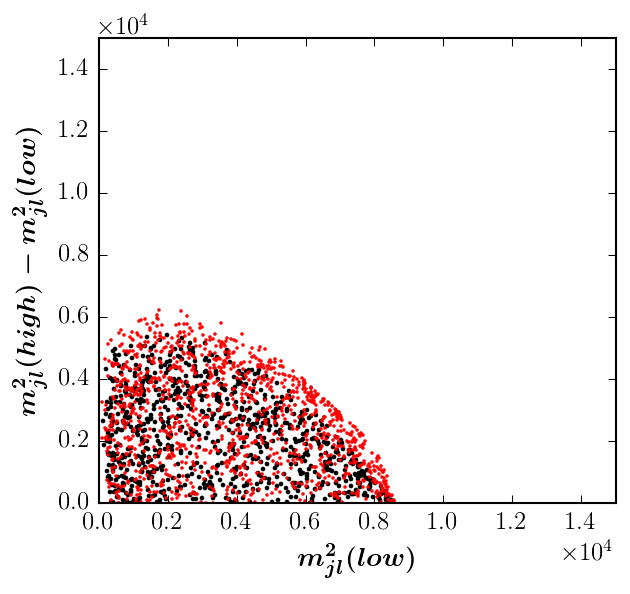} 
\includegraphics[width=3.6cm]{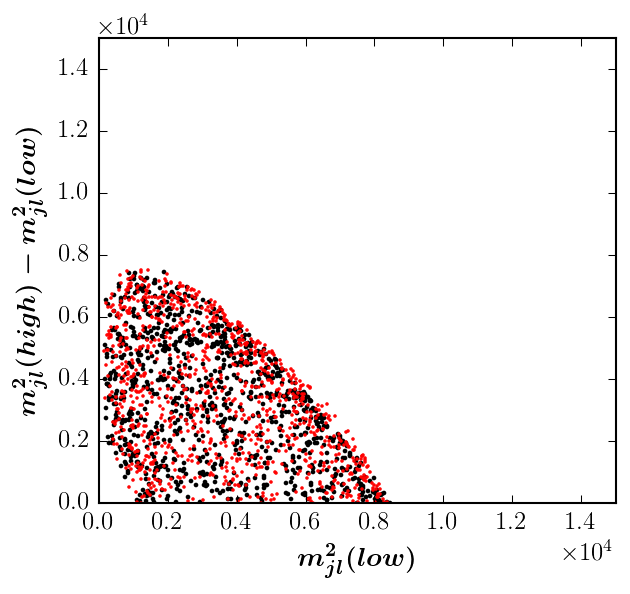} 
\includegraphics[width=3.6cm]{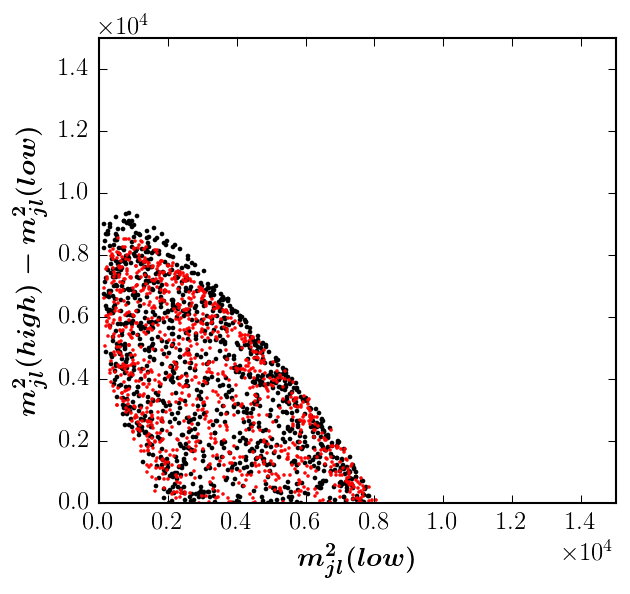} 
\includegraphics[width=3.6cm]{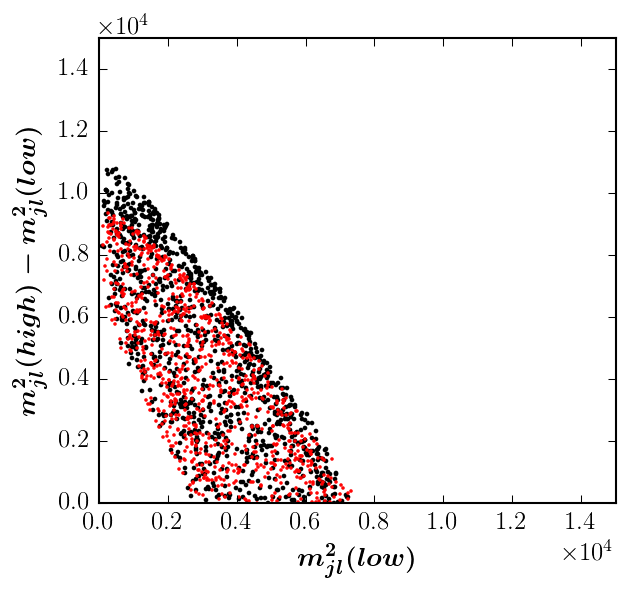} \\
\includegraphics[width=3.6cm]{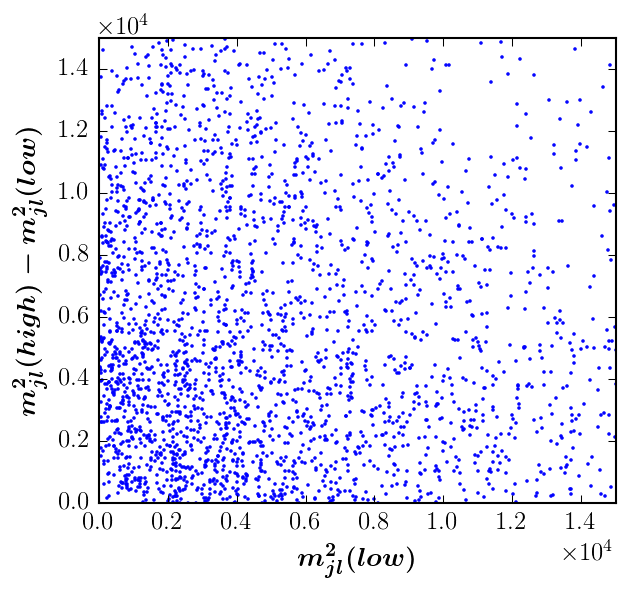} 
\includegraphics[width=3.6cm]{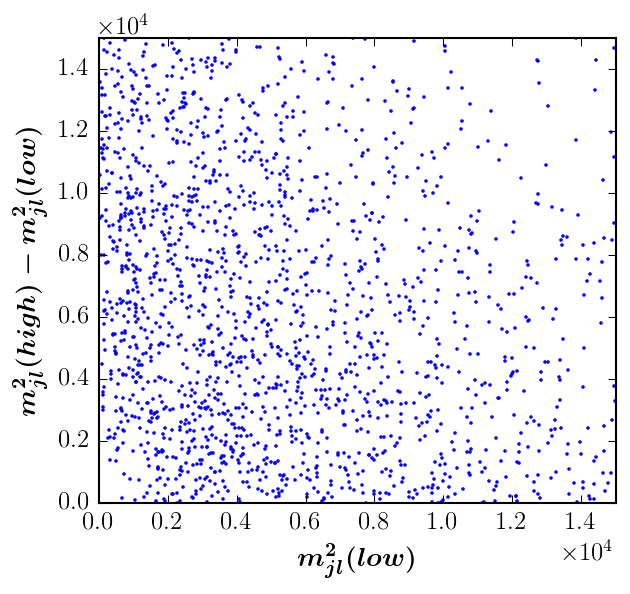} 
\includegraphics[width=3.6cm]{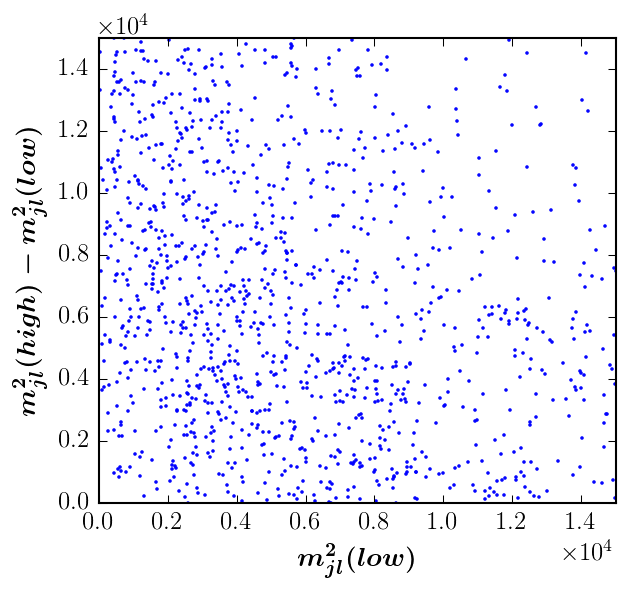} 
\includegraphics[width=3.6cm]{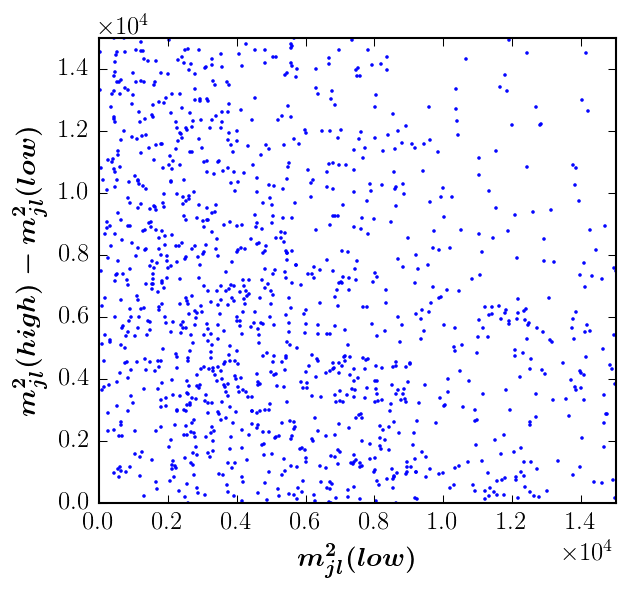} 
\caption{\label{fig:sig_ttbar} Scatter plots of signal events (black and red points, top three rows) and dilepton $t\bar{t}$ 
events (blue points, bottom row), for different ranges of the dilepton invariant mass squared:
$1,000\ {\rm GeV}^2 \le m_{\ell\ell}^2 \le 3,000\ {\rm GeV}^2$ (first column);
$3,000\ {\rm GeV}^2 \le m_{\ell\ell}^2 \le 5,000\ {\rm GeV}^2$ (second column);
$5,000\ {\rm GeV}^2 \le m_{\ell\ell}^2 \le 7,000\ {\rm GeV}^2$ (third column); and
$7,000\ {\rm GeV}^2 \le m_{\ell\ell}^2 \le 9,000\ {\rm GeV}^2$ (fourth column).
In the first row, the signal events are plotted in the plane of $(m^2_{j\ell_n}, m^2_{j\ell_f})$
and colored red (black) if $m^2_{j\ell_n} \ge m^2_{j\ell_f}$ ($m^2_{j\ell_n} \le m^2_{j\ell_f}$). 
The same points are then plotted in the planes of $(m^2_{j\ell}(low), m^2_{j\ell}(high))$ (second row)
and $(m^2_{j\ell}(low), m^2_{j\ell}(high)-m^2_{j\ell}(low))$ (third row).
The background events in the fourth row are plotted in the plane of 
$(m^2_{j\ell}(low), m^2_{j\ell}(high)-m^2_{j\ell}(low))$.
}
\end{figure}

This reordering procedure is pictorially illustrated with the first two rows of plots in Fig.~\ref{fig:sig_ttbar},
where we show scatter plots of signal events for different ranges of the third invariant mass variable (the dilepton mass): 
$1,000\ {\rm GeV}^2 \le m_{\ell\ell}^2 \le 3,000\ {\rm GeV}^2$ (first column);
$3,000\ {\rm GeV}^2 \le m_{\ell\ell}^2 \le 5,000\ {\rm GeV}^2$ (second column);
$5,000\ {\rm GeV}^2 \le m_{\ell\ell}^2 \le 7,000\ {\rm GeV}^2$ (third column); and
$7,000\ {\rm GeV}^2 \le m_{\ell\ell}^2 \le 9,000\ {\rm GeV}^2$ (fourth column).
In the first row, the signal events are plotted in the original plane of $(m^2_{j\ell_n}, m^2_{j\ell_f})$,
and the points are color-coded as follows. The points below the diagonal $45^\circ$ line, 
which have $m^2_{j\ell_n} \ge m^2_{j\ell_f}$, are colored in red, while the remaining points 
above the diagonal $45^\circ$ line with $m^2_{j\ell_n} \le m^2_{j\ell_f}$, are colored in black.
The same data is then plotted in the second row of Fig.~\ref{fig:sig_ttbar} in the plane of 
$(m^2_{j\ell}(low), m^2_{j\ell}(high))$. Notice that the effect of the reordering procedure (\ref{lowhigh})
is to leave all the black points in place, while interchanging the $x$ and $y$ coordinates 
of the red points\footnote{One can think of this operation as ``origami folding" the scatter 
plot along the diagonal $45^\circ$ line \cite{Burns:2009zi}.}. 
After the reordering (\ref{lowhigh}), half of the plane on each plot is left blank.
In order to avoid such voids, in the third row of Fig.~\ref{fig:sig_ttbar} we replot the data in
the plane of $(m^2_{j\ell}(low), m^2_{j\ell}(high)-m^2_{j\ell}(low))$, which is fully accessible.

As expected, the scatter plots in the third row of Fig.~\ref{fig:sig_ttbar} exhibit boundary lines,
which we can target with our edge-detecting method. In fact, each plot has two such boundaries where 
the signal number density sharply changes  --- one for the red points and another for the black points. 
At low values of $m_{\ell\ell}$ the ``red" (``black") boundary line is an external (internal) boundary, while
for high values of $m_{\ell\ell}$ it is the other way around. At intermediate values of $m_{\ell\ell}$ the two 
boundaries are very close to each other and that is where we expect the edge detection method to perform best.

Having thus taken care of the combinatorial problem, we now also improve our treatment of the background ---
instead of uniformly distributed background events as in section~\ref{sec:idealbknd}, we consider 
dilepton events from $t\bar{t}$ production. The corresponding scatter plots are shown in the fourth (last) row of Fig.~\ref{fig:sig_ttbar}.
Since there are 2 $b$-jets, each background event contributes two entries to the scatter plot.
We see that within the relevant range of the plotted variables $m^2_{j\ell}(low)$ and $m^2_{j\ell}(high)-m^2_{j\ell}(low)$,
the distribution of the background events is somewhat uniform, with some noticeable clustering near the origin.

\begin{figure}[t]
\centering
\includegraphics[width=4.9cm]{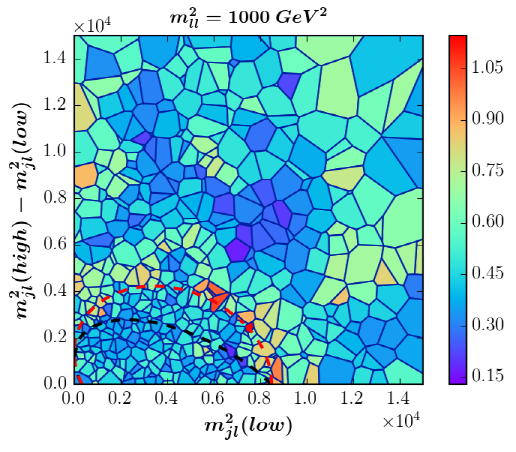} 
\includegraphics[width=4.9cm]{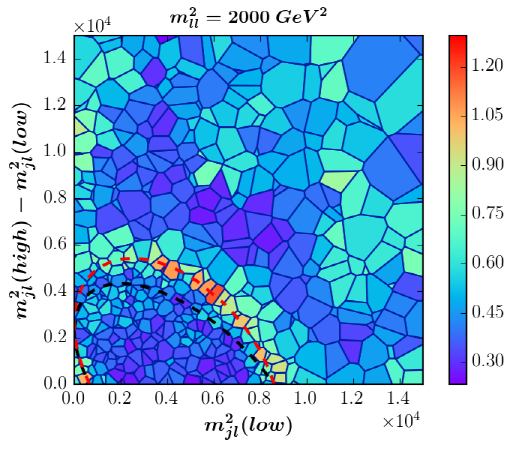} 
\includegraphics[width=4.9cm]{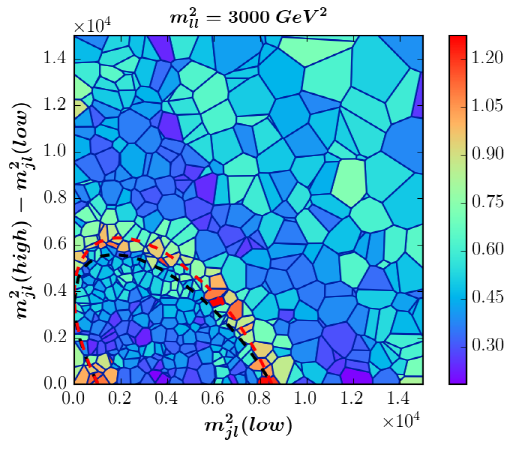} \\
\includegraphics[width=4.9cm]{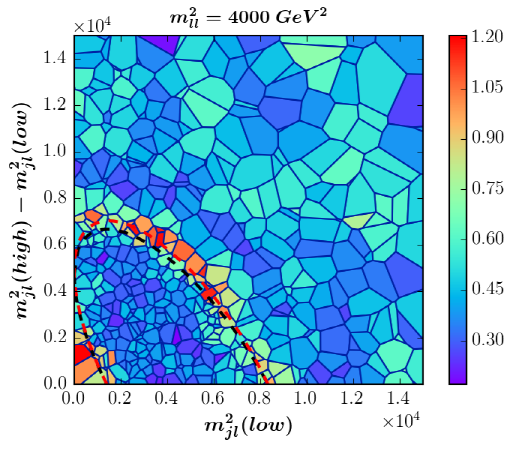} 
\includegraphics[width=4.9cm]{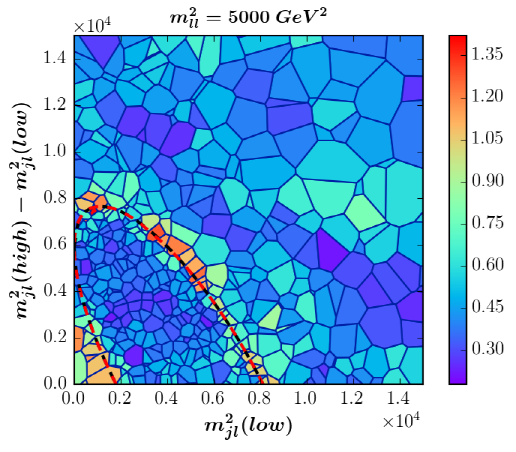} 
\includegraphics[width=4.9cm]{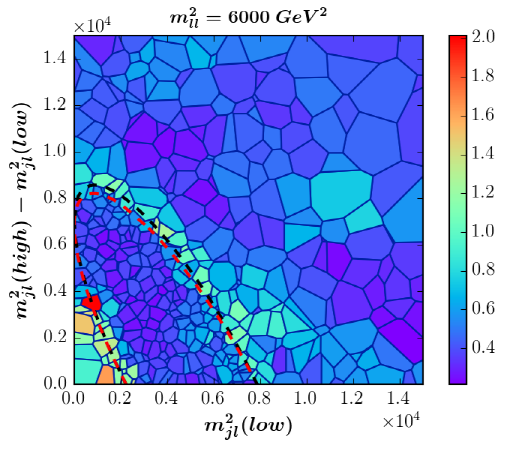}\\
\includegraphics[width=4.9cm]{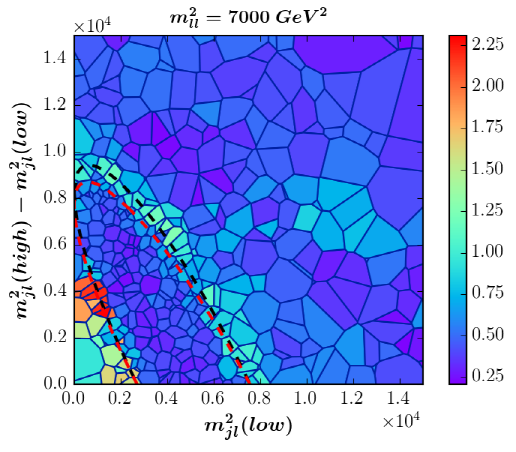} 
\includegraphics[width=4.9cm]{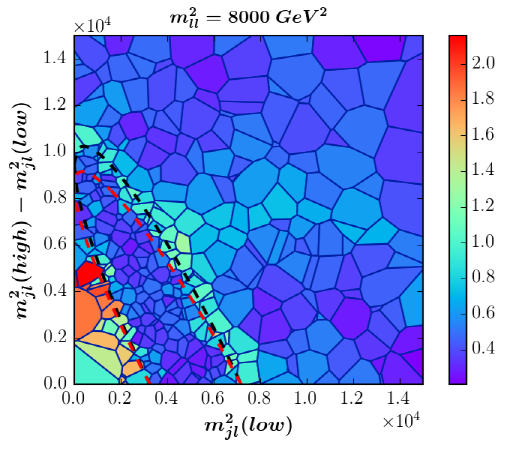} 
\includegraphics[width=4.9cm]{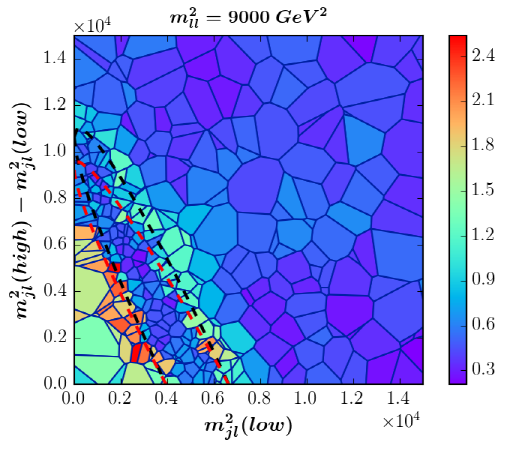} 
\caption{\label{fig:plots_ttbar} The analogue of Fig.~\ref{fig:susy_sigma} for the physics example considered in
section~\ref{sec:ttbarcomb}. We show nine slices at fixed values of $m_{\ell\ell}^2$ as indicated at the top of each panel.
The red (black) dashed line in each plot corresponds to the expected theoretical boundary implied by eq.~(\ref{eq:samosa2}) 
for the set of points with $m^2_{j\ell_n} \ge m^2_{j\ell_f}$ ($m^2_{j\ell_n} \le m^2_{j\ell_f}$) (see also the third row in Fig.~\ref{fig:sig_ttbar}).
}
\end{figure}

We are now in position to apply our Voronoi boundary detection algorithm. The result is shown in Fig.~\ref{fig:plots_ttbar},
where we present nine slices at fixed values of $m_{\ell\ell}^2$ as indicated at the top of each panel.
The red (black) dashed line in each plot corresponds to the expected theoretical boundary implied by eq.~(\ref{eq:samosa2}) 
for the set of points with $m^2_{j\ell_n} \ge m^2_{j\ell_f}$ ($m^2_{j\ell_n} \le m^2_{j\ell_f}$) 
(see also the third row in Fig.~\ref{fig:sig_ttbar}). As before, the signal and background events were normalized so that $\rho=4$.
Fig.~\ref{fig:plots_ttbar} demonstrates that the Voronoi cells with the highest values of RSD $\bar\sigma_i$ (colored in red or orange)
are indeed found near the theoretical boundaries (the red or black dashed lines). As anticipated, the method performs well for intermediate
values of $m^2_{\ell\ell}\sim(4,000-5,000)\ {\rm GeV}^2$, where the two boundaries resulting from the reordering (\ref{lowhigh}) tend to overlap.
We also observe that the boundary closer to the origin also seems to be singled out, especially at high values of  $m^2_{\ell\ell}$.

\section{Conclusions}
\label{sec:conclusions}

In this paper, we took the first steps towards developing a general method for identifying
surface boundaries in 3D phase space distributions from Voronoi tessellations.
In the case of a sequential cascade decay like the one exhibited in Fig.~\ref{fig:topology},
the surface boundary in the relevant $(m_{12}, m_{23}, m_{13})$ space is characterized 
by two properties: 1) it is the location of points where the number density is enhanced, due to 
the $\Delta_4^{-1/2}$ factor in the phase space distribution (\ref{eq:samosa}) \cite{Agrawal:2013uka};
2) it is the location of points where the number density suddenly changes 
due to the lack of signal events outside the kinematically allowed boundary.
These two properties motivate the use of the geometric variables,
$\bar\sigma_i$ and $\bar{v}_i$, derived from the Voronoi tessellation of the data.
(For other options, see \cite{Debnath:2015wra}.) We showed that the 
edge cells tend to have large values of $\bar\sigma_i$ and small values of $\bar{v}_i$,
thus we advocated empirically selected cuts in terms of $\bar{v}_i$ and $\bar\sigma_i$
for tagging edge cell candidates. We considered several examples of increasing complexity
and quantified the efficiency of those selection cuts using the language of ROC curves.

There are several directions in which this line of research can proceed from here.
\begin{itemize}
\item {\bf Statistical significance of a set of tagged edge cells.} As we have seen in Figs.~\ref{fig:susy_cut}
and \ref{fig:susy_cut_ave}, the method is not perfect, in the sense that it occasionally tags 
a few bulk cells. Therefore, we need to develop a statistical procedure for determining 
the statistical significance of a given observed set of tagged edge cell candidates.
Such a  procedure should involve not just the relative number of tagged cells, but also their 
correlations, e.g., proximity to each other, connectedness, etc. This is an interesting subject 
on its own and will be addressed in a future publication \cite{del4short}.
\item {\bf Parameter measurements from fitting to a set of tagged edge cells.} 
Having selected a set of edge cell candidates, one could imagine fitting 
to the theoretical prediction for the shape of the boundary surface (\ref{eq:samosa2}),
obtaining a measurement of the mass spectrum of the new particles $X_1, X_2, X_3, X_4$.
The actual fitting can be done in several different ways, which will also be investigated in \cite{del4short}.
\item {\bf Experimental effects.} In order to keep the discussion simple and to the point, 
in this paper we have ignored the experimental complications arising from finite particle widths, 
detector resolution, ISR jet combinatorics, fakes, etc. Our goal was to present the method as a proof of principle,
since the Voronoi approach to data analysis is still in its infancy. Once Voronoi-based 
methods have become more established and mature, it will become worthwhile to perform detailed 
and more realistic studies beyond parton level and with full detector simulation.
\end{itemize}

Here we have focused on cases where the number of signal events on the boundary is 
significant, leading to a ``step" function discontinuity in the total density 
of events as one moves across the boundary.
However, there are interesting examples of distributions where the number 
density is continuous, but exhibits a ``kink",
i.e., a discontinuity in its derivative (gradient)
~\cite{Han:2009ss,Han:2012nm,Han:2012nr}. In such cases, our methods 
can still be applied --- not directly on the initial data itself, but on a 
secondary data sample created by taking suitable derivatives. 
Indeed, while we have identified and studied a promising use of Voronoi tessellations
in the analysis of particle physics data, there are many exciting applications
yet to be developed.  We look forward with anticipation to the future development
and adoption of these novel and powerful methods.

\acknowledgments
We thank I.~Furic, M.~Klimek, and D.~Toback for useful discussions.
The research of the authors is supported by the National
Science Foundation under Grants No. PHY-1315983 and No. PHY-1316033 and
by the Department of Energy under Grants DE-SC0010296 and DE-SC0010504.
DK is in part supported by the Korean Research Foundation (KRF) through the CERN-Korea fellowship program, and also acknowledges support by the LHC-TI postdoctoral fellowship under grant
NSF-PHY-0969510.

\appendix

\section{ROC Curves and Sensitivity Measures}
\label{sec:ROC-and-ROLL}

ROC (receiver operating characteristic) curves are a useful tool for measuring 
the sensitivity of a variable\footnote{
Much of the information in this section can be found elsewhere~\cite{wiki:ROC},
however we present a unified and self-contained exposition of the main 
facts about ROC curves here for the convenience of the reader.
}.
\begin{figure}[t]
\begin{center}
\includegraphics[width=0.7\textwidth]{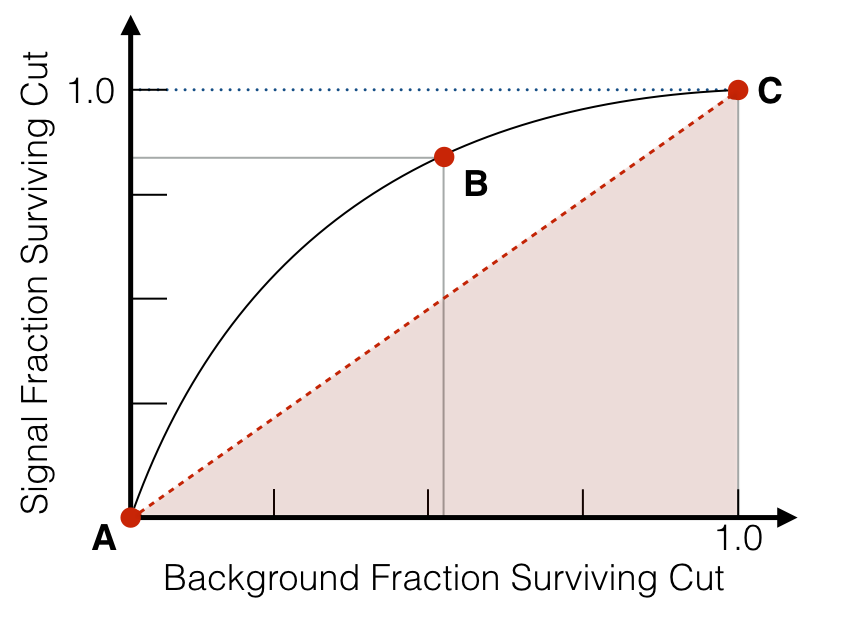}
\end{center}
\vspace{-20 pt}
\caption{Schematic illustrating the construction of the ROC curve.  Each point along the
curve indicates the fractions of signal and background events that pass a parameterizable cut.
If the cut parameter is chosen to disallow all signal and all background events,
we are at the point labelled ``A'', while if the cut parameter is chosen to allow signal and
background events, then the appropriate point is ``C''.  An intermediate point, where a certain
fraction of signal and background events are allowed is labelled by ``B''.}
\label{fig:roc-illustration}
\end{figure}
As shown in Fig.~\ref{fig:roc-illustration}, the ROC curve is found by
(i) defining an event selection procedure (``cut'') based on the variable
in question and (ii) determining the fraction, $\varepsilon_S$, of signal
events, and the fraction, $\varepsilon_B$, of background events that pass 
the given cut.
The coordinates of each point along the curve are then provided by 
$(\epsilon_B, \epsilon_S)$.
It is easy to see that the ROC curve must include the point $(0,0)$,
``A'' in Figs. \ref{fig:roc-illustration} and \ref{fig:gini-illustration},
where all events, signal and background, have been disallowed by the cut.
The point $(1,1)$, where all events pass the cut
(``C'' in Figs. \ref{fig:roc-illustration} and \ref{fig:gini-illustration}),
is also part of every ROC curve.  ROC curves have a number of important and
useful properties which we shall explore in the remainder of this section.

\begin{figure}[t]
\begin{center}
\includegraphics[width=0.7\textwidth]{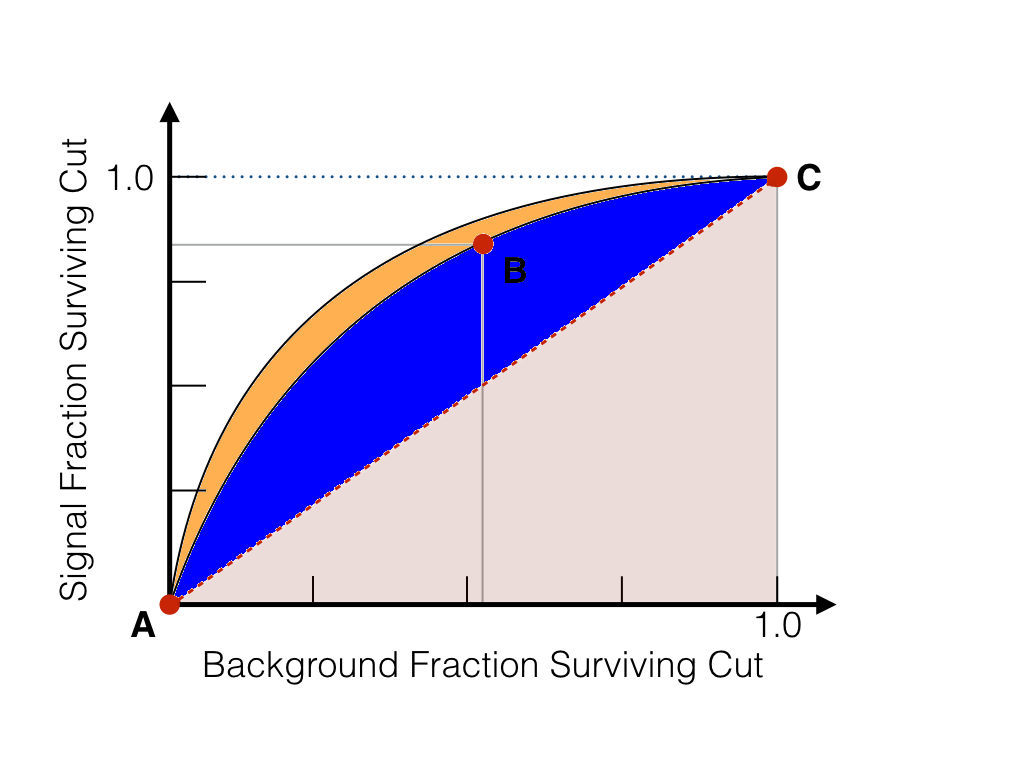}
\end{center}
\vspace{-20 pt}
\caption{Schematic illustrating the connection between the Gini coefficient (\ref{eq:Gini}) and sensitivity.
Points ``A'', ``B'', and ``C'' from Fig.~\ref{fig:roc-illustration} together with the
ROC curve from Fig.~\ref{fig:roc-illustration} are reproduced here.  We have also indicated
the ROC curve of a perfectly insensitive cut, which is shown by the line from point ``A'' to
point ``C''.  Finally we have shown a ROC curve for a cut which gives a higher signal fraction
for every choice of background fraction than the ROC curve from Fig.~\ref{fig:roc-illustration}.
We see that the area under the ROC curve of the perfectly insensitive variable (shown in pink)
is $1/2$, that there is additional area under the ROC curve shown in Fig.~\ref{fig:roc-illustration}
(shown in blue), and that there is even an even greater area under the ROC curve for the
more sensitive variable.
}
\label{fig:gini-illustration}
\end{figure}

\subsection{Comparing ROC curves}
\label{subsec:Comparing-ROC-curves}

If we consider two ROC curves, $R_1(\varepsilon_B)$ and $R_2(\varepsilon_B)$, 
obtained for the same
signal and background processes using different choices of variable and/or 
the cut procedure, then if
\begin{equation}
\label{eq:optimal}
R_1(\varepsilon_B) \geq R_2(\varepsilon_B)~\mathrm{for}~\mathrm{all}~
\varepsilon_B,
\end{equation}
the variable/ cut
combination used to produce $R_1$ is clearly more sensitive than the 
variable/cut combination used
to produce $R_2$.  This statement is uncontroversial, but the comparison 
is not always applicable, as a pair of ROC curves,
$R_1$ and $R_2$ may have points $\varepsilon_{B,1} \ne \varepsilon_{B,2}$ 
such that
\begin{equation}
R_1(\varepsilon_1) > R_2(\varepsilon_1),
\label{eq:}
\end{equation}
but
\begin{equation}
R_1(\varepsilon_2) < R_2(\varepsilon_2).
\label{eq:2}
\end{equation}
We will therefore need to develop other procedures to compare ROC curves; 
we will present several approaches in 
section~\ref{subsec:Measures-of-ROC-curve-sensitivity}.  
First, however, we must investigate the connection between ROC curves 
and likelihood and explore some of the more important consequences of 
this relationship.

\subsection{ROC curves and likelihood}
\label{subsec:ROC-curves-and-likelihood}

There are many ways to perform an analysis using a given variable, and hence
many ROC curves that may be constructed with no other information than the
value of a given variable for signal and background events.
We note that the ratio of signal and background likelihoods is an optimal
test statistic, i.e., choosing to accept an event, $e$, when
\begin{equation}
\label{eq:neyman}
\frac{L_S(e)}{L_B(e)} \geq l({\varepsilon_B}),
\end{equation}
where $L_S$ is the signal likelihood and $L_B$ is the background likelihood,
is the procedure which accepts the maximum fraction of signal events 
($\varepsilon_S)$ for a given choice of $\varepsilon_B$ (which in turn 
determines the numerical value of $l(\varepsilon_B)$ in eq.~(\ref{eq:neyman})).
This fact is known as the Neyman-Pearson lemma~\cite{neyman1992problem} 
and is equivalent to the statement that the likelihood ratio produces the 
optimal ROC curve.

We now provide a heuristic proof of this assertion in terms of ROC curves,
which hopefully provides useful insights into both ROC curves and the 
Neyman-Pearson lemma.  We consider the situation where signal and background 
likelihoods are calculated numerically from samples of signal
and background events, $S$ and $B$, which can be assumed arbitrarily 
large in order to approximate analytic expressions with any desired accuracy.
First, we create a set of ``variable bins'' in our variable
of interest, i.e.,
\begin{equation}
\label{eq:v-bins}
V_i = [v_{min, i}, v_{max,i}),
\end{equation}
such that for every event, $e$, in our signal and background samples,
\begin{equation}
v(e) \in [v_{min,i}, v_{max,i})
\end{equation}
for exactly one choice of $i$, where $v(e)$ refers to the value of the variable
under consideration obtained from the event, $e$.
We then divide our signal and background events into bins
using the values of the variable, $v$, under consideration.
I.e., we obtain ``signal bins''
\begin{equation}
\label{eq:signal-bins}
S_i = \{ e~|~v(e) \in V_i \}~\mathrm{for}~e \in S
\end{equation}
and ``background bins''
\begin{equation}
\label{eq:background-bins}
B_i = \{ e~|~v(e) \in V_i \}~\mathrm{for}~e \in B.
\end{equation}
Since every signal and every background event must be in some bin, by 
construction, we have
\begin{equation}
\label{eq:S-norm}
\sum_i |S_i|  = |S|
\end{equation}
and
\begin{equation}
\label{eq:B-norm}
\sum_i |B_i|  = |B|.
\end{equation}
Hence the estimators of the signal and background likelihood,
\begin{equation}
\label{eq:L_S}
L_{S,i} = L_S (v_i) \approx |S_i|/|S|~\mathrm{for}~v_i \in V_i
\end{equation}
and
\begin{equation}
\label{eq:L_B}
L_{B,i} = L_B (v_i) \approx |B_i|/|B|~\mathrm{for}~v_i \in V_i,
\end{equation}
are automatically normalized.
Further, we note that the contribution to the ROC curve from bin $i$ is a 
line segment from initial point
$(\varepsilon_{B,0},~\varepsilon_{S,0})$ to $(\varepsilon_{B,0} + |B_i|/|B|,
~\varepsilon_{S,0} + |S_i|/|S|)$,
i.e. from $(\varepsilon_{B,0},~\varepsilon_{S,0})$ to $(\varepsilon_{B,0} 
+ L_{B,i},
~\varepsilon_{S,0} + L_{S,i})$.
We therefore see from eqs.~(\ref{eq:L_S}) and (\ref{eq:L_B}) that the slope 
of the line segment corresponding to a particular bin is given by the ratio 
of signal and background likelihoods calculated for that bin.  
An interesting corollary, since $L_{S,i}$ and $L_{B,i}$ are always 
non-negative, is that the ROC curve must be monotonically increasing, i.e.,
\begin{equation}
\frac{d \varepsilon_S}{d \varepsilon_B} \geq 0.
\label{eq:ROC-derivative}
\end{equation}

We now assert that any physically reasonable cut procedure\footnote{We may 
have to choose different, and, in particular, smaller
bins, e.g., we may not be able to model the cut function $v > v_0$ if $v_0$ 
is not on a bin boundary.  In principle this limitaiton can be circumvented 
by choosing the variable bins to be smaller than the detector resolution, 
which must be finite.
Since physically reasonable cut functions can only use the output of such a 
detector, we can eliminate pathological cut functions, like only accepting 
events with rational $v$.
}
accepts a set of variable bins
\begin{equation}
V_A = { v \in V_i} ~\mathrm{for}~ i \in \{ i_1, i_2, ... \},
\end{equation}
signal events
\begin{equation}
S_A = { e \in S_i} ~\mathrm{for}~ i \in \{ i_1, i_2, ... \},
\end{equation}
and background events
\begin{equation}
B_A = { e \in B_i} ~\mathrm{for}~ i \in \{ i_1, i_2, ... \},
\end{equation}
where $\{ i_1, i_2, ... \}$ are the indices of the variable bins which pass 
the cut.  Further if we have a cut
that can be parameterized and that ranges from a cut in which all events are 
accepted to a cut in which
no events are accepted, then we can express the cut function as a permutation, 
$p$ of $1, 2, ..., n$, where $n$ is the number of bins, such that the 
variable bin labelled by $p(1)$ is the last to be eliminated,
the variable bin eliminated by $p(2)$ is the next-to-last to be 
eliminated, etc.

The optimal cut function identified by the Neyman-Pearson lemma, i.e., 
eq.~(\ref{eq:neyman}) corresponds to the
case where $p(1)$ labels a bin with the maximum value of signal to background 
likelihood ratio and
\begin{equation}
\frac{L_S(v_{p(i)})}{L_B(v_{p(i)})} \geq \frac{L_S(v_{p(j)})}{L_B(v_{p(j)})}
\end{equation}
if $i > j$.  We claim that this gives the ROC curve for which 
$\varepsilon_S(\varepsilon_B)$ is maximized
for any choice of $\varepsilon_B$.  This is geometrically obvious given that 
the likelihood ratio gives the slope of the line segment in the ROC curve; 
the optimal ROC curve is one in which the steepest segments occur first.  
A more rigourous proof notes that the ROC curve based on the likelihood 
ratio has
\begin{equation}
\label{eq:eps-S}
\varepsilon_S =  \sum_i^m \frac{L_{S,p(i)}}{L_{B,p(i)}} L_{B,i} + f_{m+1}
    \frac{L_{S,p(m+1)}}{L_{B,p(m+1)}} L_{B,(m+1)},
\end{equation}
where
\begin{equation}
\label{eq:eps-B}
\varepsilon_B = \sum_i^m L_{B,i} + f_{m+1} L_{B,(m+1)}.
\end{equation}
and $f_{m+1} \in [0,1]$ is the fraction of bin $m+1$ that must be included
to give the desired value of $\varepsilon_B$.

We now consider the value of $\varepsilon_S$ at the same value of 
$\varepsilon_B$ in a different ROC curve.  Since this ROC curve differs 
from the optimal ROC curve described by eqs.~(\ref{eq:eps-S}-\ref{eq:eps-B}), 
it must include some different bins or different fractions of bins.
These ``new bins'' give a contribution to $\varepsilon_B$ of $\Delta 
\varepsilon_B$, which is also the contribution to $\varepsilon_B$ from 
the ``replcaed original bins'' they replace.
The contribution to $\varepsilon_S$ from the replaced original bins was 
$\Delta \varepsilon_S$, and can be found by multiplying $\Delta 
\varepsilon_B$ by the (appropriately weighted) average slope
of the replaced original bins.  The contribution to $\varepsilon_S$ from 
the new bins is also the weighed average slope of the bins.  
However every bin in the new bins has a slope less than or equal
to the slope of any bin in the replaced original bins.  Thus the value 
of $\varepsilon_S$ in the new ROC curve is less than or equal to its 
value in the ROC curve based on the likelihood ratio.

\subsection{Subdividing variable bins}
\label{subsec:Subdividing-variable-bins}

We now consider the effect on our ROC curve of subdividing a bin, by 
which we mean that we take a bin, $V_i$, and break it into two bins, 
which we label $V_i$ and $V_{n+1}$ as follows:
\begin{equation}
\label{eq:breaking-V}
V_i = [v_{i,min}, v_{i,max}) \to [v_{i,min}, v_{i^\prime,max})~\mathrm{and}~
[v_{i^\prime, max}, v_{i,max}) = V_{i^\prime}~\mathrm{and}~
V_{i^{\prime\prime}}.
\end{equation}
We then obtain the signal bins $S_{i}$ and $S_{n+1}$ following
eq.~(\ref{eq:signal-bins}) and the background bins $B_i$ and $B_{n+1}$
following eq.~(\ref{eq:background-bins}).
Using these new bins will always allow for the construction of a ROC curve 
that is as good as or better than the curve before subdivision (in the 
sense of eq.~(\ref{eq:optimal})).  This is clear as if the permutation of 
bins that led to the original ROC curve is $p$, then the permutation
\begin{eqnarray}
p^\prime(j) = p(j) ~\mathrm{for}~ j < i \\
p^\prime(i) = i~(\mathrm{which}~\mathrm{now}~\mathrm{refers}~\mathrm{to}~
\mathrm{a}~\mathrm{fraction}~\mathrm{of}~\mathrm{the}~\mathrm{original}~
\mathrm{bin,}~i) \\
p^\prime(i+1) = i^\prime = n+1 \\
p^\prime(j) = p(j-1)~\mathrm{for}~ j > i+1
\end{eqnarray}
yields a ROC curve which is the same everywhere except in the region 
corresponding to the subdivided
bin.  If the two ``daughter'' bins have the same slope (likelihood ratio), 
then this curve will be exactly the same as the original ROC curve.  
However, if this is not the case, then if we choose (without loss of 
generality) the daughter bin with the greater slope to come 
first (i.e., to be labelled as $i$), we have a ROC curve which has strictly 
greater signal values than the original ROC curve in the regime corresponding 
to the subdivided bin.  
Finally, it is possible that daughter bin $i$ may have a greater slope than 
some of the other bins, that daughter bin $n+1$ may have a lesser slope 
than some of the other bins, or both of these 
possibilities may be realized.  
In all of these cases the ROC curve may be further imporoved by sorting the 
bins by slope.

As a consequence of this result considering smaller bins will increase 
sensitivity; the extreme limit is to use an unbinned likelihood rather 
than binned likelihood.  While we have not proved it,
this fact also motivates the (true) statement that considering additional 
relevant, but still independent, variables will improve sensitivity 
(as subdividing a bin using the new variable may yield daughter
bins with different slopes).

However, it should be noted, when using actual samples of signal and 
background events to construct likelihoods, that one can reach false 
conclusions about sensitivity by subdividing bins too far.
An extreme example of this is provided by the case where all the signal 
and background events we are considering have their own bin.  In this case, 
the likelihood ratio is infinite for bins with one signal event and zero 
for bins with one background event and the ROC curve is vertical from
$(0,0)$ to $(0,1)$, and then horizontal from $(0,1)$ to $(1,1)$.  
Obviously this is too good to be true.  What has happened is an example 
of posterior statistics and is clearly an invalid inference.  
Clearly such a situation should be avoided, e.g., by using bins large enough 
to be robust with respect to statistical fluctuations.

\subsection{Measures of ROC curve sensitivity}
\label{subsec:Measures-of-ROC-curve-sensitivity}

In section~\ref{subsec:Comparing-ROC-curves}, we noted that sometimes a 
ROC curve will indicate that one variable/ cut procedure is absolutely better 
than another variable/ cut procedure.  However, this may not always be 
the case.  Thus it is important to have procedures for comparing ROC curves 
(and hence sensitivity) that always allow the comparison to be made.
All of these procedures can be viewed as functionals, i.e., they describe 
a ROC curve by a number that corresponds directly to the quality of a variable.

Simple functionals that can be used are (i)
the value of $\varepsilon_S$ at a fixed value of $\varepsilon_B$ and
(ii) the value of $\varepsilon_B$ at a fixed value of $\varepsilon_S$.
We can view (i) as the fraction of signal events chosen with a fixed
false positive rate, while in (ii) we demand that the cut accept a fixed
fraction of signal events.  A better variable will then reject a larger
fraction of background events.

In particle physics, we are generally interested in the significance with
which we can claim the discovery (or the exclusion) of some process.
Therefore we can describe ROC curves by
(i) the maximum value of $S/\sqrt{B}$ attained on the ROC curve,
(ii) the maximum value of $S/B$ attained on the ROC curve, or
(iii) the maximum discovery significance on the ROC curve.
Choice (i) is appropriate as a measure of significance in the (Gaussian) limit
of a large number of events in situations where systematic uncertainties do not
have a large effect.  Choice (ii) is appropriate in the situation where
systematic errors dominate.  There are different appropaches to implementing
choice (iii) to model the statistical and systematic components of the 
significance.  The difference between the Gaussian and Poisson distributions 
for small numbers of events may also be modelled.  While choice (iii) 
represents the actual quantity to be maximized for a specifc analysis, 
it has the drawbacks of (i) being complicated and (ii) being
dependent on specific experimental conditions, such as the luminosity gathered,
systematic uncertainties, etc.

To make comparisons between ROC curves in a general way, we note that the 
worst possible likelihood-based ROC curve is the ROC curve that reflects 
throwing away signal and background events indiscriminately, i.e. a straight 
line from (``A'' to ``C'' in Figs. \ref{fig:roc-illustration} and 
\ref{fig:gini-illustration}).  On the other hand, a perfect variable 
would take us from ``A'' to $(0, 1)$ to ``C''.
This ROC curve contains the entire range of $\varepsilon_B$ and 
$\varepsilon_S$ underneath it, suggesting the use of the integral of 
the ROC curve as a sensitivity measure.
Further, as indicated in Fig.~\ref{fig:gini-illustration},
variables which lead to ``better'' ROC curves, in the sense of 
eq.~(\ref{eq:optimal}) in section~\ref{subsec:Comparing-ROC-curves}
have more area under their ROC curves.  In order for the worst possible 
likelihood ROC curve to have a value of $0$ and the best possible ROC curve 
to have a value of $1$, we multiply the integral under the ROC curve by $2$ 
and subtract $1$.  The resulting expression,
\begin{equation}
G_1\equiv 2\, {\rm AUROC} - 1 = 2\int_0^1 d\varepsilon_B \times 
\varepsilon_S(\varepsilon_B) -1,
\label{eq:Gini}
\end{equation}
where AUROC referes to the ``area under the ROC curve'', e.g., under 
curve $ABC$ in Figs.~\ref{fig:roc-illustration} and~\ref{fig:gini-illustration}
gives the measure of sensitivity known as the Gini coefficient; 
it is our main quantifier of ROC curves (and hence of variable/ 
cut procedure sensitivity) in the bulk of this work.

\end{document}